\documentclass[twocolumn,trackchanges]{aastex7}
\usepackage{amsmath}
\usepackage{CJKutf8}

\begin{document}
\begin{CJK*}{UTF8}{gbsn}

\title{How well is the local Large Scale Structure of the Universe known? \\ CosmicFlows vs. Biteau's Galaxy Catalog with Cloning}

\author[0009-0005-4065-9744]{Yifei Li (李宜霏)}
\affiliation{Department of Physics, New York University Shanghai, Shanghai 200122, China}
\email{yl10327@nyu.edu}

\author[0000-0003-2417-5975]{Glennys R. Farrar}
\affiliation{Center for Cosmology and Particle Physics, New York University, New York, NY 10003, USA
} 
\email{gf25@nyu.edu}

\begin{abstract}
Knowledge of the actual density distribution of matter in the local universe is needed for a variety of purposes -- for instance, as a baseline model for ultrahigh energy cosmic ray sources in the continuum limit and for predicting the diffuse Dark Matter annihilation signal.  Determining the local mass density and velocity distribution is the aim of the CosmicFlows project.  An alternate approach is based on catalogs of galaxies, supplemented with some scheme for filling in for unseen galaxies. Here, we compare the density field proposed by~\citet{Biteau21} with the quasi-linear density field of CosmicFlows2~\citep{2018NatAs...2..792H} and the mean posterior field of CosmicFlows4~\citep{Valade2026inprep}.  We find factor-two level differences in some regions and even greater in regions toward the Galactic center zone of avoidance (ZoA) ($|\ell|<30^\circ, |b|<20^\circ$) as filled by Biteau using ``cloning". Within 11 Mpc the density field is well-determined by the Local Volume catalog~\citep{karachentsev_morphological_2018} (which Biteau directly incorporates) and is to be preferred over the CosmicFlows modeling.  At larger distances,~\citep{Biteau21} is also to be preferred over CosmicFlows for the direction and integrated mass of non-obscured structures, but the radial distribution of mass in~\citep{Biteau21} is less robust due to line-of-sight peculiar velocities.  Within the ZoA the ``galaxies" of~\citep{Biteau21} are entirely fictitious and thus their use is ill-advised except in circumstances where they do not contribute to the observable of interest anyway.  Moreover, as evidenced by the South Pole Wall which was discovered by CosmicFlows but is not found in the galaxy catalogs, obscuration can be a problem even outside the ZoA.  Unexpectedly, we find that the angular positions of structures in CosmicFlows are sometimes not correctly aligned with their true positions as manifest in the galaxy catalog outside the ZoA. 

\end{abstract}

\keywords{\uat{Cosmology}{343} --- \uat{Large-scale structure of the universe}{902} --- \uat{Galaxies}{573} --- \uat{Cosmic rays}{329} }

\section{Introduction} \label{sec:intro}
Recent works have shown that the direction and magnitude of the dipole anisotropy of ultrahigh energy cosmic rays (UHECRs) can be explained by UHECR sources which trace the Large Scale Structure (LSS) of the matter distribution as given by the CosmicFlows2 density field~\citep{Globus+19,DGF21,BF24}.  Interestingly, subsequent work~\citep{bfu24} showed that the dipole magnitude -- and not just its direction -- is a sensitive probe of the coherent magnetic field of the Galaxy, if the UHECR source distribution within $\sim 200$ Mpc is known.  When both the local baryon density and the Galactic magnetic field (GMF) are sufficiently well determined, the magnitude of the UHECR dipole will enable deviations due to individual sources to be robustly identified~\footnote{The recent work of~\citet{Allard+2026} explicitly demonstrates the sensitivity to source and GMF models.}.  We also note a significant literature on the local structure for cosmological applications, an incomplete listing of which includes~\citet{2010MNRAS.406.1007L, 2022PhRvD.106j3526B,2025MNRAS.540..716M,2026arXiv260117125W,2026MNRAS.545f1960S}. These and other applications motivate a careful comparison of present models of the local matter distribution. 

Here, for the first time, we provide a quantitative comparison of the density fields derived by directly using catalogs of galaxies with distance estimates \citep{Biteau21}, versus densities inferred from a combined analysis of the observed galaxies' redshift-independent positions and peculiar velocities as carried out by the CosmicFlows collaboration \citep{Tully2008, Tully2013, Tully2016, Tully2023}. The strength of the CosmicFlows approach is that it self-consistently exploits the information inherent in the relative motions of galaxies and hence is less sensitive to individual galaxy distance determinations and in principle provides constraints on matter concentrations and voids in directions where galaxies are obscured by dust.  On the other hand, the direction of galaxies is perfectly known so where obscuration is not a problem galaxy catalogs give better angular information.

Early methods for reconstructing the mass density field from peculiar velocities were developed in the linear regime, starting with the POTENT algorithm \citep{1990ApJ...364..349D}, followed by Bayesian approaches such as the Wiener Filter \citep{Zaroubi1995} and constrained realizations \citep{Hoffman1991}. These techniques were later extended to the quasi-linear regime to produce the CosmicFlows-2 (CF2) density fields \citep{2018NatAs...2..792H}, which provided data for approximately 8,000 galaxies \citep{Tully2013}. In this work, we utilize the significantly expanded CosmicFlows-4 (CF4) catalog, which increases the sample size to 55,877 galaxies \citep{Tully2023}. For the latest CosmicFlows-4 (CF4) reconstruction, we adopt the mean posterior density field derived by \citet{Valade2026inprep}, in preparation. This field was generated using the HAMLET algorithm \citep{Valade2022}, which employs Hamiltonian Monte Carlo sampling to iteratively update initial conditions to match the observed $z=0$ velocity field. Here, we adopt the mean density fields of the simulations in real space to plot the CF4 maps. 

 Another approach to determining the local matter distribution is to look directly at galaxies as tracers of the matter distribution. For this, having good distance measures that are not distorted by peculiar velocities is critical; these have been provided for a varying fraction of galaxies in catalogs such as the Local Volume sample \citep{karachentsev_morphological_2018}, the Extragalactic Distance Database \citep{Tully2016EDD} and the NASA/IPAC extragalactic database - Redshift-Independent Distances \citep{Steer2017NEDD}.  Additional challenges are that 1) galaxy catalogs are necessarily flux-limited, so that the sampling of structure becomes very sparse at large distances, and 2) dust obscuration in the Galactic plane leads to a ``zone of avoidance", where galaxy surveys are not complete. 
If the triple challenges of accurate distance determination, the zone of avoidance, and sparse sampling at large distance can be overcome, using the direct information on the mass distribution from observed galaxies should give an inherently better-resolved, higher-fidelity density map.   

\citet{Biteau21} tackled these challenges and produced a semi-empirical mass catalog, designated B21 below. It provides a selected sample of 489,000 galaxies within 350 Mpc. Each galaxy is assigned a mass-upscaling factor, calculated based on its luminosity as inferred from its flux, distance and latitude, using the well-measured luminosity function of galaxies.  This procedure should do a good job recovering galaxy clusters at large distances, which typically contain multiple bright galaxies, but misses secondary structures such as filaments composed of less-luminous galaxies.  Depending on their scale and the application at hand, such secondary structures may or may not be important.  
More problematic is how to treat the unseen galaxies in the zone of avoidance where dust obscuration must be taken into account.  \citet{Biteau21} takes the zone of avoidance to be $|b|<20^\circ$ for $|\ell|< 30^\circ$ and $|b|< 3^\circ$ for $30^\circ< |\ell|<180^\circ$ and addresses obscuration by replacing observed galaxies in this region by ``cloning" galaxies above and below the region, following~\cite{Lavaux2016}.  Specifically, he mirrors galaxies across the ZoA boundary at $|b|=b_{\mathrm{ZoA}}$:
\begin{equation}
    \sin b_{\mathrm{clone}} = \sin b_{\mathrm{ZoA}} - (\sin b-\sin b_{\mathrm{ZoA}}),
\end{equation}
maintaining the galaxies' other properties such as distance, mass and luminosity. This ensures continuity in galaxy density at the $b = \pm \,b_{\mathrm{ZoA}}$ boundaries (but not at $b=0$ or on the sides of the ZoA at $\ell=\pm 30^\circ$).  Unfortunately, as we shall see in Sec.~\ref{sec:result}, the cloning procedure creates artificial large scale structure where none may exist.  
Thus, B21 cannot {\it a priori} be expected to be a reliable mass-density model in ZoA directions, except within 11 Mpc where ~\citep{Biteau21} just uses the Local Volume catalog~\citep{karachentsev_morphological_2018} which is much more complete including in the ZoA. 

This paper is organized as follows. 
Section \ref{sec:methods} details the smoothing method we apply to the galaxies in Biteau's catalog for comparison to CF2 and CF4. Section \ref{sec:result} reports the solid-angle-averaged mean densities as a function of distance and provides skymaps of the density fluctuations in shells of different distances, to exemplify the differences and similarities between catalogs. 
Section \ref{sec:Discrepancies} explores possible origins of the differences we have uncovered. A summary and conclusions are in Sec.~\ref{sec:discussion}.

\section{Galaxy smoothing method} \label{sec:methods}
We divide 3D space into spherical shells centered on the observer, each with fixed radial width $\Delta d \approx 1.43$ Mpc. On each shell, we use \textsc{Healpix} at NSIDE = 32 to divide the sky into equal-area pixels of approximately 3.36 deg$^2$. Each voxel is therefore identified by its shell index and its \textsc{Healpix} pixel index, providing a common basis for both CF and Biteau maps.

For CosmicFlows2 we use the voxelization as provided by N.~Globus and used in~\citet{DGF21,BF24,bfu24}.  For CosmicFlows4, the density field is given on a $512^3$ Cartesian grid. To compare with the shell-voxel representation, we assign to each voxel the density of the nearest CF4 grid point. 
Since the CF4 grid spacing ($\sim1.40$~Mpc) closely matches the shell width ($\sim1.43$~Mpc), this simple one-to-one mapping preserves the CF4 dynamic range and provides consistent skymaps.  

To enable comparison with the Cosmic Flows density fields, we smooth Biteau’s galaxy catalog as follows. 
For technical reasons we factorize the smoothing into radial and angular directions.  Beyond the nearest distance shells, a galaxy at distance $r_0$ is modeled by a Gaussian density profile,
\begin{equation}
f(r)=\frac{1}{\sqrt{2\pi\sigma^2}}
\exp\!\left[-\frac{(r-r_0)^2}{2\sigma^2}\right],
\end{equation}
where $r$ is the distance from the galaxy, $\sigma$ is the standard deviation, same as the physical smoothing scale $R_s$. We consider $R_s=5$ and $10$ Mpc, the same values used by Biteau in his 3D visiualization \citep{Biteau21}. The fraction of a galaxy’s mass contributing to a given shell is obtained by integrating $f(r)$ over the shell's thickness. This fraction of the galaxy mass is assigned to the radial voxel in the Healpix pixel corresponding to the galaxy’s position.

Angular smoothing for each galaxy is applied independently on each shell using the public Python code \texttt{hp.sphtfunc.smoothing} and set the standard deviation $\sigma$ as angular smoothing radius $\theta$. We set 
\begin{equation}
    \theta=\frac{\sqrt{R_s^2 - |d_\mathrm{galaxy}-d_\mathrm{shell}|^2}}{d_\mathrm{shell}}
\end{equation}
for a galaxy at distance $d_\mathrm{galaxy}$ and shell whose center is at the distance $d_\mathrm{shell}$. This expression holds provided that the term within the square root is positive; if not, $\theta$ is set to be zero. 

When the distance of the galaxy is smaller than the smoothing scale, radial smoothing redistributes galaxy mass to $d < 0$. 
Since we set $R_s=5$ or $10$ Mpc, this error is significant mainly within $10$ Mpc. To fix this, we adopt a 2D nested pixel smoothing scheme for 0-10 Mpc, which only smooths the galaxies along angular directions. In each step, the mass in a pixel is evenly redistributed among the pixel itself and its eight neighboring pixels, and this redistribution is applied twice, with the output of the first application serving as the input to the second. As a result, the mass initially contained in a single pixel is spread over a two-ring neighborhood of 33 pixels, corresponding to an effective angular smoothing of 10.53 degrees wide. This method is used for the rightmost $0-20$ Mpc skymap in Fig.~\ref{skymap}.

\section{Comparison of the mass distributions} \label{sec:result}
In this section we analyze and compare the density fields constructed from CosmicFlows2, CosmicFlows4 and Biteau's catalog.\footnote{The ``mother" galaxy catalog developed by~\citet{Allard_2022} is not to our knowledge publicly available and is not described in sufficient detail for us to include it in our comparison.}  First we familiarize ourselves with the magnitude and character of the differences by averaging over angles and then we consider skyplots of the various density fields in fixed distance shells.  

\subsection{Average Mass Density as a function of distance}
\label{subsec:ave}
\begin{figure}[ht!]
\includegraphics[width=0.47\textwidth]{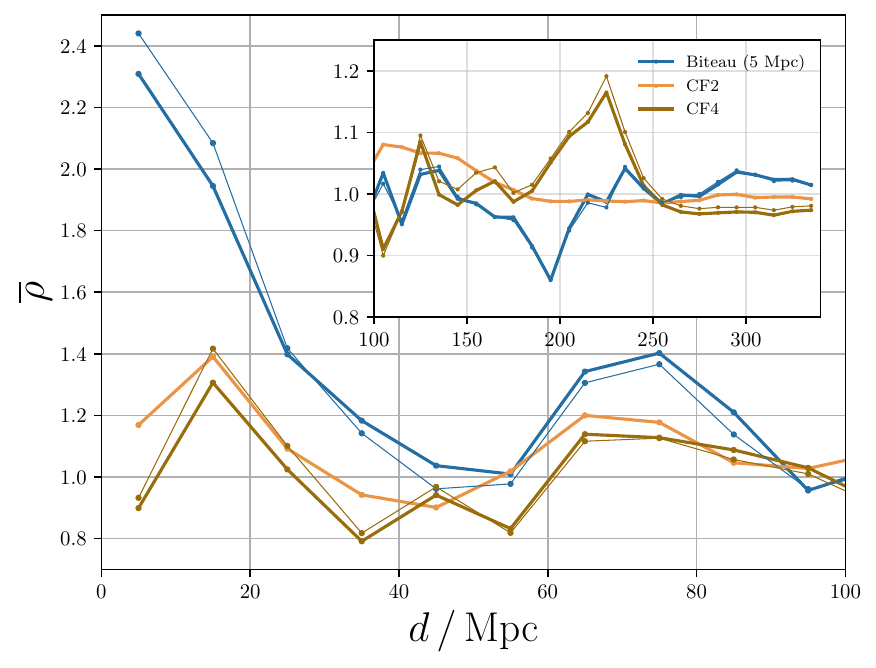}
\caption{Mean over-density in 10-Mpc thick spherical shells centered at distance $d$, 
derived from CosmicFlows2 (gold), CosmicFlows4 (brown) and the Biteau Catalog (blue), with the main plots extending from $0-100$ Mpc and the inset covering the range $100-350$ Mpc. The thin lines show the same, but excluding the Zone of Avoidance (not inluding CF2, for legibility). }
\label{ave}
\end{figure}

Figure~\ref{ave} compares the mean over-density in 10-Mpc thick shells (7 shells) as a function of radius, for the three density models. The thin lines show the same, but excluding the Zone of Avoidance.  
Each curve is normalized by its global mean over $0-350$ Mpc.

All three catalogs have a peak in the density near 70 Mpc, from the concentrated mass of the Great Attractor, but there are substantial quantitative differences between the different methods.  There are also  significant discrepancies at large distance.  Excluding the ZoA (thin lines in Fig.~\ref{ave}) shows that cloning has a visible impact on the density excursion in some radial shells, even averaged over solid angle.  

A major discovery of the Cosmic Flows project is the South Polar Wall~\citep{PomaredeTully20} which is responsible for the difference between CF2 and CF4 in the $\approx 180-250$ Mpc region; the deficit in galaxy catalogs in that direction is due in part to obscuration by foreground gas clouds, according to~\cite{PomaredeTully20}. The systematic excess with respect to the the cosmic mean density for B21 seen in Fig.~\ref{ave}, from about 270 Mpc to 340 Mpc, is due at least in part to cloning, as can be seen in Fig.~\ref{skymapratio} in the Appendix.    

\subsection{Skymaps} \label{subsec:skymaps}
\begin{figure*}[ht!]
\includegraphics[width=0.245\textwidth]{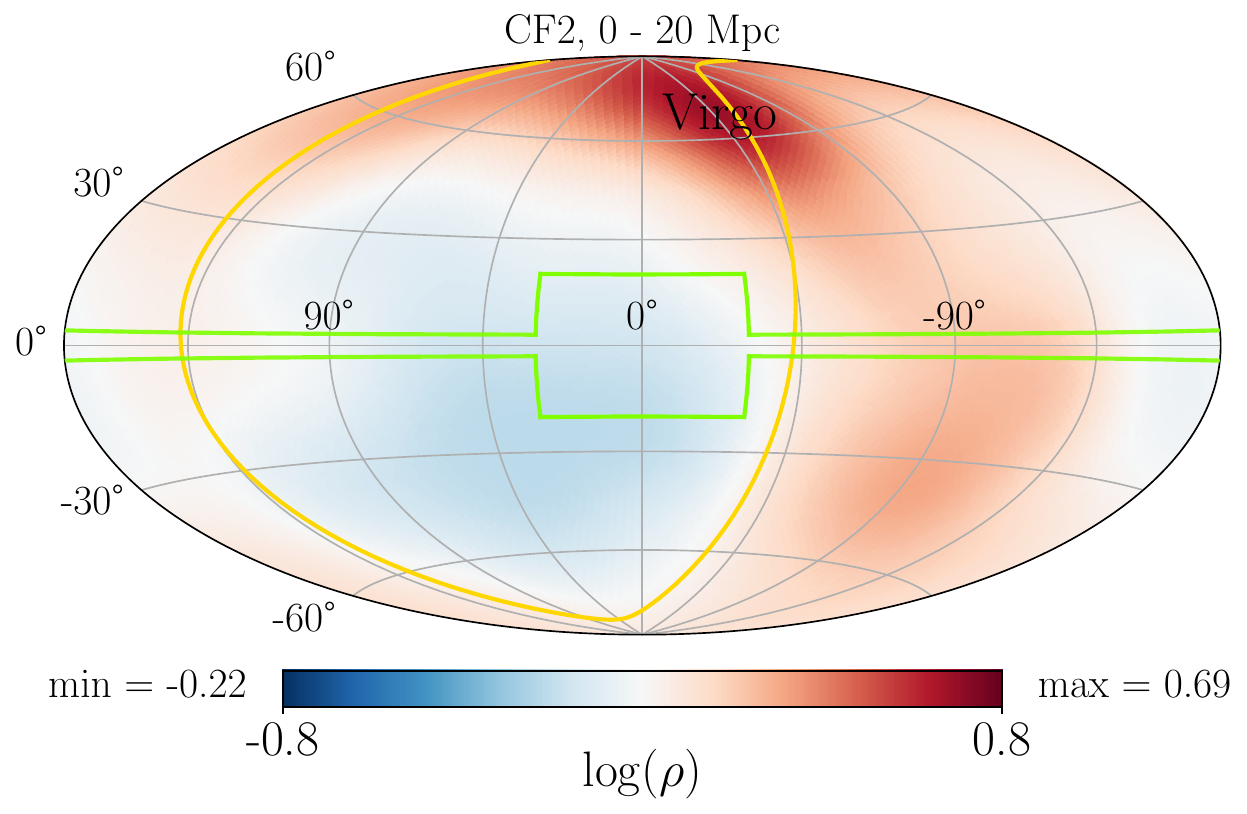}
\includegraphics[width=0.245\textwidth]{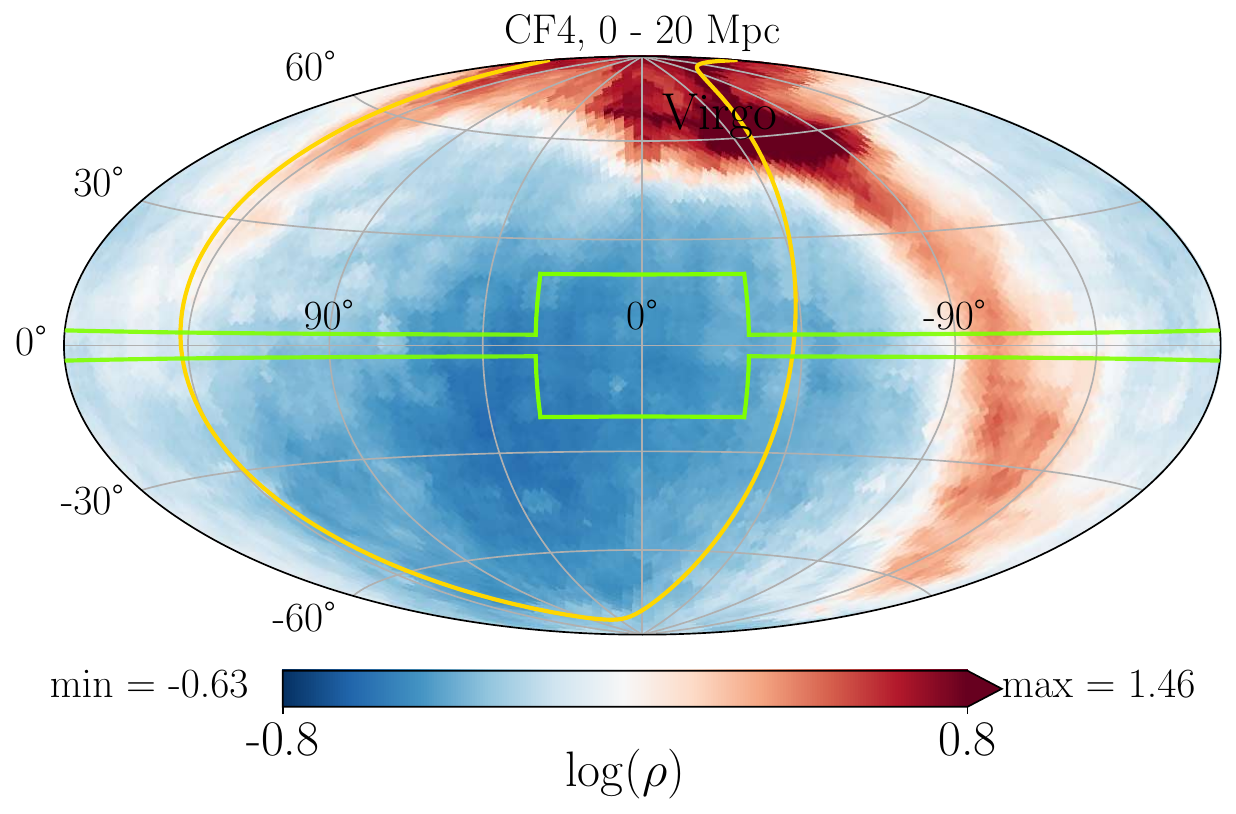}
\includegraphics[width=0.245\textwidth]{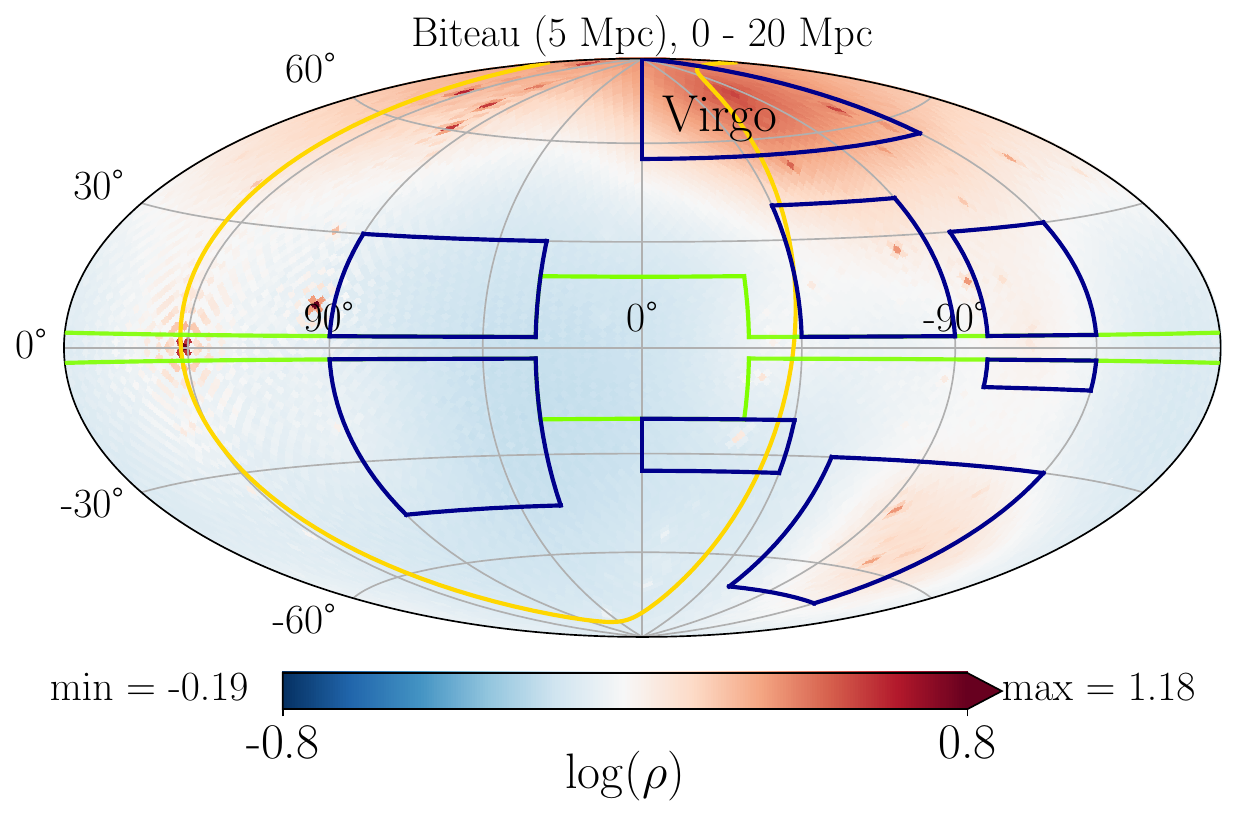}
\includegraphics[width=0.245\textwidth]{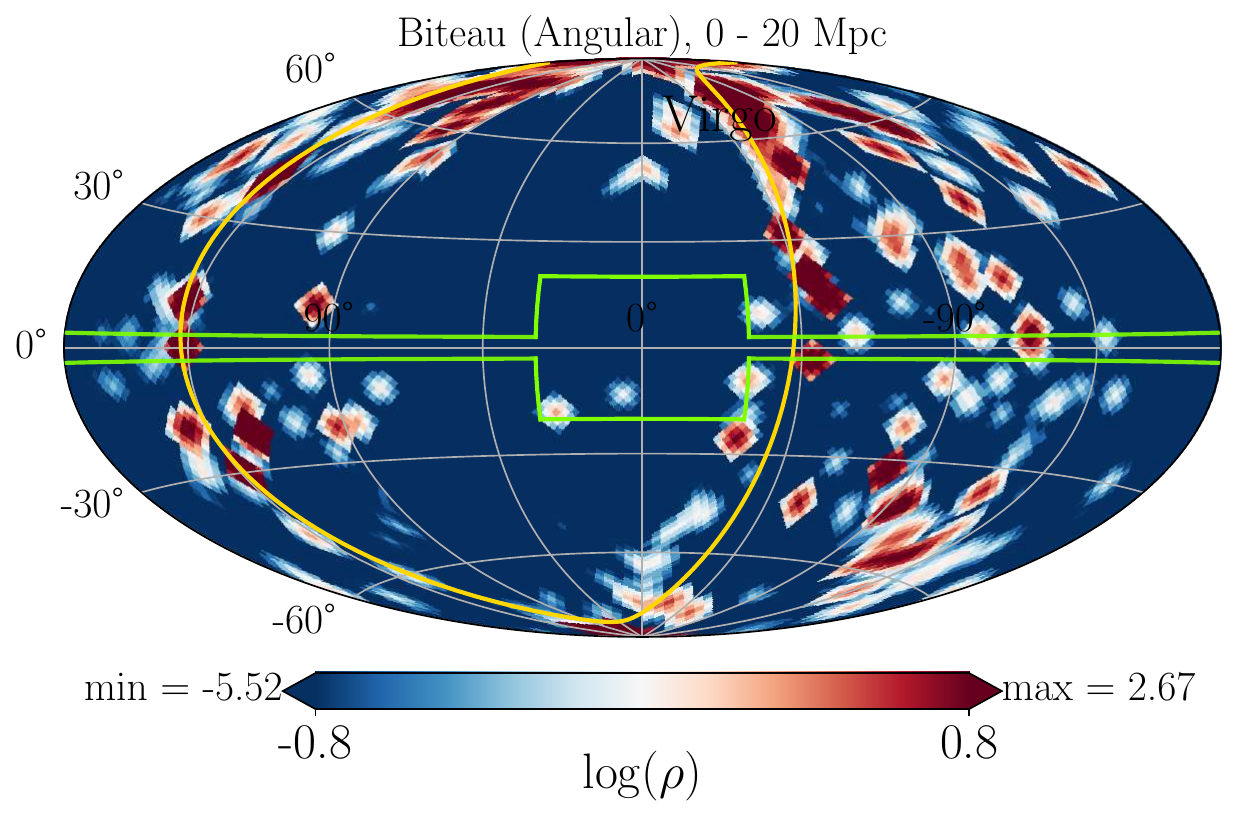}

\includegraphics[width=0.245\textwidth]{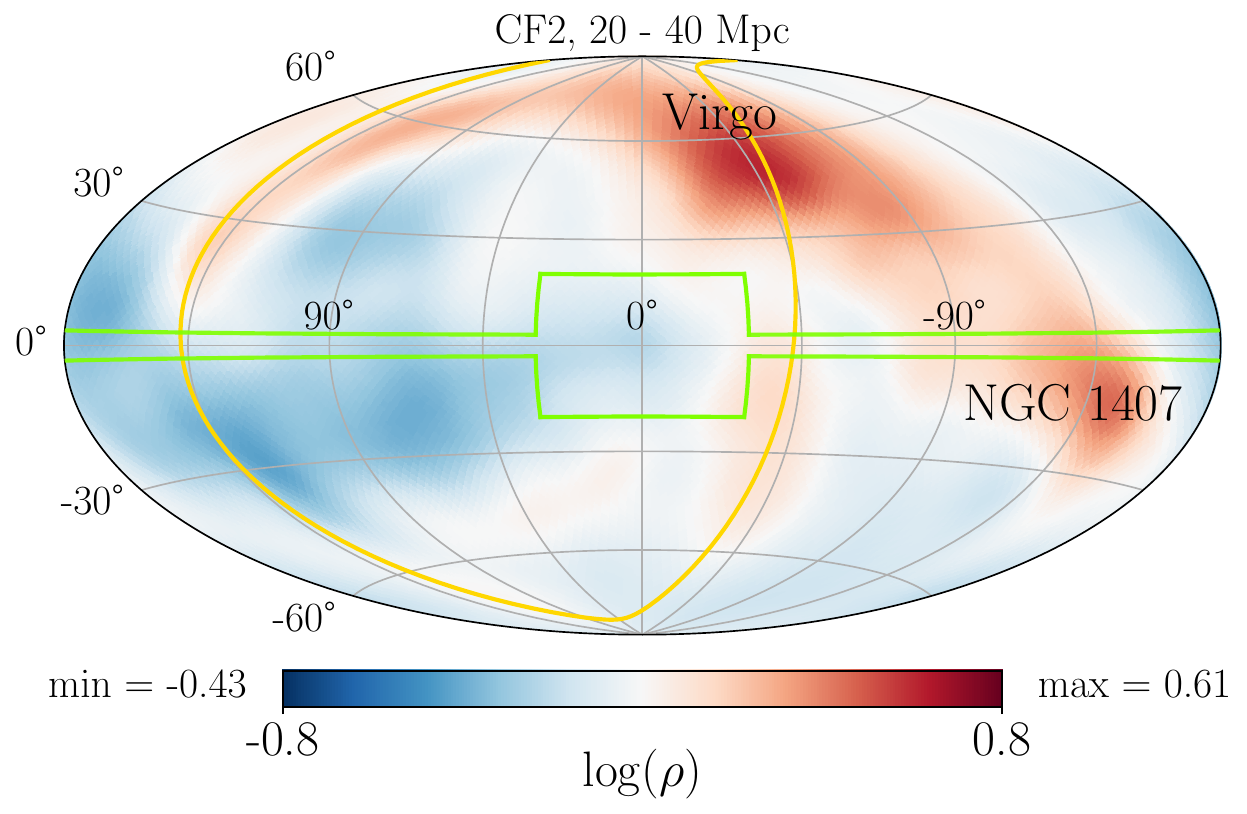}
\includegraphics[width=0.245\textwidth]{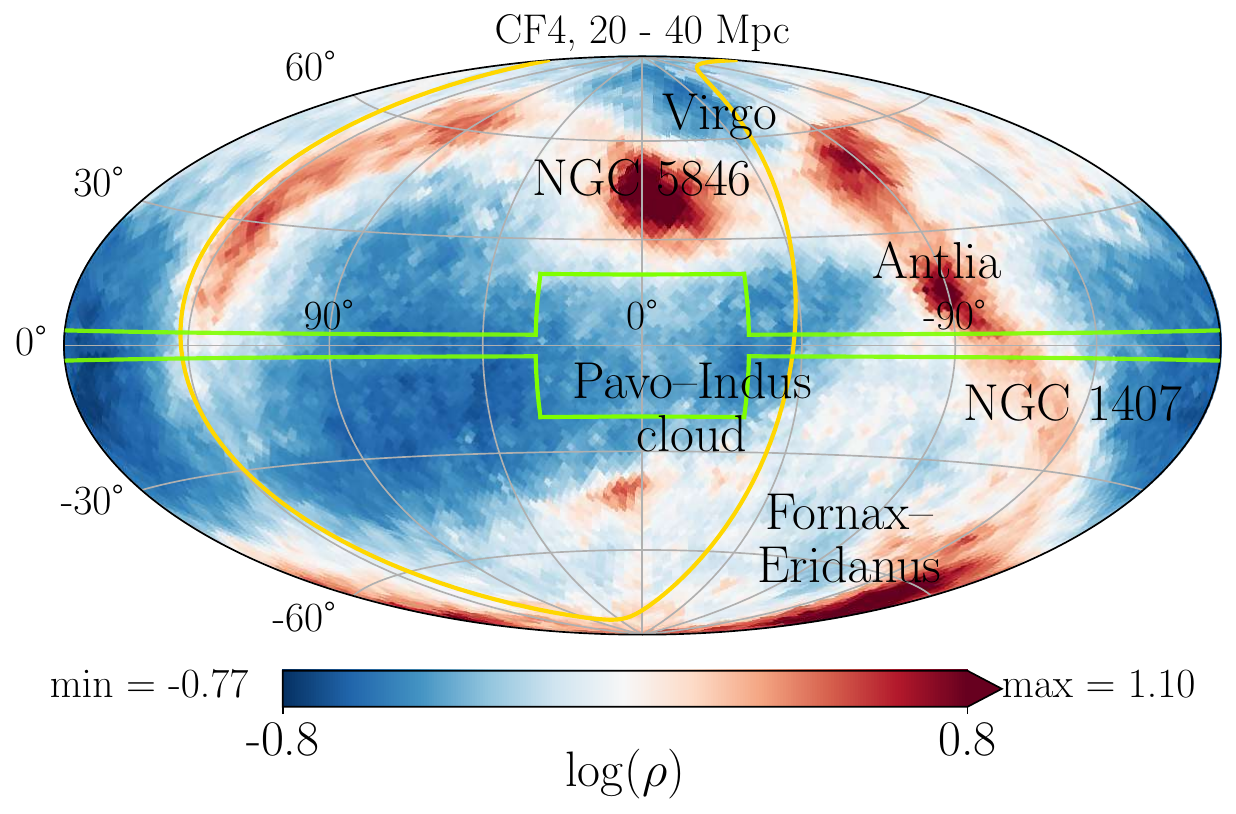}
\includegraphics[width=0.245\textwidth]{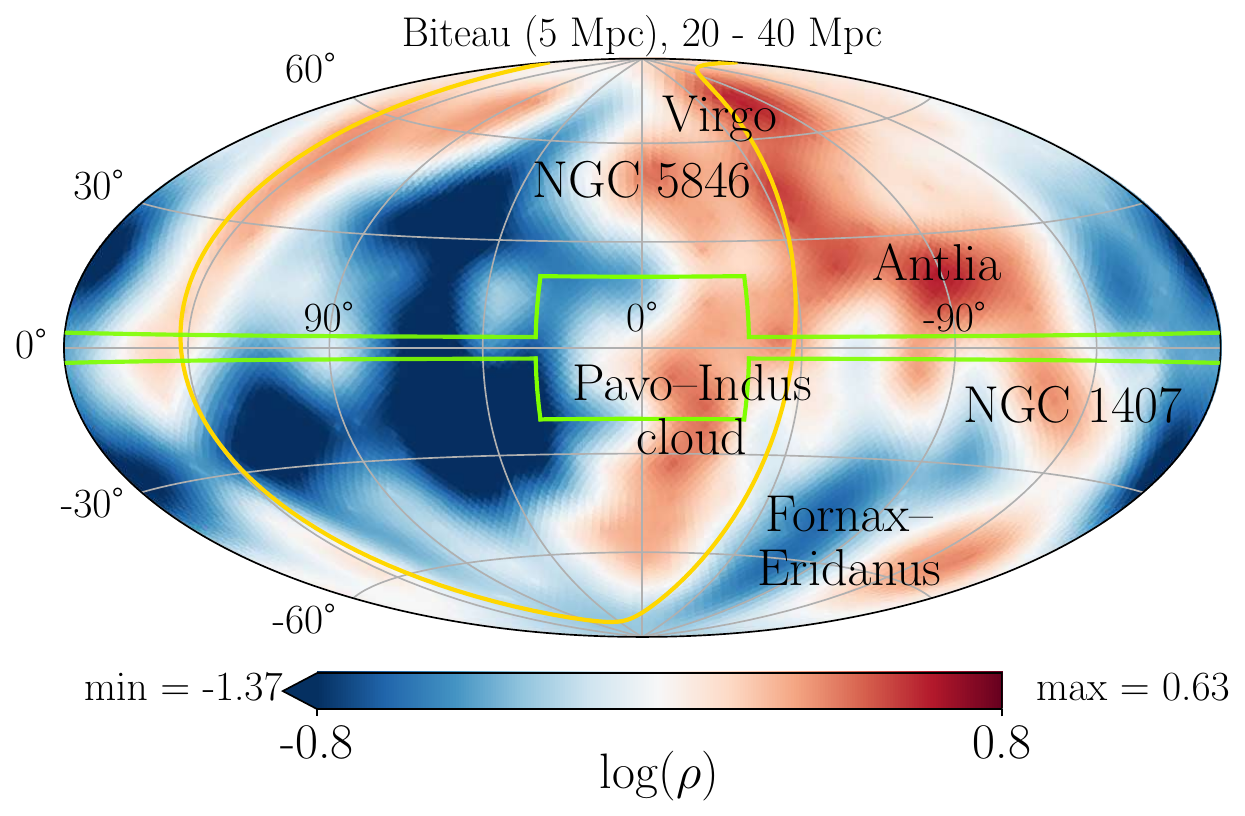}
\includegraphics[width=0.245\textwidth]{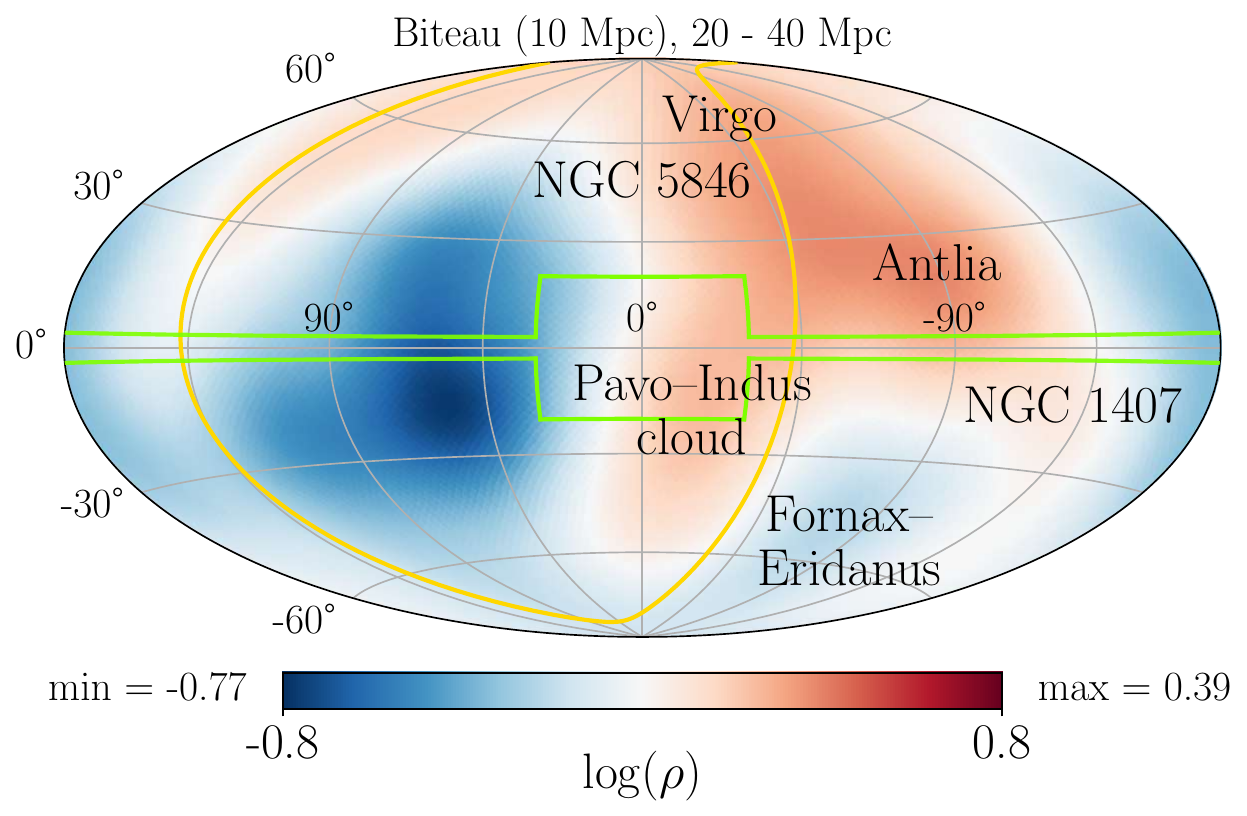}

\includegraphics[width=0.245\textwidth]{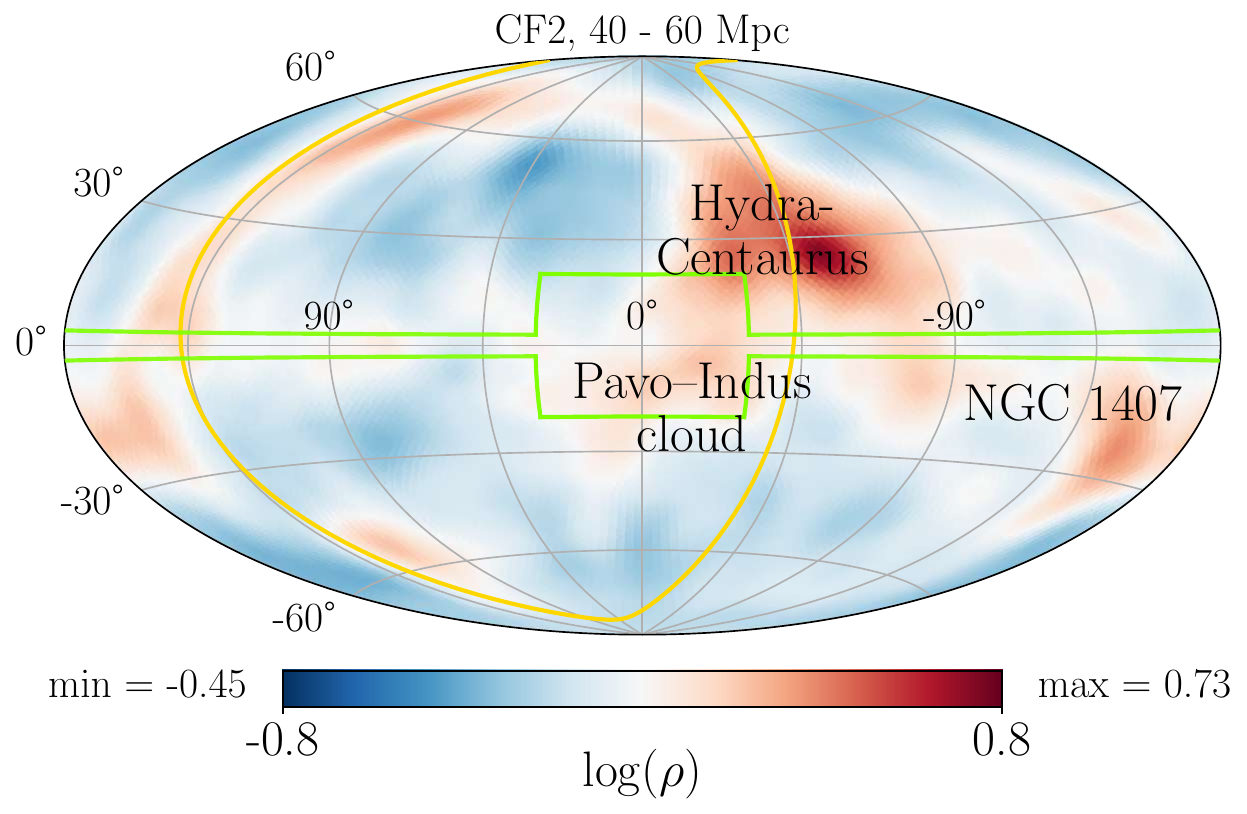}
\includegraphics[width=0.245\textwidth]{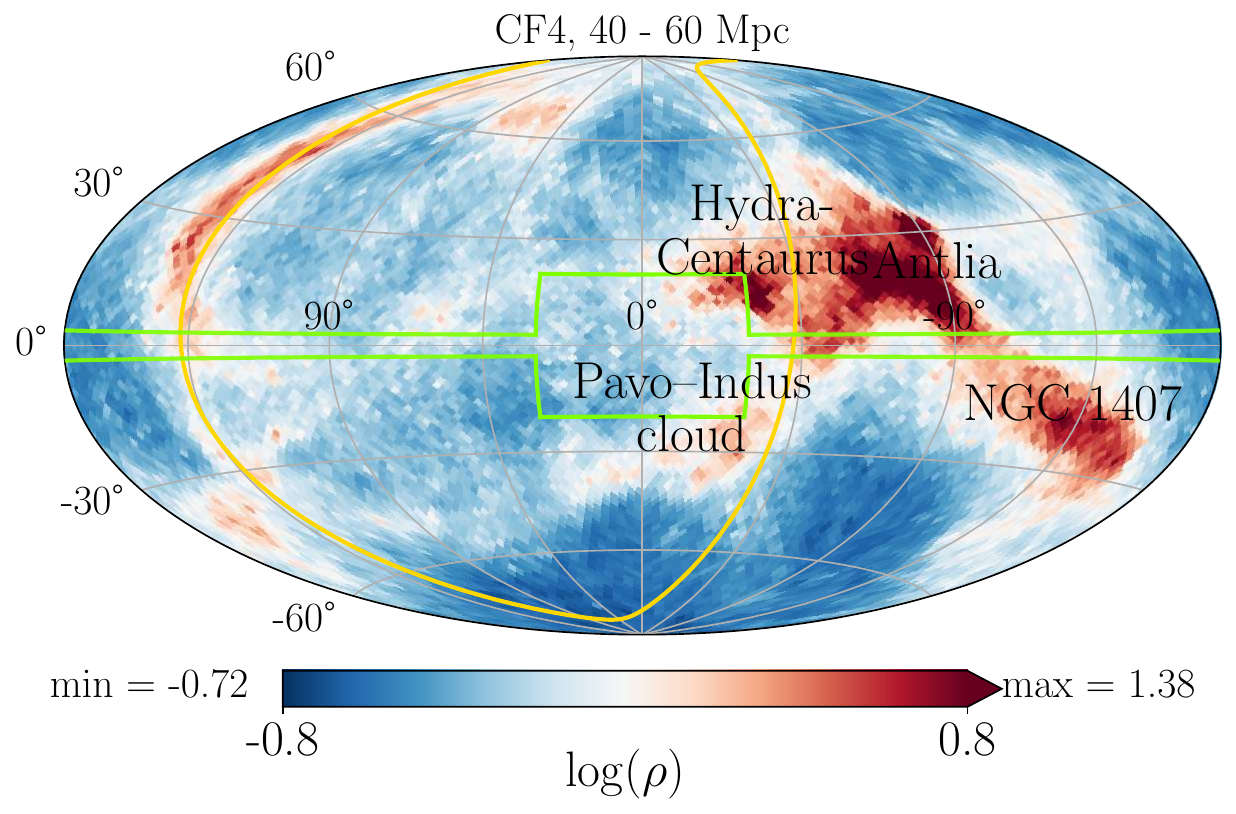}
\includegraphics[width=0.245\textwidth]{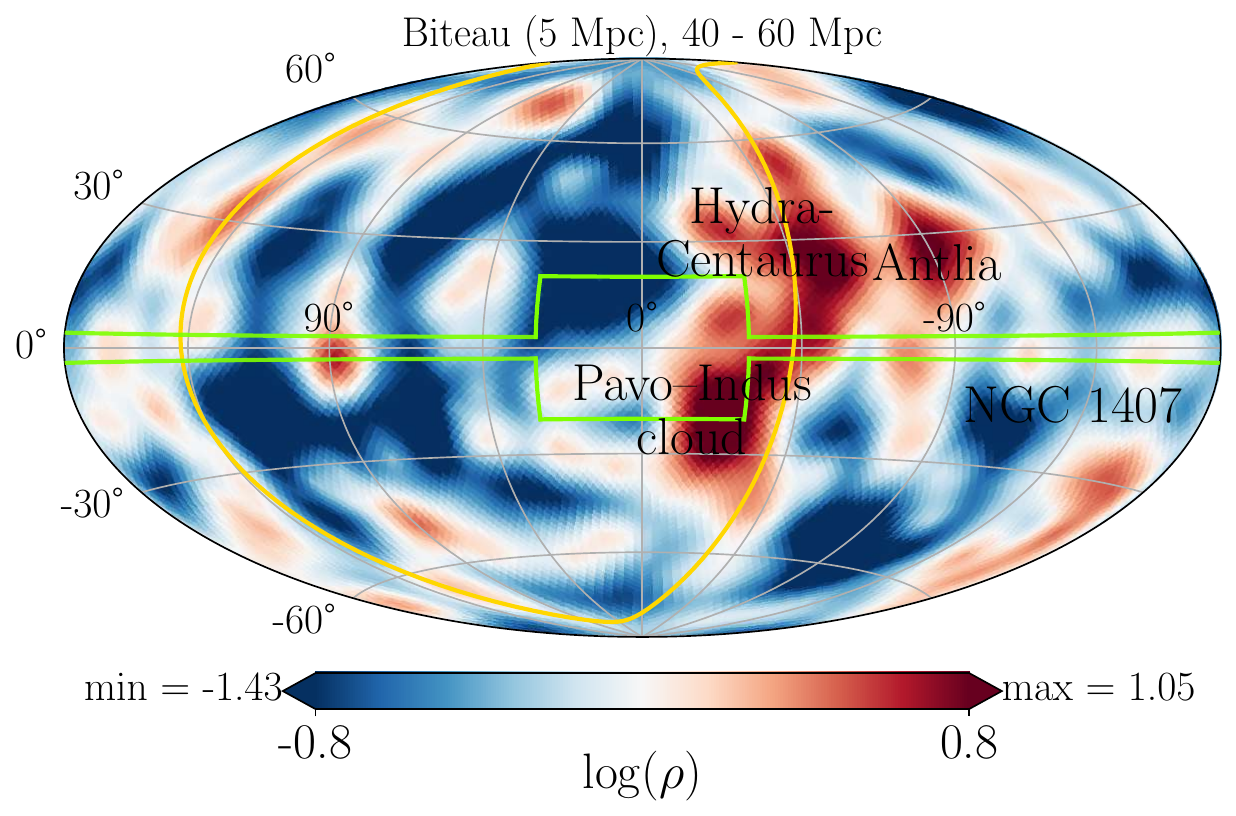}
\includegraphics[width=0.245\textwidth]{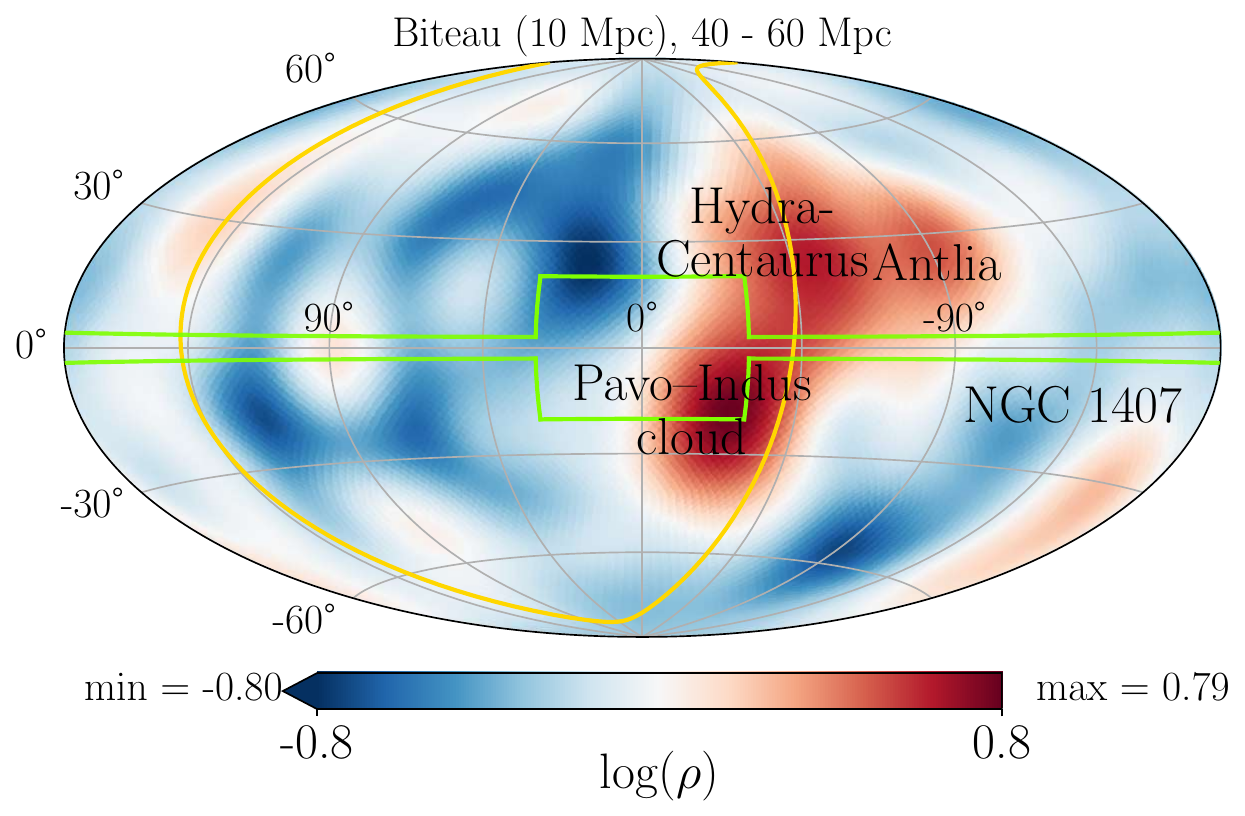}

\includegraphics[width=0.245\textwidth]{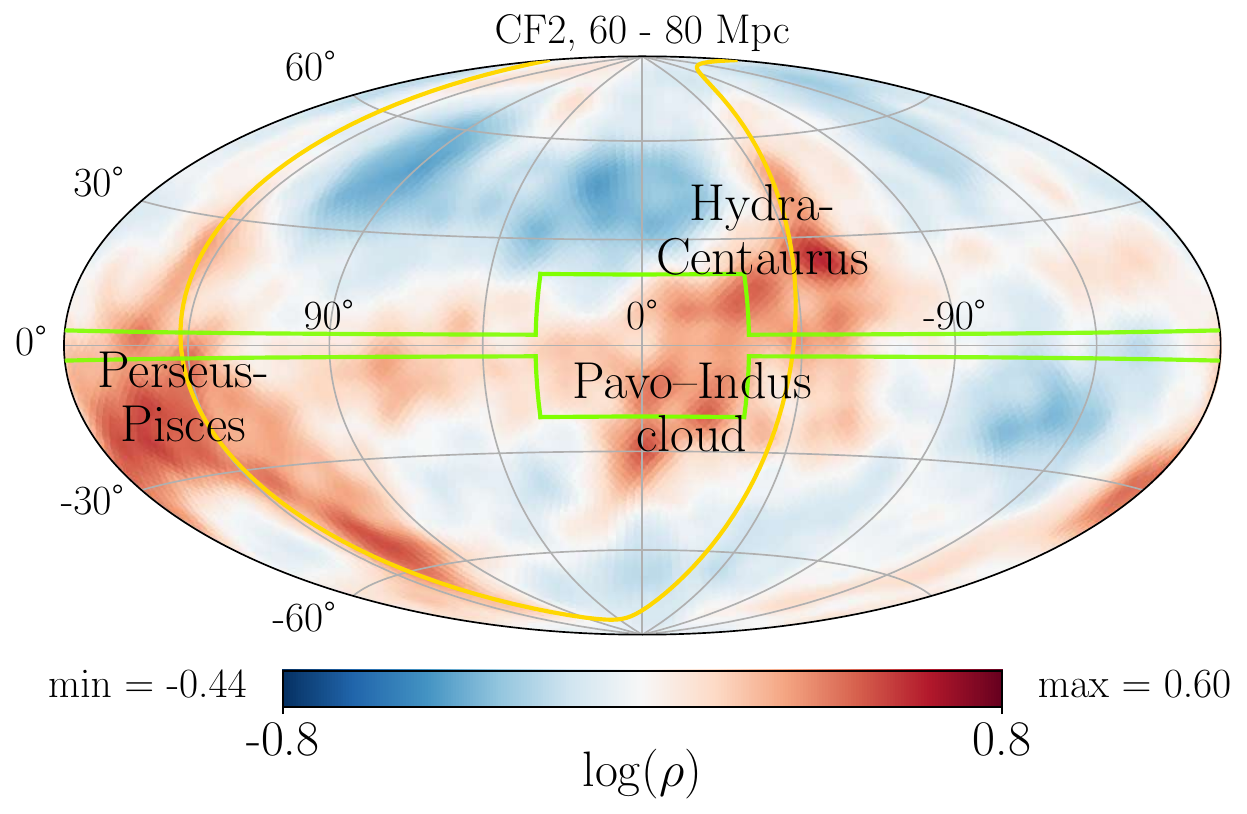}
\includegraphics[width=0.245\textwidth]{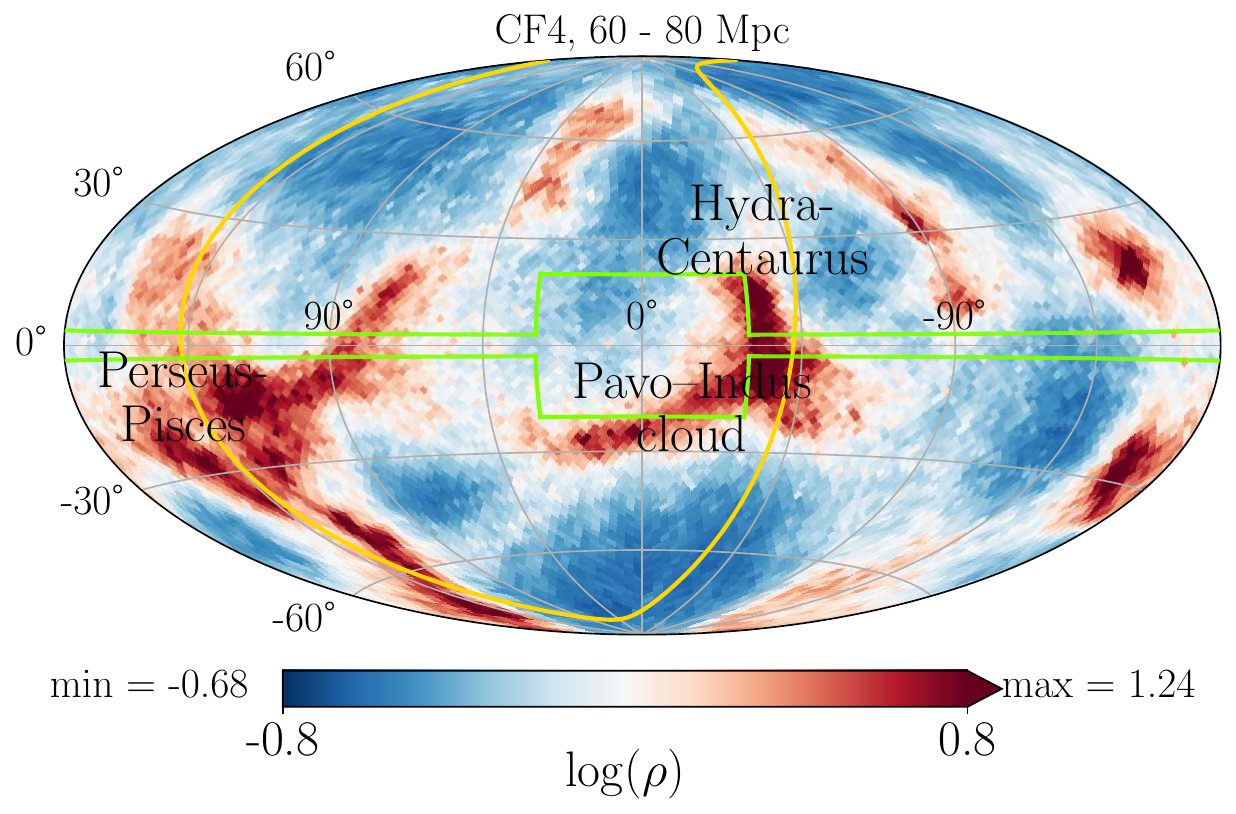}
\includegraphics[width=0.245\textwidth]{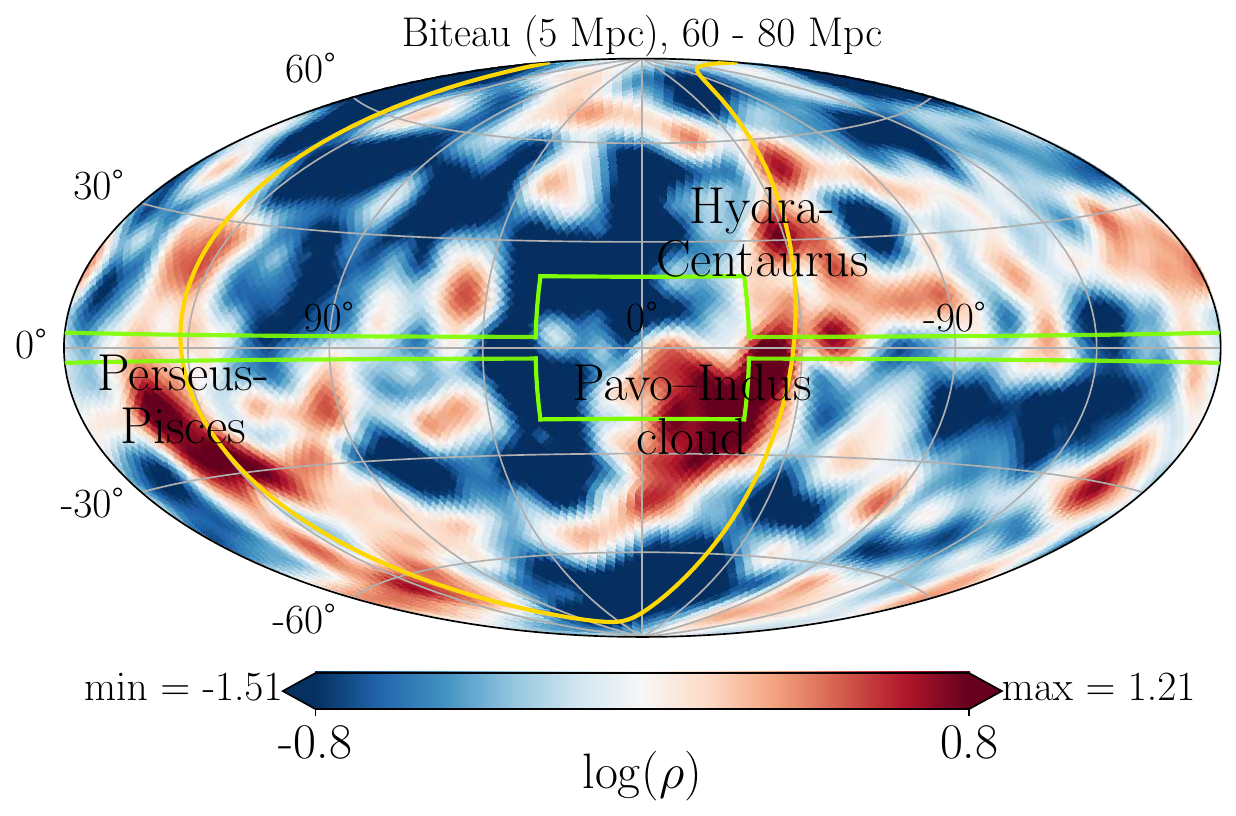}
\includegraphics[width=0.245\textwidth]{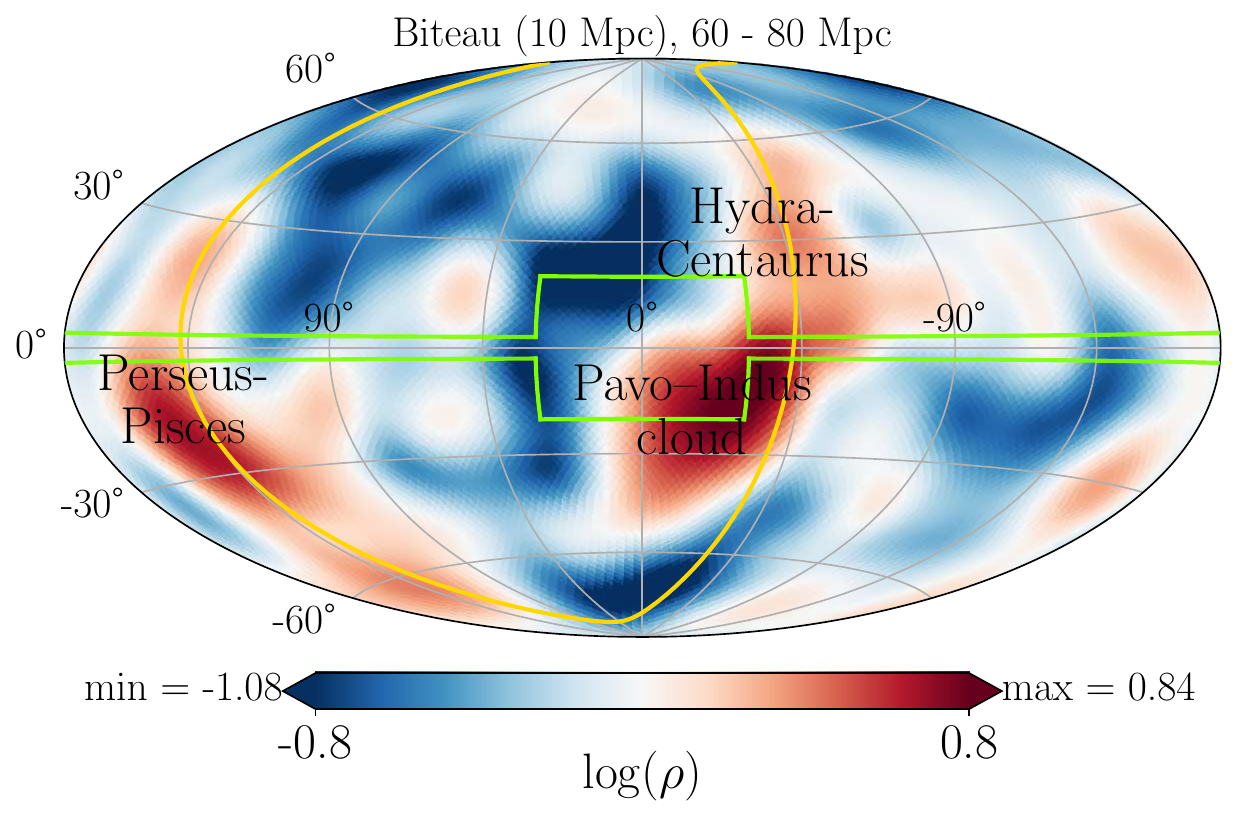}

\includegraphics[width=0.245\textwidth]{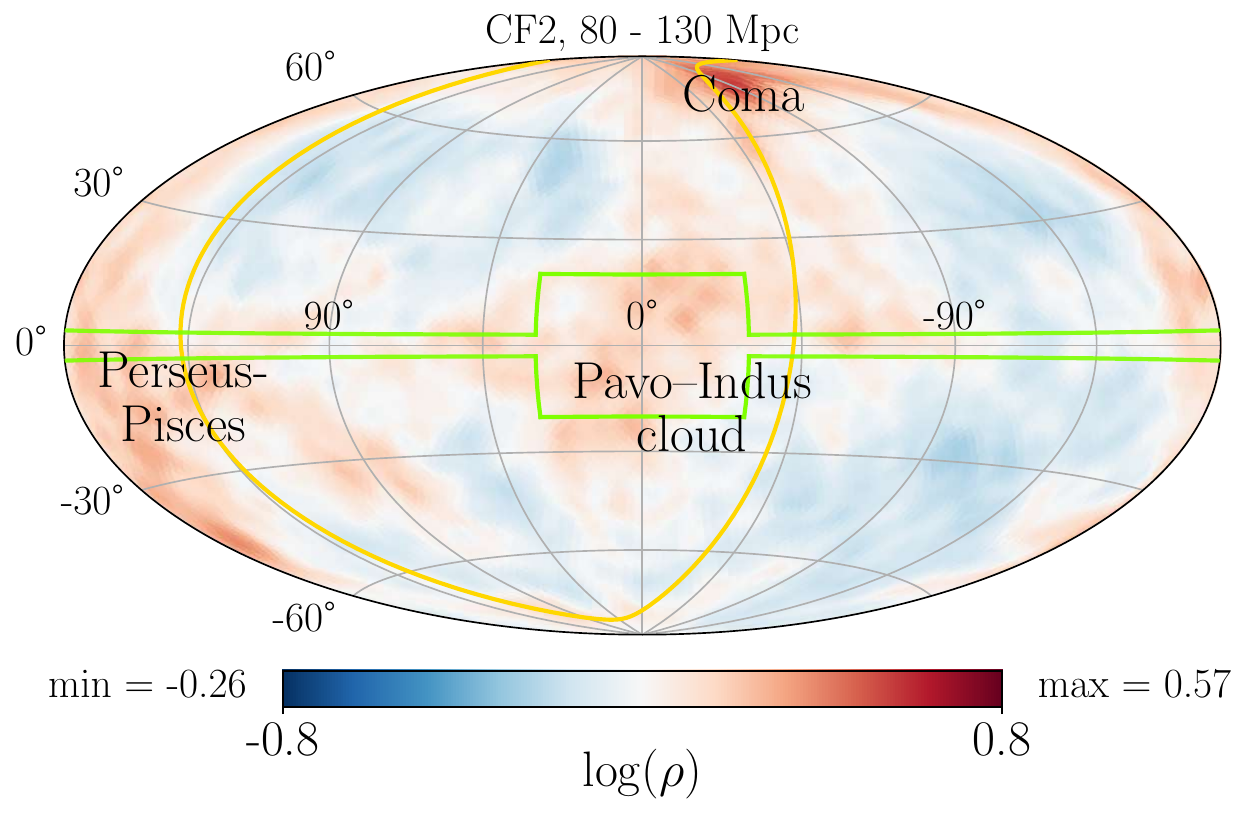}
\includegraphics[width=0.245\textwidth]{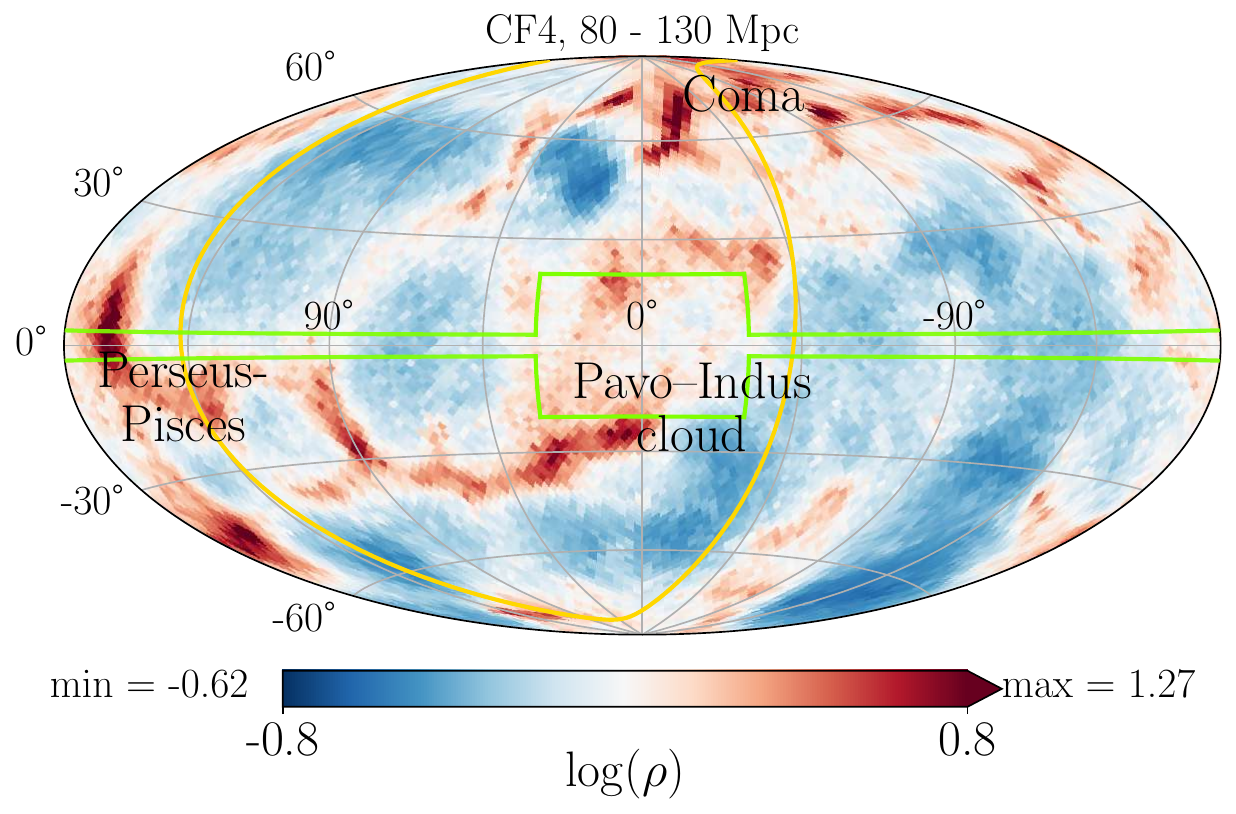}
\includegraphics[width=0.245\textwidth]{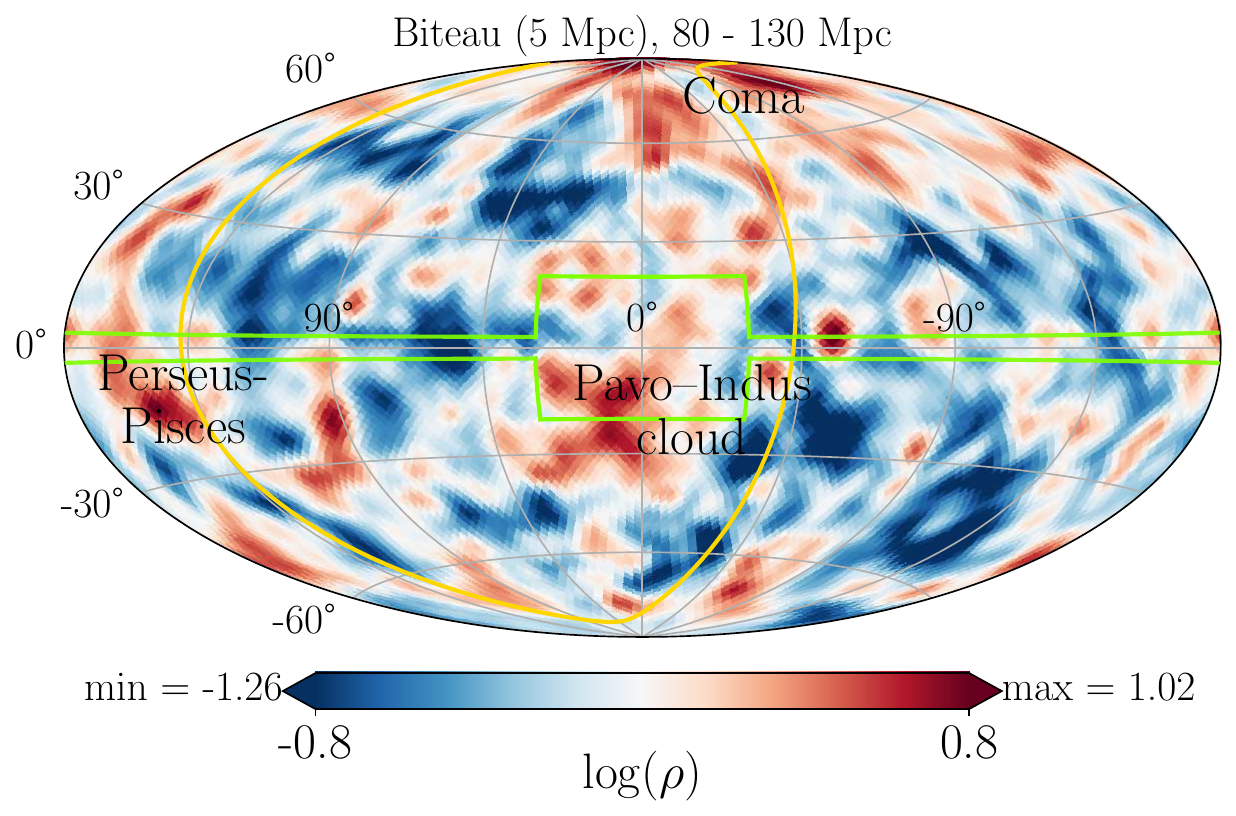}
\includegraphics[width=0.245\textwidth]{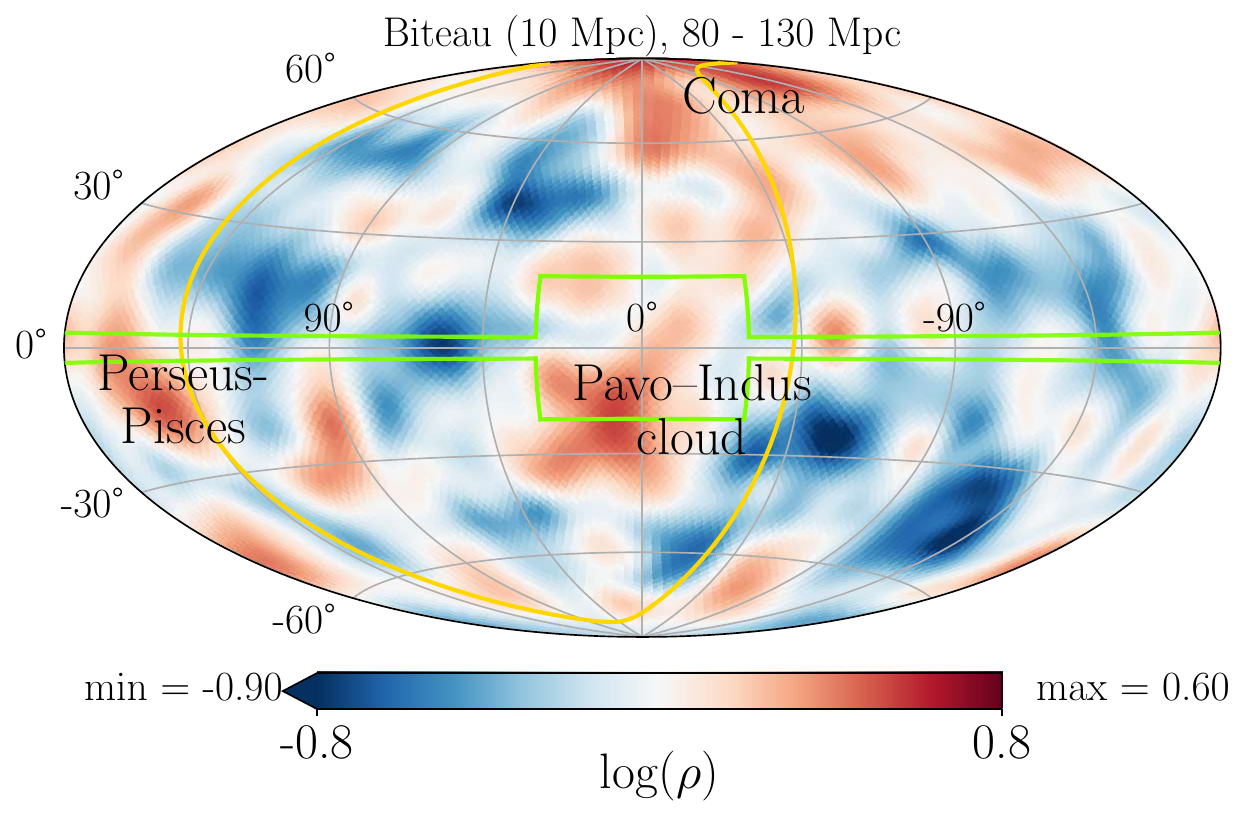}

\includegraphics[width=0.245\textwidth]{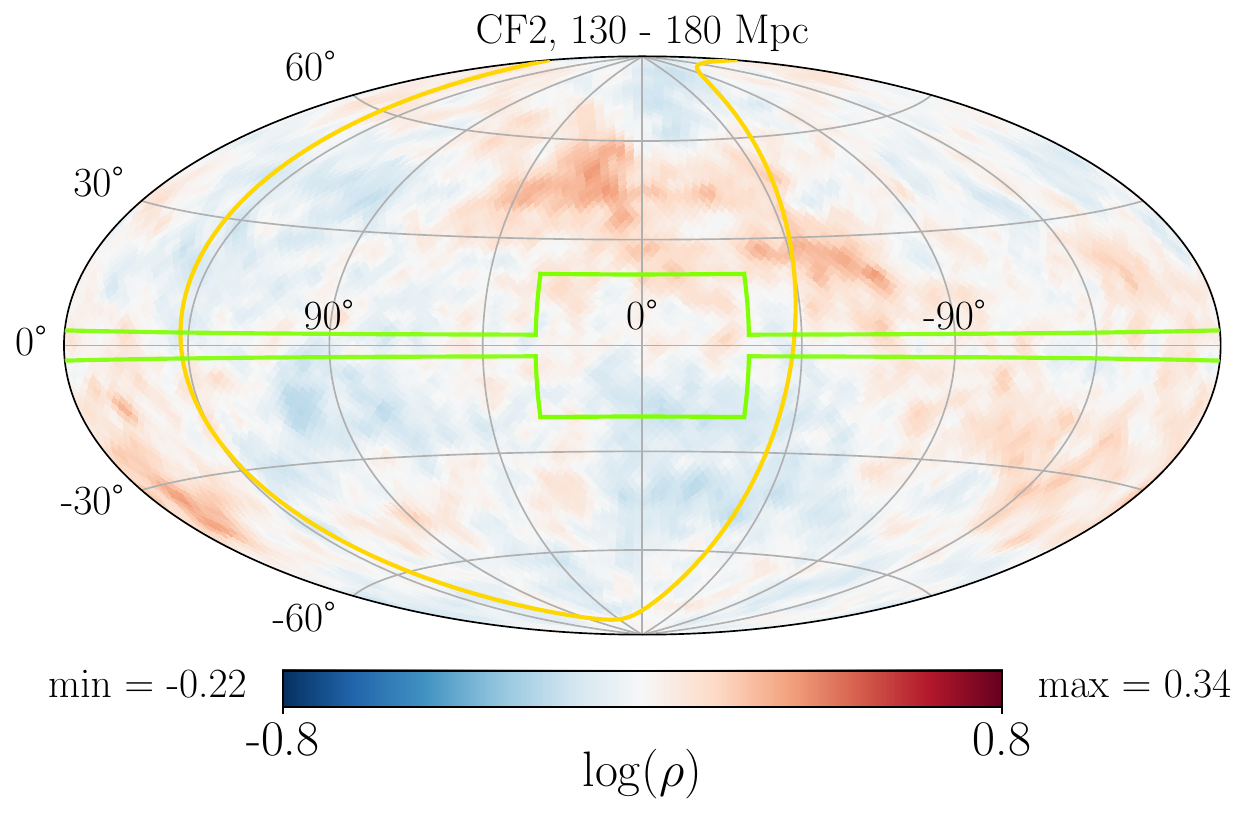}
\includegraphics[width=0.245\textwidth]{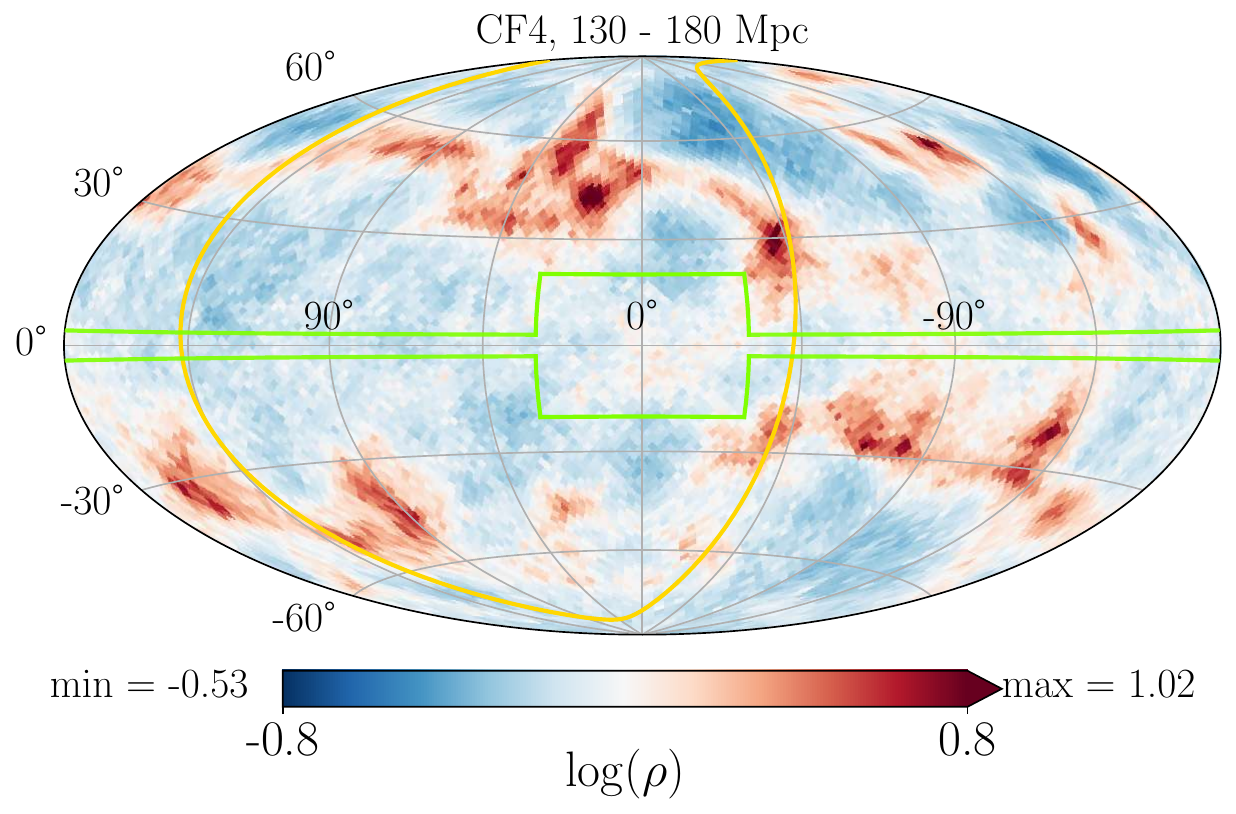}
\includegraphics[width=0.245\textwidth]{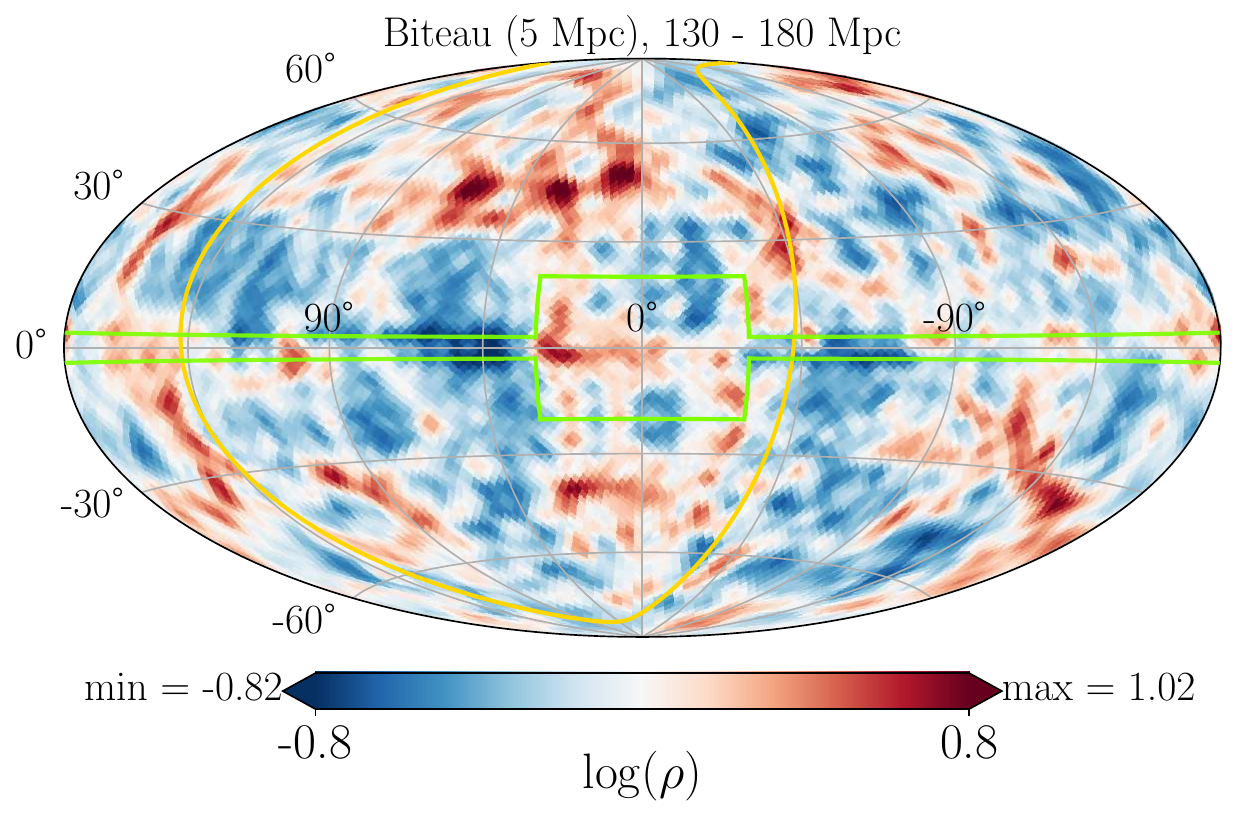}
\includegraphics[width=0.245\textwidth]{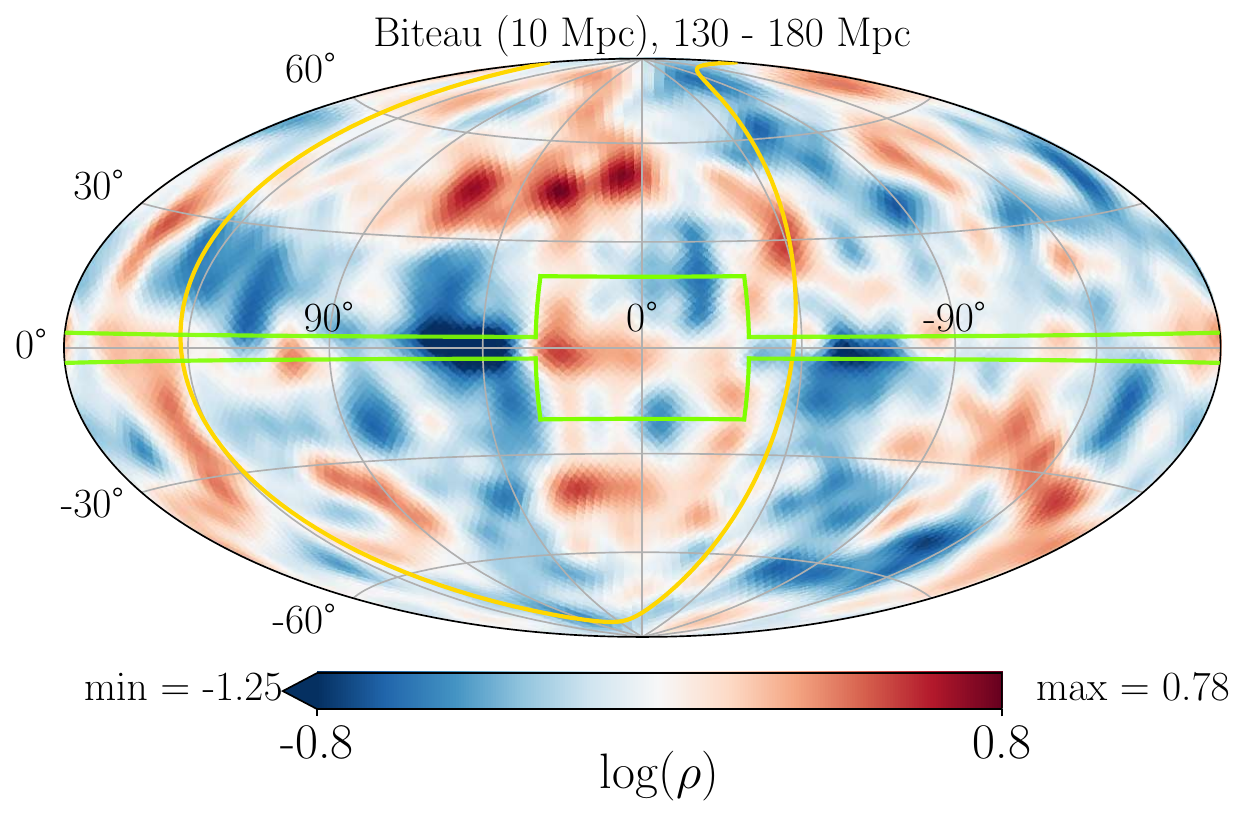}

\includegraphics[width=0.245\textwidth]{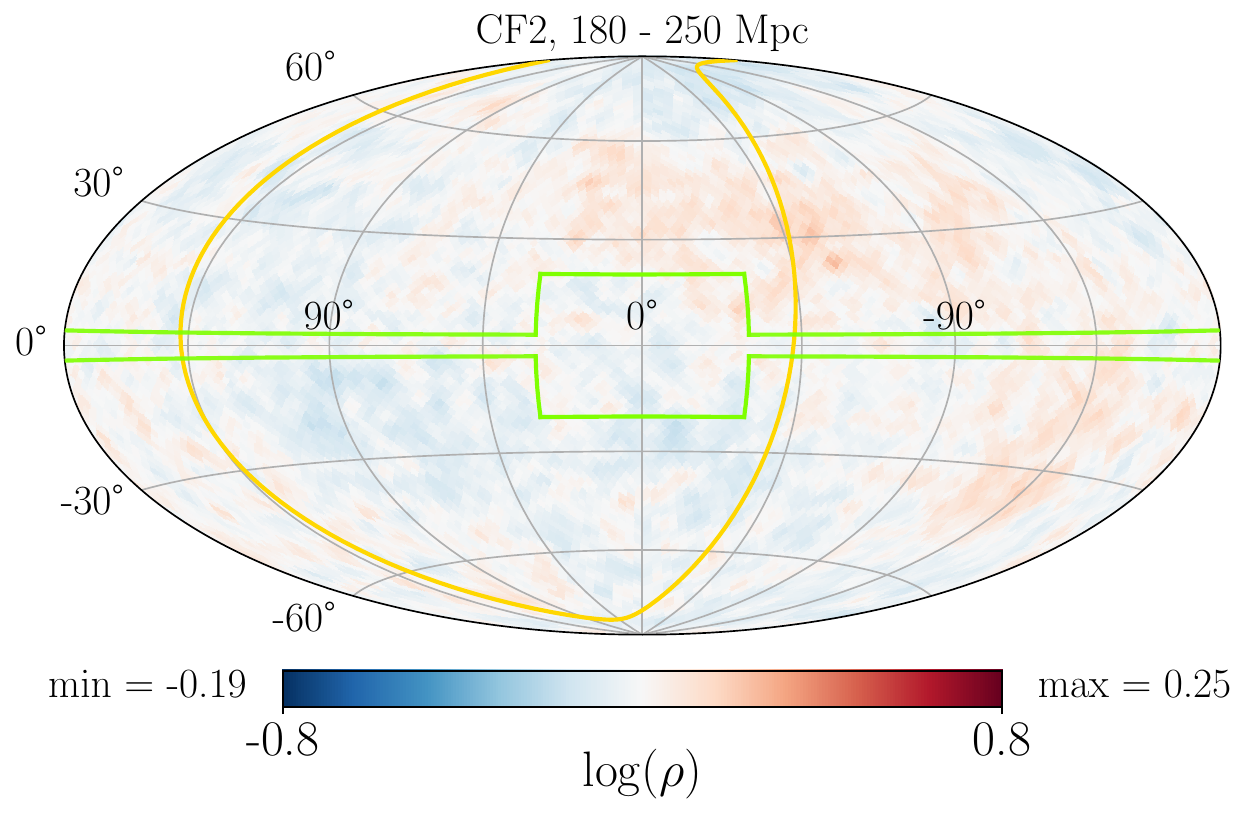}
\includegraphics[width=0.245\textwidth]{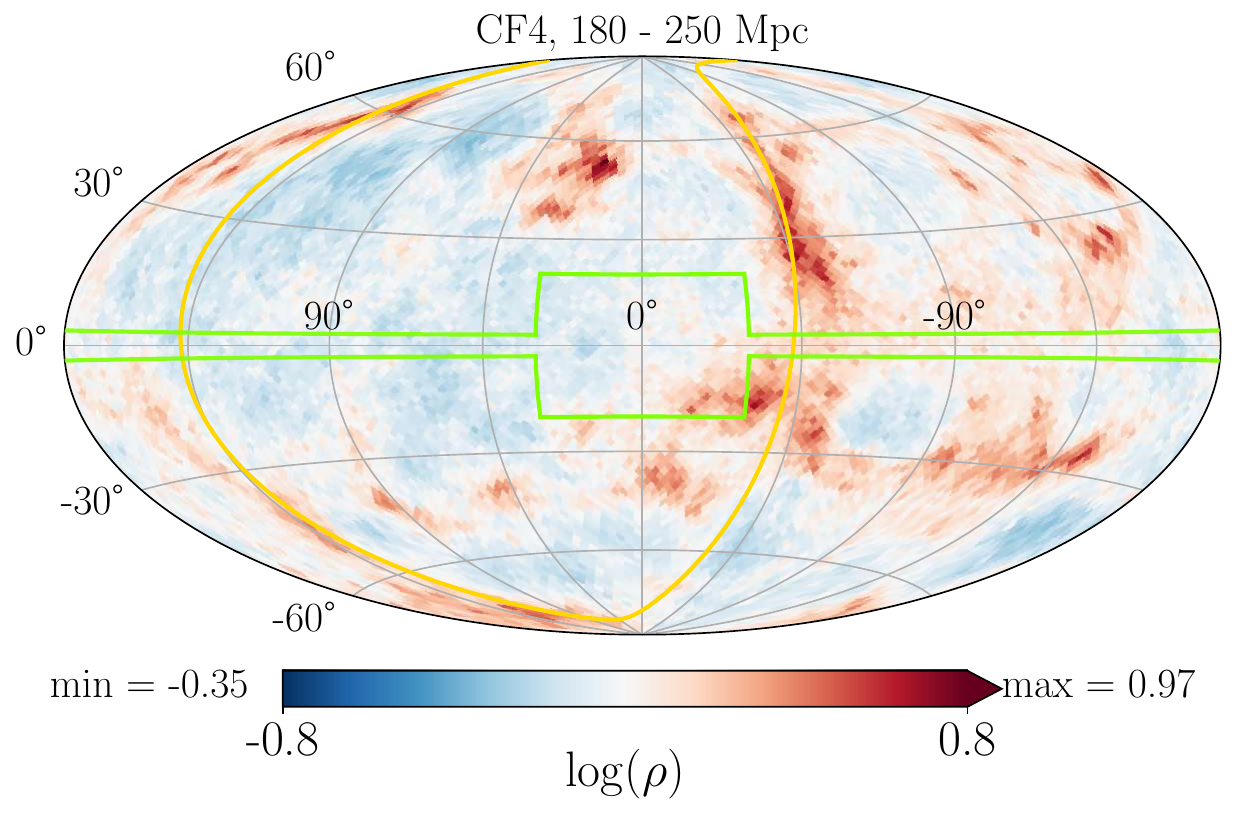}
\includegraphics[width=0.245\textwidth]{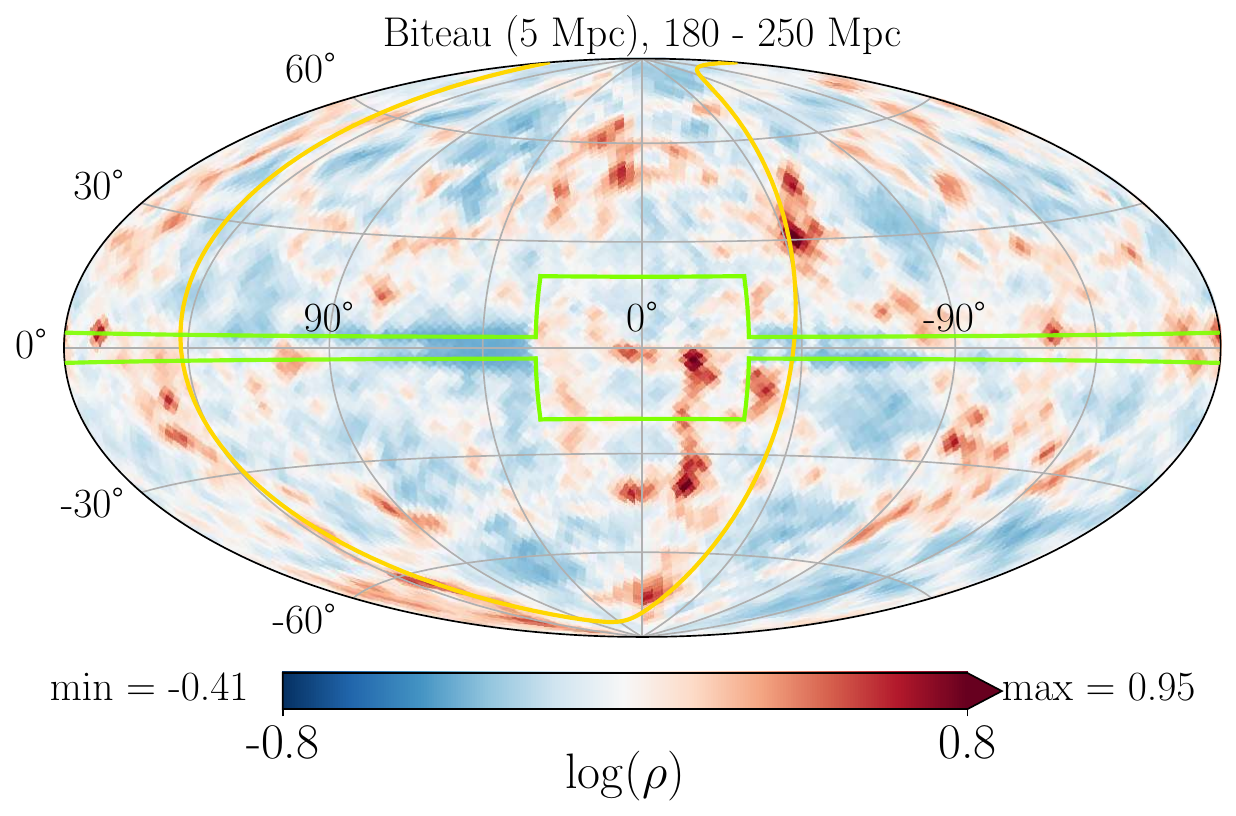}
\includegraphics[width=0.245\textwidth]{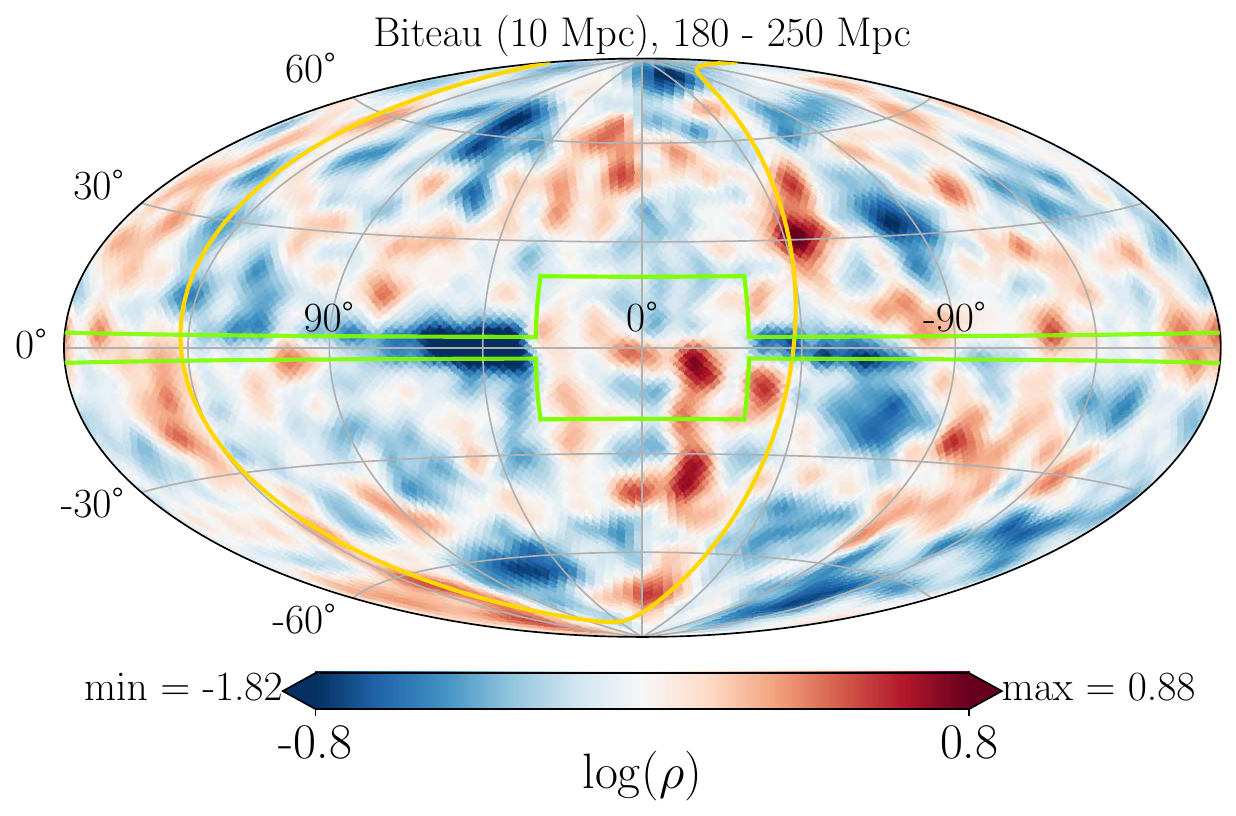}

\caption{Skymaps of logarithmic mass density excursion derived from CosmicFlows2, CosmicFlows4, and Biteau's catalog smoothed by 5 and 10 Mpc and averaged over the indicated radial shell, except that the rightmost entry in the first row (0 - 20 Mpc) uses ``angular smoothing" to better reveal the actual data. In all plots the supergalactic plane is indicated in yellow and the cloning region is outlined in green -- inside, Biteau's density model simply mirrors the density in the observed region immediately above or below the green line, as described in the text and seen by eye. The boxes in dark blue in the third panel in the top row indicate the angular regions discussed in Fig.~\ref{fig:selectLoSden}. Some major clusters are labeled at their corresponding positions. }
\label{skymap}
\end{figure*}

Figure~\ref{skymap} shows the logarithm of the density relative to its global mean value, as a function of direction in the sky, for CF2, CF4 and Biteau with smoothing radii of 5 Mpc and 10 Mpc, in mostly 20 Mpc thick shells up to 250 Mpc, except that the rightmost column in the nearest shell is used to show the nested-pixel angular smoothing scheme for Biteau. 
The mass within every voxel is calculated and then integrated along the $\boldsymbol{r}$ direction in the given distance range to find the mean density for the given direction.  The colorbar values, showing the logarithm of the mass over-densities in each pixel, are cutoff at $\pm 0.8$ for all skymaps, with the maximum value shown in small characters. 

When comparing the skymaps from CosmicFlows, one sees that although CF2 and CF4 display similar large-scale structures, they show significant differences. Both CF2 and CF4 include major clusters such as Virgo, NGC 1407, Hydra-Centaurus, Pavo-Indus, Perseus-Pisces, and Coma, but CF4 further shows substructures like Antlia, NGC 5846, Fornax-Eridanus. The Pavo-Indus regions in CF4 are more concentrated than in CF2 within $60-80$ Mpc. CF4 also exhibits larger variances and more detailed void structures than CF2. 

Biteau’s galaxy catalog produces skymaps that are grossly similar to those from CF4, especially at the locations of several major clusters, when using 10 Mpc smearing for Biteau (fourth column). However, the galaxy catalog shows mass concentrated along the super-galactic plane (yellow line) in the $0-20$ Mpc shell which is absent in CF4. In the $20-40$ Mpc shell, and also the $40-60$ Mpc shell, one sees structure in the central part of the Biteau maps that is absent in CosmicFlows. 
To a considerable extent, this is an artifact of the cloning procedure which for $|\ell|<30^\circ$ mirrors structures from  $b = 20^\circ-40^\circ$ around $b = 20^\circ$ to populate $b=20^\circ-0^\circ$ and similarly structure below the Galactic plane is mirrored about $b=-20^\circ$.  The perimeter of the  region filled by cloning is shown by the green lines. The mirrored cloned structure toward the inner Galaxy can be clearly seen in all distance bins beyond 10 Mpc by blowing up the skymaps in the 3rd column. 
Aside from the central cloned region, Biteau’s catalog and CF4 resemble each other fairly well beyond 40 Mpc, as is most easily seen comparing the second and fourth columns of Fig.~\ref{skymap}; both capture NGC 5864, Antlia, and Fornax-Eridanus, which do not appear clearly in CF2.

In Biteau’s maps, the Pavo-Indus Clouds appear more concentrated within $40-60$ Mpc. While they are almost absent in the CosmicFlows maps at the same range. This is the largest difference between the results from two methods, aside from the cloning. The best agreement lies in the $60-80$ Mpc range, where both Biteau and CosmicFlows show concentrated Pavo-Indus clusters. CF4 is more concentrated here than CF2, similar to Biteau's galaxy maps, although their exact shapes differ. Also, CF4 shows almost no Hydra-Centaurus structure in the $60-80$ Mpc shell, while CF2 and Biteau both show some weak features in that region.  As we discuss in Sec.~\ref{sec:dist} below, these differences may be due to incorrect redshift-based distance assignments for galaxies in clusters with large peculiar velocities.

In the $80-130$ Mpc range, both CF4 and Biteau show a concentration near Pavo-Indus, but it appears larger and more symmetric in Biteau’s maps. This is due to the cloning procedure Biteau used to fill the zone of avoidance (ZoA), which duplicates the structures near ZoA. Since the Pavo-Indus regions lie close to the ZoA boundary, their appearance in Biteau’s maps is amplified by cloning. The mirrored look of Pavo-Indus Cloud in $40-60$ Mpc and in $80-130$ Mpc also comes from this, whereas CF4 does not show this strong vertical reflection symmetry about $b=\pm 20^\circ$.

A voxel-level comparison of the densities in the various models in the $20-120$ Mpc distance range, is presented in the corner plots of Fig. \ref{corner} in the Appendix. These plots compare CF2 vs. Biteau (10 Mpc), CF4 vs. Biteau (10 Mpc), CF2 vs. Biteau (5 Mpc), CF4 vs. Biteau (5 Mpc), and CF4 vs. CF2. 
Consistent with what is visible by eye in the skymaps, CF4 shows a larger density range than CF2, while Biteau’s catalog exhibits a much larger range, even after smoothing. 

Another perspective on the difference between Biteau and CF4 is given by the skyplots in Fig.~\ref{skymapratio} of the Appendix, showing the (log of) the ratio of Biteau to CF4 in shells of 10 Mpc thickness, or 20 Mpc at the largest distances.

\section{Discrepancies between CosmicFlows and Biteau}
\label{sec:Discrepancies}
The approaches to determining the local mass distribution adopted by CosmicFlows and Biteau are quite complementary, in principle.  In this section we review issues and discrepancies in the resultant mass models which our analysis has uncovered, discuss their origins and consider possible improvements.

\subsection{Cloning}
\label{sec:cloning}  
The procedure of ``cloning" galaxies to fill the zone of avoidance can introduce entirely fictitious structure in the Biteau mass density model, especially in the expansive central ZoA adopted there: $-20^\circ < b < 20^\circ$ and $-30^\circ < \ell < 30^\circ$.  Cloning doubles the size of already major coherent mass structures in the $20-80$ Mpc range, introducing mass not actually present according to the Cosmic Flow analyses (Fig.~\ref{skymap}).  
  
In spite of the ZoA containing only 10\% of the solid angle (and the central $40^\circ \times 60^\circ$ portion only 6\%), at some distances the cloning makes significant distortions in the overdensity.  For instance in the shell centered on 85 Mpc, cloning doubles the overdensity (see Fig.~\ref{ave}).  Examples of explicit unphysical structure due to cloning can be seen by comparing the second and third columns of skyplots of Fig.~\ref{skymap}, where the reflection symmetry about the $b=\pm 20^\circ$ boundaries of the ZoA seen in the Biteau model is not present in the CF4 model.  

Therefore, for applications sensitive to the ZoA direction, better techniques need to be developed for determining the mass distribution in the ZoA volume when using a catalog-based approach.  Analogously to the way that Biteau compensates for the loss of distant dimmer galaxies in a flux limited catalog, by up-weighting the observed brighter galaxies as called for by the known luminosity function, one could up-weight galaxies in the ZoA according to their observed spectra, to compensate statistically for the reddening and dust obscuration.  A sophisticated implementation would make use of machine learning and the myriad of detailed studies of Galactic absorption. 

The CosmicFlows approach exploits self-consistency of large scale motions of clusters of galaxies under their mutual gravitational attractions in the Hubble flow to mitigate the inadequate  information on individual galaxies.  Thus, for the time being, CF4 gives the best available information on the mass density in the ZoA region.  

\subsection{Local overdensity}
\label{sec:local} 
 A second discrepancy between the B21 and CF mass models is the factor-two larger local mass density of B21 compared to the CF models (see Fig.~\ref{ave}).  Cloning plays only a small role in this discrepancy, as is seen from the small difference between thick and thin lines in Fig.~\ref{ave}, which respectively include and do not include the ZoA contribution.  The skyplots for $d<20$ Mpc in the upper row of Fig.~\ref{skymap}, also show that cloning is not contributing significantly to the mass density in this shell.  
 
 Within 11 Mpc, Biteau uses the 1029 local galaxies from the Local Volume (LV) catalog of ~\citet{karachentsev_morphological_2018}.  These galaxies are well-surveyed even in the ZoA and have distance measures from tip-of-the-red-giant-branch, Cepheids and other techniques that should be accurate to better than 25\%. ~\citet{karachentsev_morphological_2018} provide the K-band luminosity of each galaxy, which Biteau converts to stellar mass using $M_{\star,\rm LV} = 0.6\,(M_\odot/L_\odot) \times L_K$, based on~\citet{Ducoin+2020}. Thus, modulo extinction and the precision of the conversion of luminosity to stellar mass over the relevant range of galaxy types, we expect that within 11 Mpc the LV/B21 mass model is robust and is to be preferred over CF4, which reconstructs the mass by fitting the large scale as well as local mass and velocity distributions. 
 
 What, then, can be responsible for the factor-two discrepancy between the local mass density as recovered in the CF modeling, and the true mass distribution as derived from the LV catalog given by Biteau?  We propose that the difference may primarily result from the fact that CosmicFlows derives the mass density field based on observed relative velocities, without putting a prior that we are inside the Milky Way.  The origin of our coordinate system is by construction inside Milky Way, whereas in CF modeling there is no bias on the local density from the observer being in a Milky Way-like galaxy.  

\begin{figure}[ht!]
\includegraphics[width=0.45\textwidth]{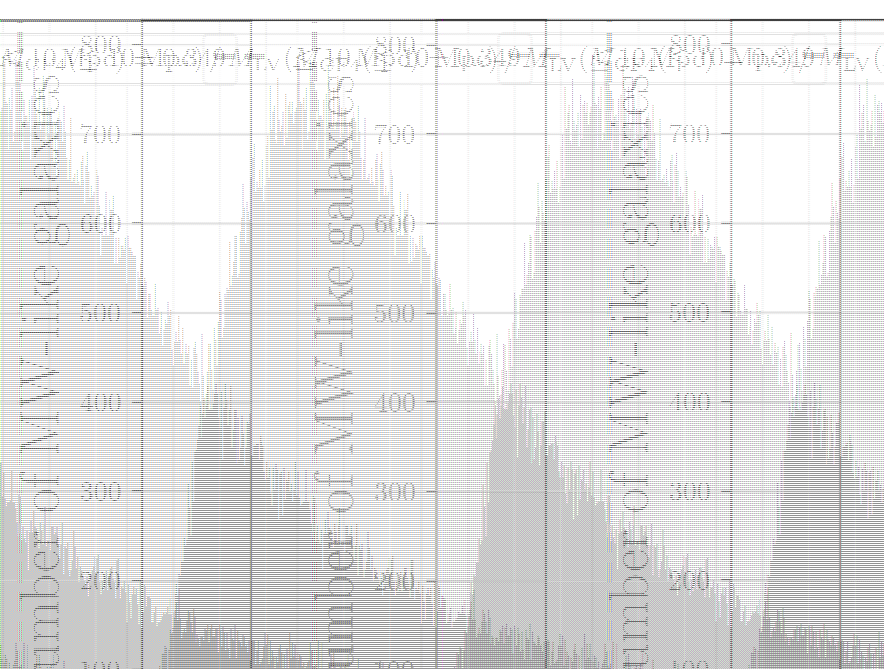}
\caption{Total mass of galaxies within 10 Mpc of galaxies whose mass is (0.5-2)$M_{\rm MW}$, relative to the mass within 10 Mpc of the Milky Way, according to the Biteau catalog.}
\label{fig:Local}
\end{figure}

\begin{figure}[ht!]
\includegraphics[width=0.45\textwidth]{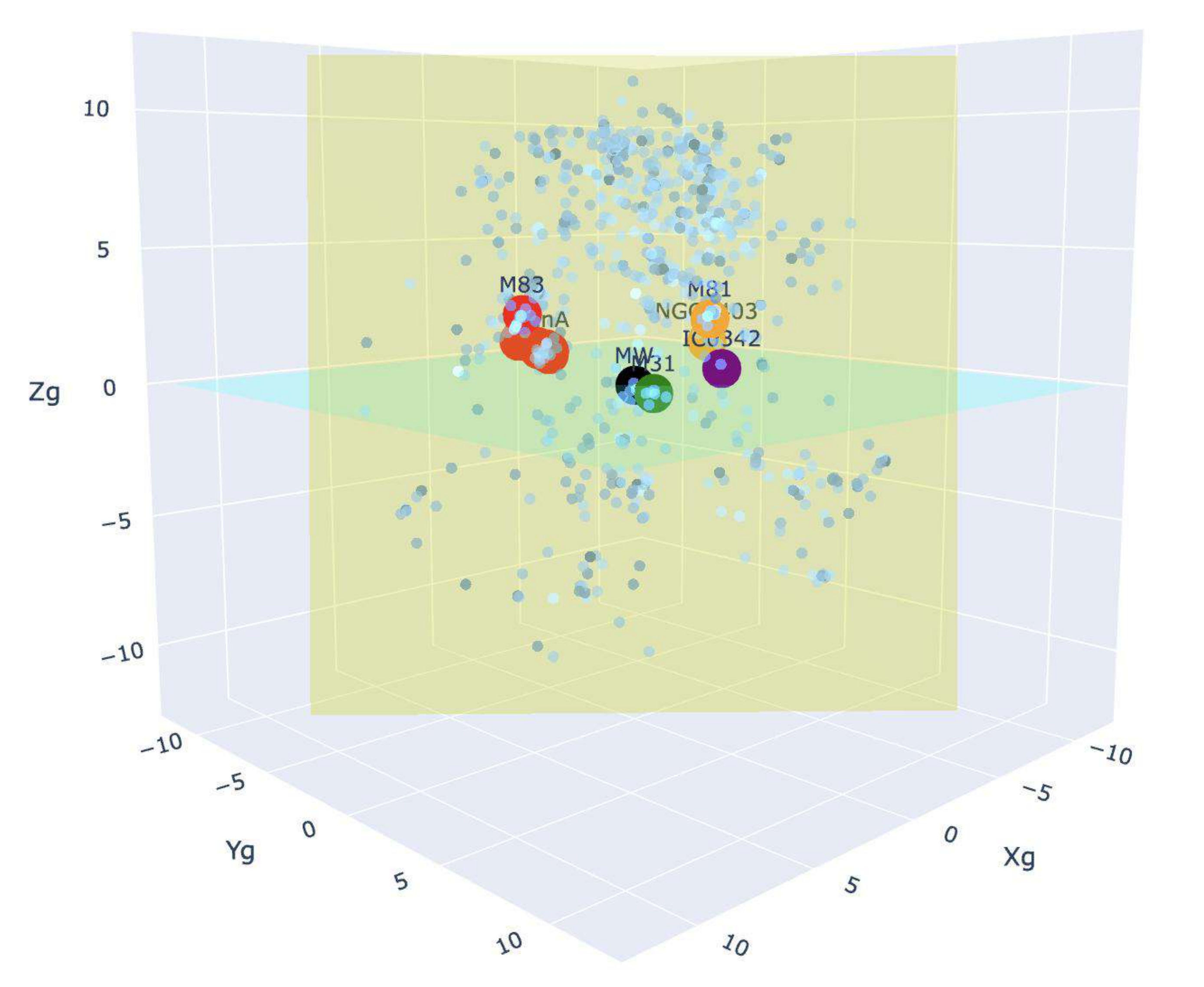}
\caption{3D visualization of galaxies in the Local Volume. Points are shown in Galactic Cartesian coordinates, with the $x$-axis pointing toward the Galactic Center. The cyan plane indicates the Galactic plane, and the yellow plane shows the Supergalactic plane. Colors denote the main galaxy groups: Milky Way (black) and M31 (green), M81 and NGC~2403 (orange), Centaurus~A and M83 (red), and IC~342 (purple). Point opacity encodes stellar mass, scaled linearly with logarithmic mass. The interactive version of this figure is available online at \url{https://yl10327.github.io/LV-3D/LV_plotly.html}. The figure without planes is available at \url{https://yl10327.github.io/LV-3D/LV_plotly_noplane.html}.}
\label{fig:LV}
\end{figure}

 The likelihood of observing a given overdensity within a 10 Mpc sphere surrounding a MW-like galaxy can be assessed as follows: Find all galaxies with $0.5\, M_{\rm MW}<M<2\, M_{\rm MW}$ in the Biteau catalog, excluding the ZoA and the 10 Mpc surrounding it.  For each of these MW-like galaxies, measure the mass within 10 Mpc.  Figure~\ref{fig:Local} shows the resulting mass distribution relative to the actual value for the Milky Way, using the Biteau catalog assigned masses.  The red dashed line shows the mass wtihin 10 Mpc of the Milky Way according to CF4.  Thirty percent of the galaxies in the distribution are to the right of 1.0, i.e., they exist in a local over-density as large or larger than ours according to the Biteau catalog.  Hence the lack of a prior on the local density is indeed a good explanation for the CF4 model's underestimate of the local density.  

 To assure ourselves that there is not a significant ZoA deficit in the Local Volume catalog, we produced a 3D visualization which shows that the low density regions do not align with the Galactic Plane.  Since the visualization may be useful to readers, we provide it in an interactive figure, with a fixed perspective shown in Fig.~\ref{fig:LV}.

\subsection{Inaccurate distance assignments}
\label{sec:dist}
A third difference between the CosmicFlows models and Biteau's is the method of assigning galaxy distances. The CosmicFlows approach combines non-redshift-based distances, galaxy redshifts, self-consistent modeling of peculiar velocities within a cluster, and large scale bulk flows, to obtain a 3D model of the density and velocity fields. By contrast, apart from galaxies within 11 Mpc which Biteau takes from the Local Volume catalog which have direct distance measurements~\citep{karachentsev_morphological_2018}, the vast majority of Biteau's galaxy distances are assigned assuming their redshift is entirely due to Hubble-flow recession, so $d = z c/ H_0$.  (Moreover, the redshifts of the majority of galaxies have additional uncertainty due to being determined photometrically rather than spectroscopically.) 
However using redshift and the Hubble formula ignores peculiar velocities.  The problem is most severe for galaxies in a massive cluster whose infall velocity toward the center of mass of the cluster has a line-of-sight component comparable to or even larger than the Hubble recession velocity of the cluster as a whole.  (The infall velocity in a massive cluster can exceed 1000 km/s.)  \citet{RobertsFarrar13} examined galaxies within 100 Mpc for which robust distance measures were available and compared the actual distance to the Hubble-flow-based distance, finding more than a factor-10 discrepancy in some cases.  Such inaccurate distance assignments can move mass belonging in a cluster in a given distance shell, to an adjacent or more distant shell, and thereby spread or otherwise distort the derived radial mass distribution. 

\begin{figure*}[ht!]
\includegraphics[width=0.47\textwidth]{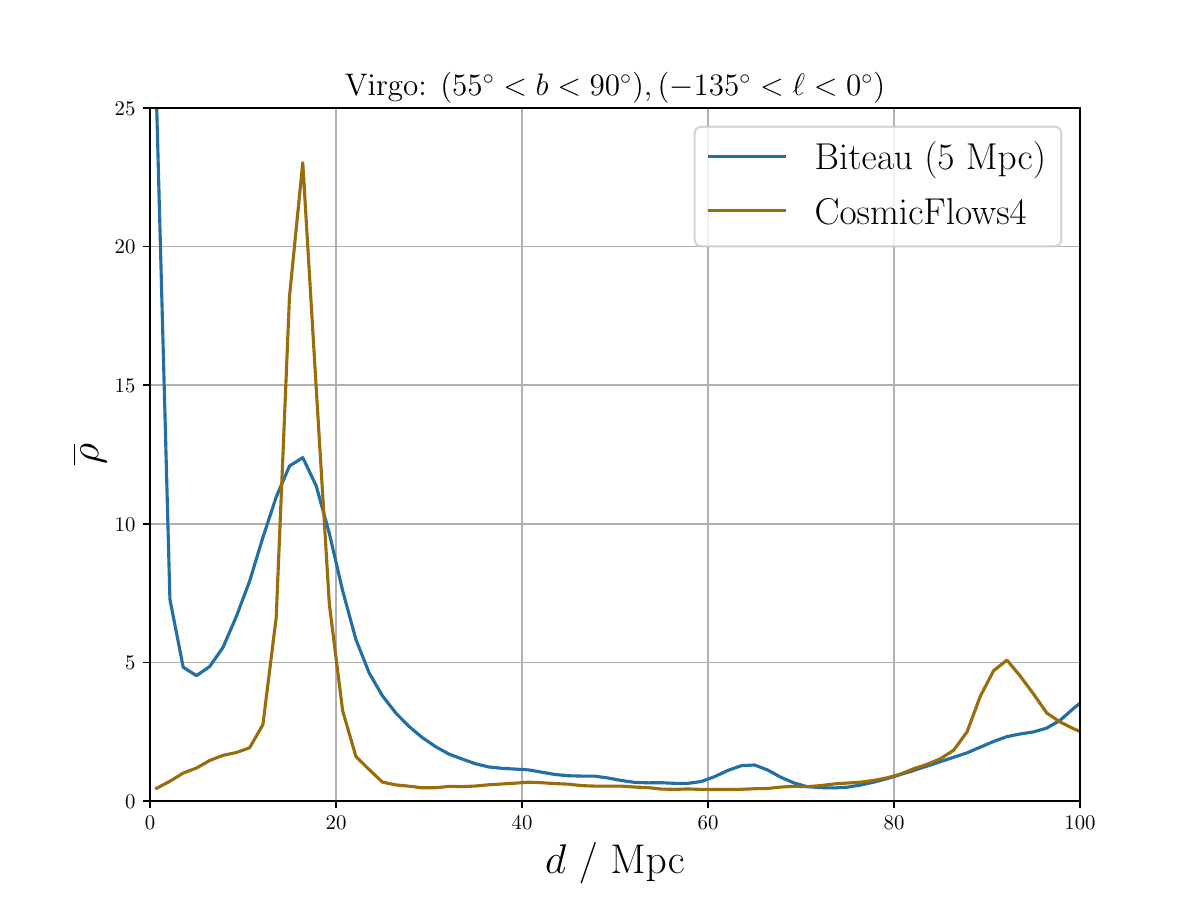}
\includegraphics[width=0.47\textwidth]{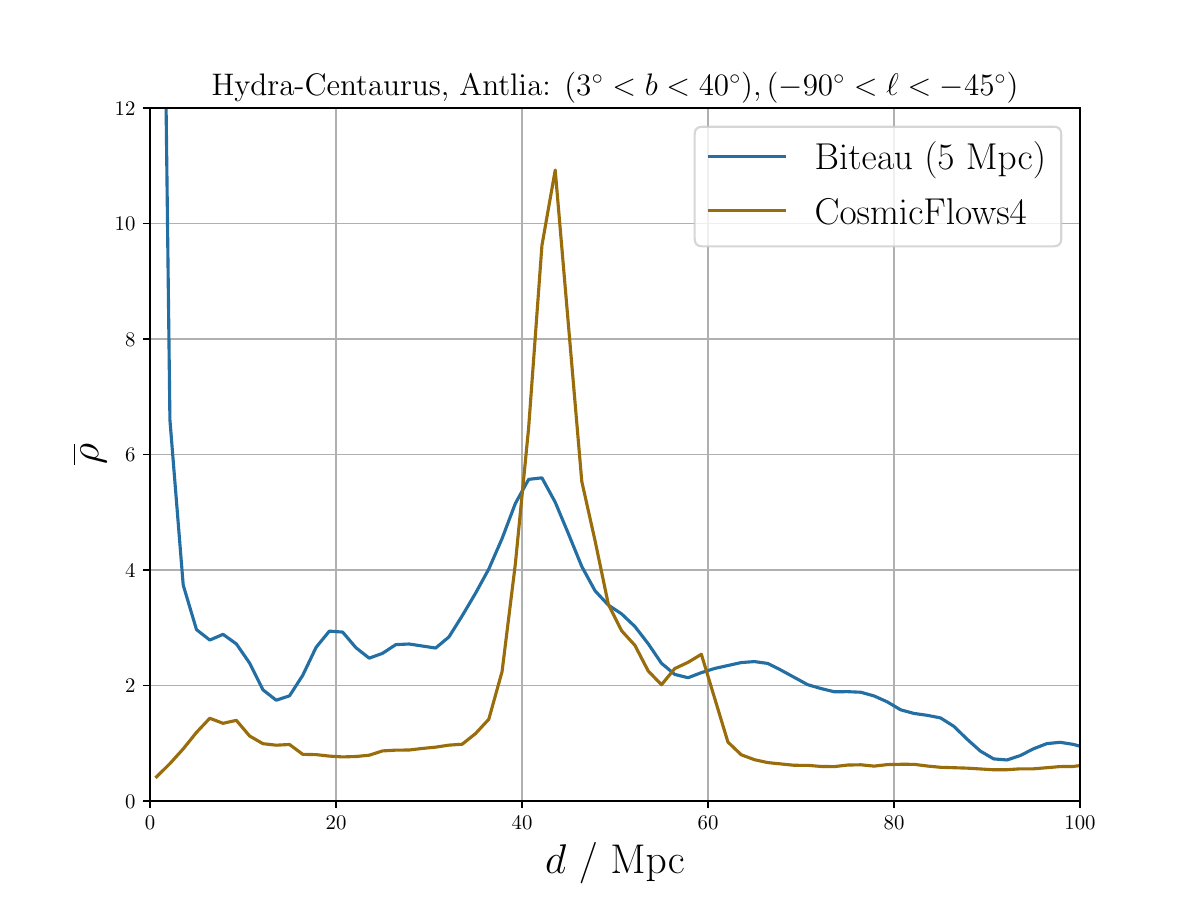}

\includegraphics[width=0.47\textwidth]{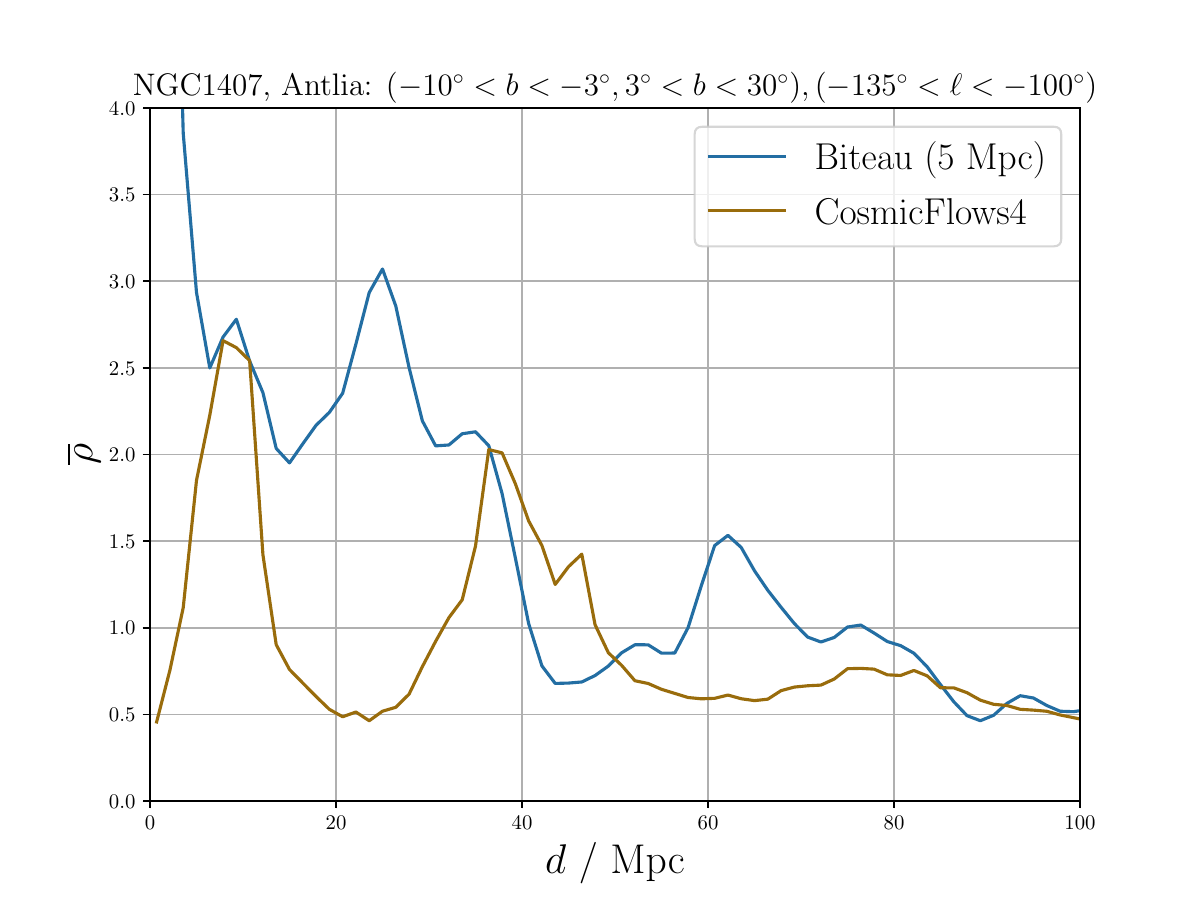}
\includegraphics[width=0.47\textwidth]{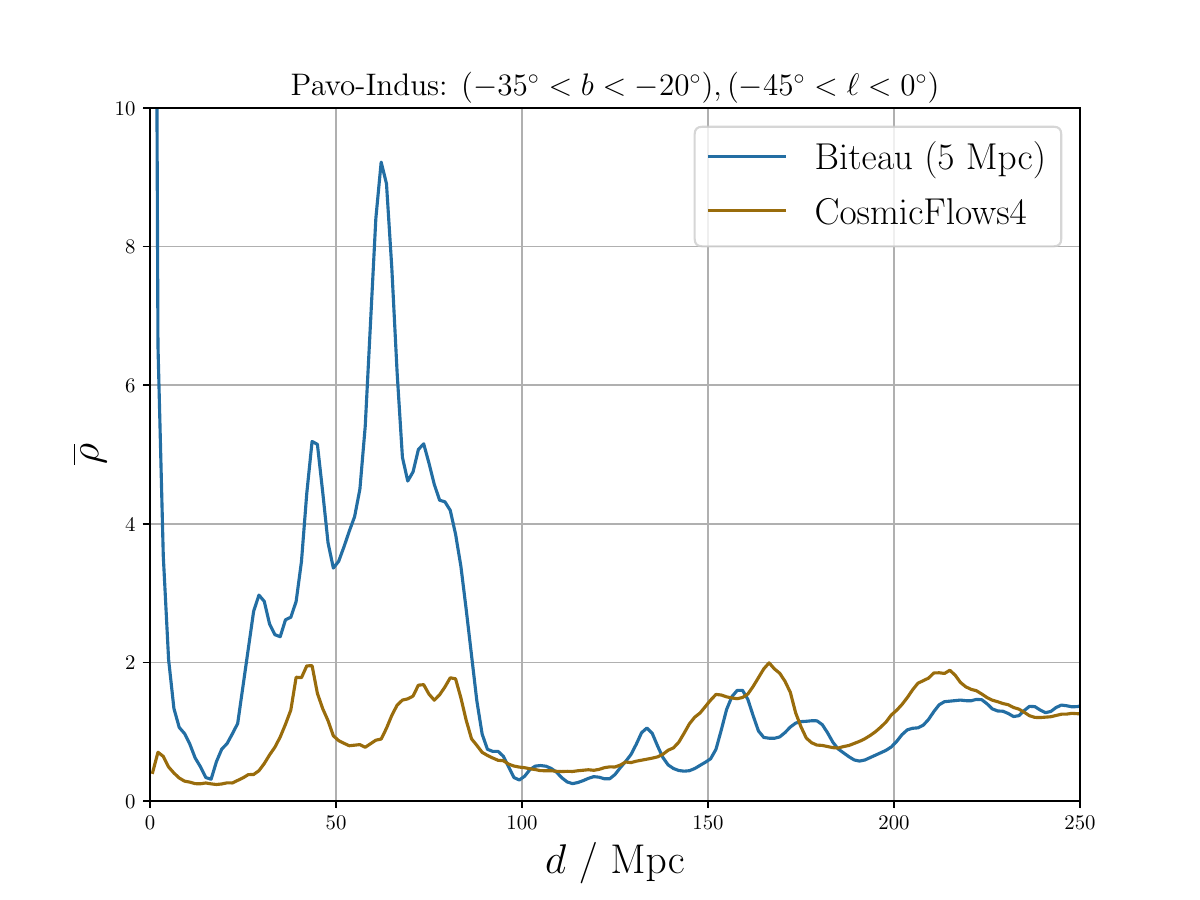}

\includegraphics[width=0.47\textwidth]{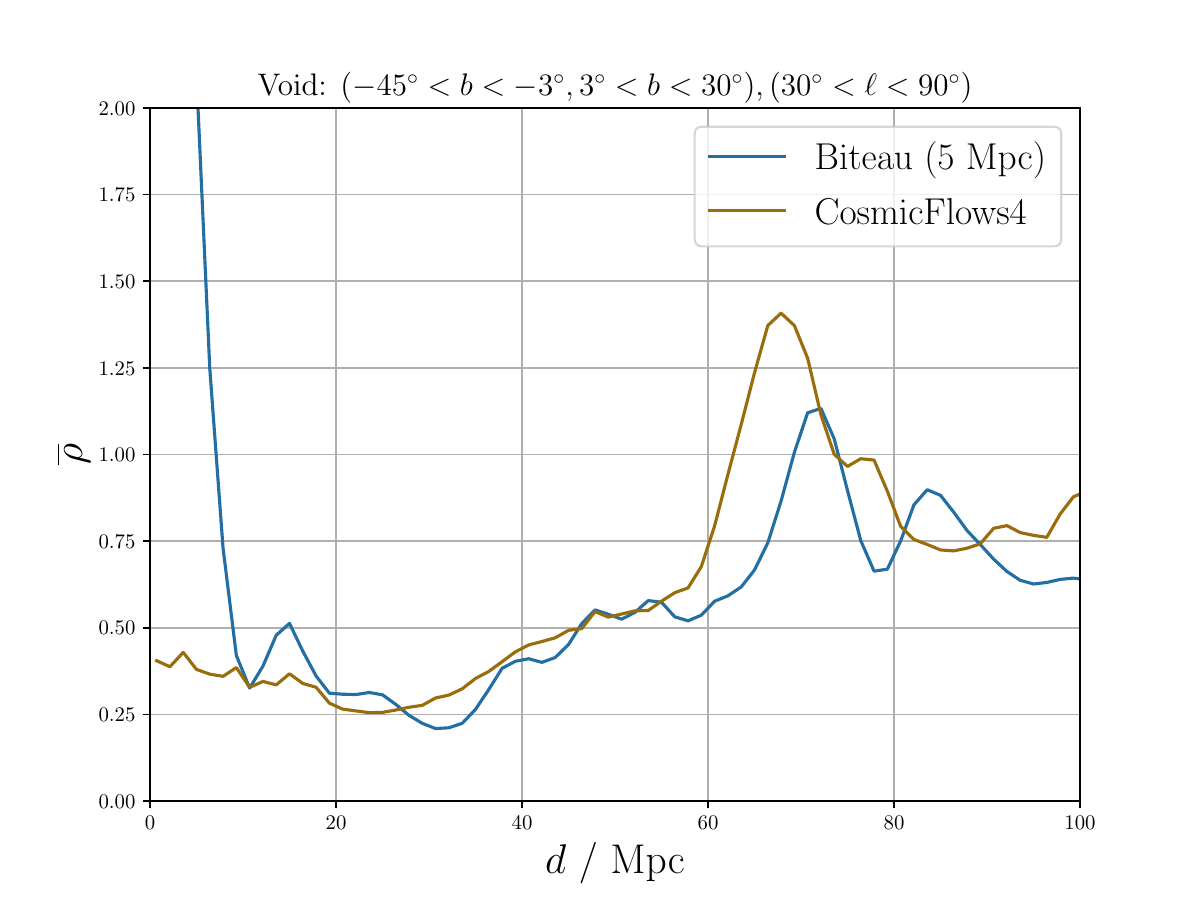}
\includegraphics[width=0.47\textwidth]{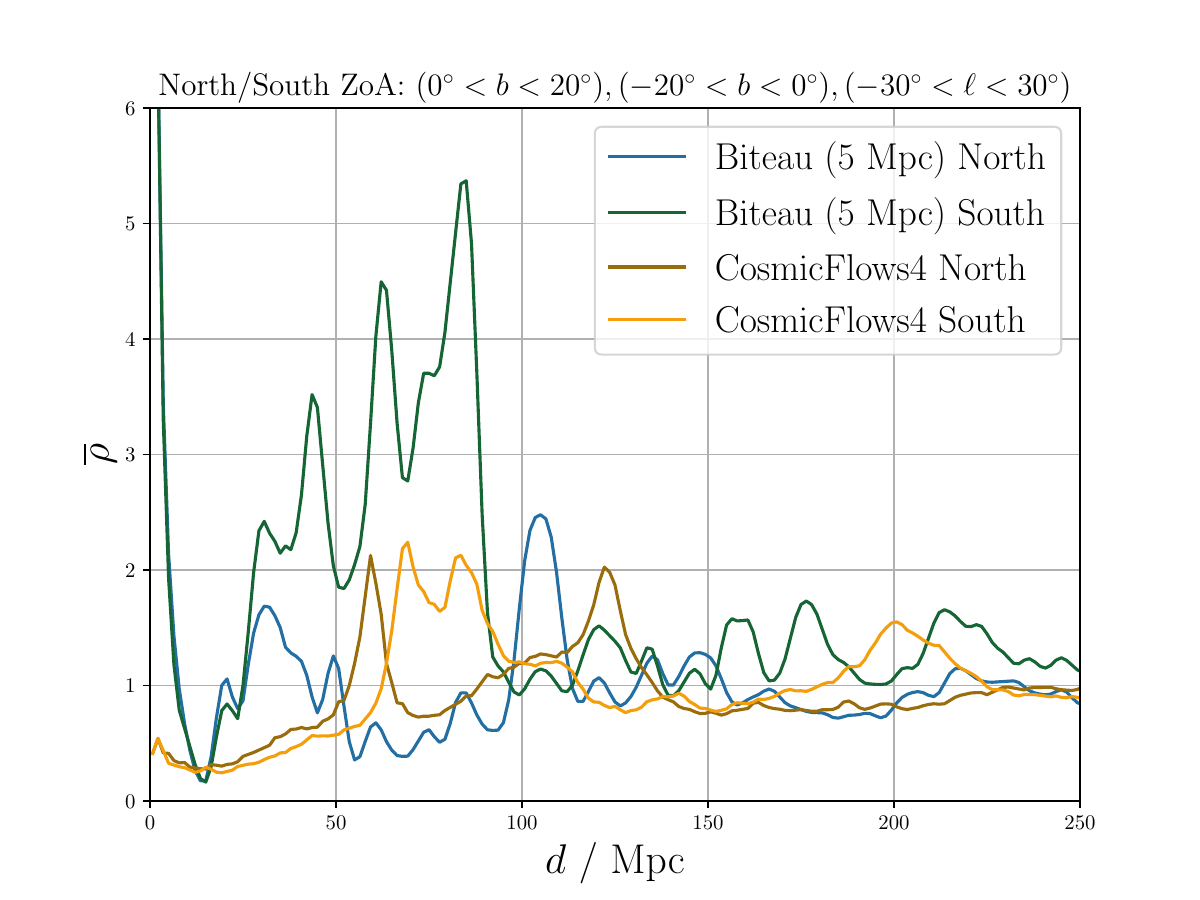}
\caption{Upper four panels: Overdensity as a function of radius in 4 illustrative angular regions containing clusters.  Lower left: Overdensity as a function of radius in a major void direction, showing that although the CF4 density field is smoother it accurately matches the 5 Mpc Biteau density field in the void distance range, beyond the local volume.  The last plot shows the mass density in the North and South ZoA regions as modeled in CF4 (brown, gold) compared to Biteau's cloning scheme (green, blue).  The cloning significantly distorts the mass density in the ZoA, commonly disagreeing with CF4 by a  factor $2-5$. }
\label{fig:selectLoSden}
\end{figure*}

We assess the possibility that use of Hubble-flow distances may spread mass between shells in the Biteau modeling, as follows.  We select four regions at different positions in the sky containing major clusters and plot the over-density along each of their lines of sight according to CF4 and B21; these are shown in Fig.~\ref{fig:selectLoSden}.  In the first row plots -- Virgo and Hydra-Centaurus/Antlia -- the CF4 mass reconstruction is much more concentrated radially than in B21, as is expected for the portion of the galaxies in B21 having redshift-based distance assignments which spread their masses over a larger apparent radial distance range. 
A more non-trivial mis-assignment of distance might be able to explain the apparent mass redistribution for NGC1407/Antlia.  

Distance inaccuracies, however, cannot explain all of the differences between CF and B21 outside the ZoA.  An example is the Pavo-Indus region shown in the righthand plot in the middle row of Fig.~\ref{fig:selectLoSden}.  The integrated mass from about $20-80$ Mpc is simply a lot larger in this angular region according to Biteau than according to CosmicFlows. One might conjecture that  the lack of good constraining data in the adjacent ZoA may lead CosmicFlows to incorrectly ``redistribute" mass into the ZoA, but the skymaps in the relevant distance region in Fig.~\ref{skymap} do not support this idea.  It is a mystery why there is such a clear overdensity in this region according to the Biteau catalog, but no identifiable corresponding mass in CF4.  Our tentative assessment is that the projected mass density from the Biteau catalog in this angular region may be more reliable than that of CosmicFlows,  while the radial distribution as deduced from the Biteau approach may be unreliable in detail due to large peculiar velocities expected if indeed such a large over-density is present.

\subsection{Bias}
The CosmicFlows approach is sensitive to the total mass distribution whereas Biteau's galaxy-catalog approach measures the stellar mass distribution.  Baryonic processes such as AGN and SNe feedback directly impact baryons but only indirectly dark matter.  Hence locally the overdensity in stellar mass does not track the overdensity of total mass.  The $5-10$ Mpc distance scales over which CF4 is resolved and over which we smear the Biteau galaxies is so large that one would expect the baryonic and total mass densities to track one another, however for completeness, we tried fitting the distributions allowing for a non-linear relationship between the DM and stellar mass densities.  There was no such effect.  

\section{Summary and Conclusions}
\label{sec:discussion}

We have compared the CosmicFlows and Biteau models of the  distribution of mass within about 300 Mpc. The models show agreement in the grossest of terms, but have significant disagreements as well.  The CosmicFlows2 density distribution we have used, derived by N. Globus from the work of~\citet{2018NatAs...2..792H}, is significantly smoother than the current CosmicFlows4 model. CF4 should presumably be considered the state-of-the-art model using the CosmicFlows approach and studies of the UHECR dipole anisotropy based on CF2 such as ~\citet{BF24} and ~\citet{bfu24}, should be repeated with CF4.

We found significant differences between CF4 and the Biteau mass density model smoothed on both 5 and 10 Mpc scales, although qualitatively they are generally similar.  
We concluded that the portion of the BF21 model based on the Local Volume catalog~\citep{karachentsev_morphological_2018} with distance measures not relying on redshift, i.e., distances $\lesssim 11$ Mpc, can be expected to give a superior description of the local density distribution than CF4; it shows a significantly larger local overdensity than CF4. 

At larger distances, we saw that the line-of-sight structure in the mass density distribution is in general different between CosmicFlows and the Biteau model, with mass spread out radially in B21 relative to its true distance distribution due to ``finger-of-God" artifacts in Hubble-flow distance determinations.  The CosmicFlows approach has the benefit of imposing a physical prior on the character of mass concentrations, so may in general be superior in regions where few galaxy distances are directly measured.  However there are some regions where we found differences between CosmicFlows' collective-mass-flow modeling and direct observation of the line-of-sight total mass, which we were unable to explain; see our discussion of the Pavo-Indus region at the end of Sec.~\ref{sec:dist}.  

Different applications of the local mass density model are more or less sensitive to having accurate information on the radial density distribution.  At the highest energies, the UHECR flux is very sensitive to the distance of the source, because the composition and energy of a UHE cosmic ray evolves rapidly during propagation due to interactions with the extragalactic background light.  But at lower energies where the ``horizon" is hundreds of Mpc, the lack of robust distance information is less of a liability. 

An important difference between the Cosmic Flows and Biteau density fields is that beyond 11 Mpc the latter has large scale ``fake'' structure due to the cloning procedure used to fill-in the zone of avoidance.  This is mosst problematic in the central ZoA ($|\ell|<30^\circ$) where the region $|b|<20^\circ$ is filled by mirroring galaxies across $b=\pm20^\circ$. Due to the large size of the ZoA and the presence in some distance ranges of major overdensities which become mirrored into the regions near the Galactic plane, cloning may significantly impact predictions for the UHECR dipole anisotropy, especially its magnitude~\cite{bfu24}.  However flux from behind the Galactic center is de-magnified by lensing in the Galactic magnetic field~\citep{fComptRend14,fsCRdefs17,bfu24}, so the dipole may be insensitive to artifacts of cloning. The sensitivity to artifacts of cloning can be seen by calculating the anisotropy with and without mass from the ZoA; if the results differ the predictions cannot be trusted. It is impossible to know without actually carrying out that analysis, how much the results will differ.  

For galaxy catalogs to be more broadly useful for finding the mass density, methods should be developed to fill in the zone of avoidance by reweighting observed galaxies based on modeling the impact of dust and gas, as discussed in Sec.~\ref{sec:cloning}. Potentially a method can also be developed to combine the benefits of CosmicFlows' prior on the spatial structure, with information from real galaxies with robust distance measures to obtain a yet more accurate model of the distribution of mass within a couple-hundred Mpc of the Milky Way. Then, anisotropies in the arrival directions of UHECRs will become a more powerful tool for inferring properties of the sources, as well as constraining cosmic magnetic fields.

\begin{acknowledgments}
We acknowledge and thank Teresa Bister for her comments and contributions to this work, and note the forthcoming joint paper (Bister, Farrar and Li, in preparation) which compares predictions for the UHECR anisotropy from the CF2, CF4 and B21 mass models.  We are grateful to Aurélien Valade for generously sharing the MCMC mean posterior field from his simulations using CosmicFlows4 data, and for clarifying the Markov Chains Monte Carlo (MCMC) reconstruction methods. His contributions of data and expertise were essential for the analysis in this work.  We also thank N. Globus for sharing the voxelized CosmicFlows2 density field she developed, J. Biteau for clarification about his procedures, and D. Allard, D. Fielding, I. Karachentsev and M. Unger for useful information.  The research of G.R.F. has been supported by NSF-PHY-221045.

\end{acknowledgments}


\bibliography{sample7}{}
\bibliographystyle{aasjournalv7}

\onecolumngrid
\appendix

\begin{figure}[ht!]
\includegraphics[width=0.33\textwidth]{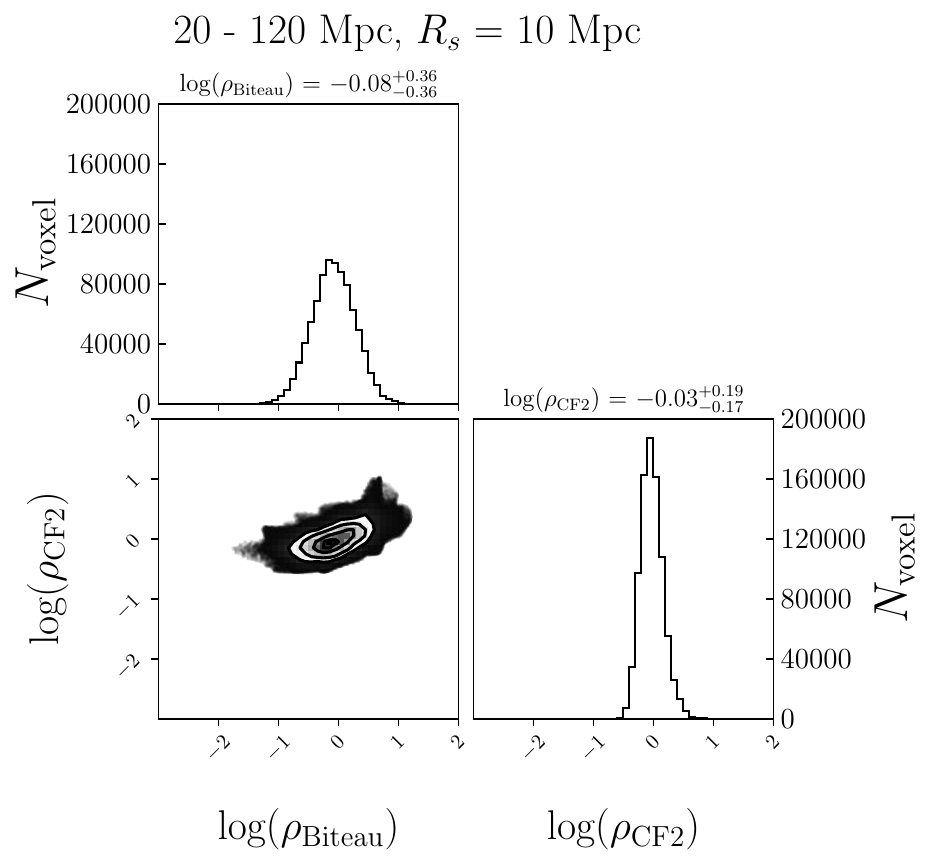}
\includegraphics[width=0.33\textwidth]{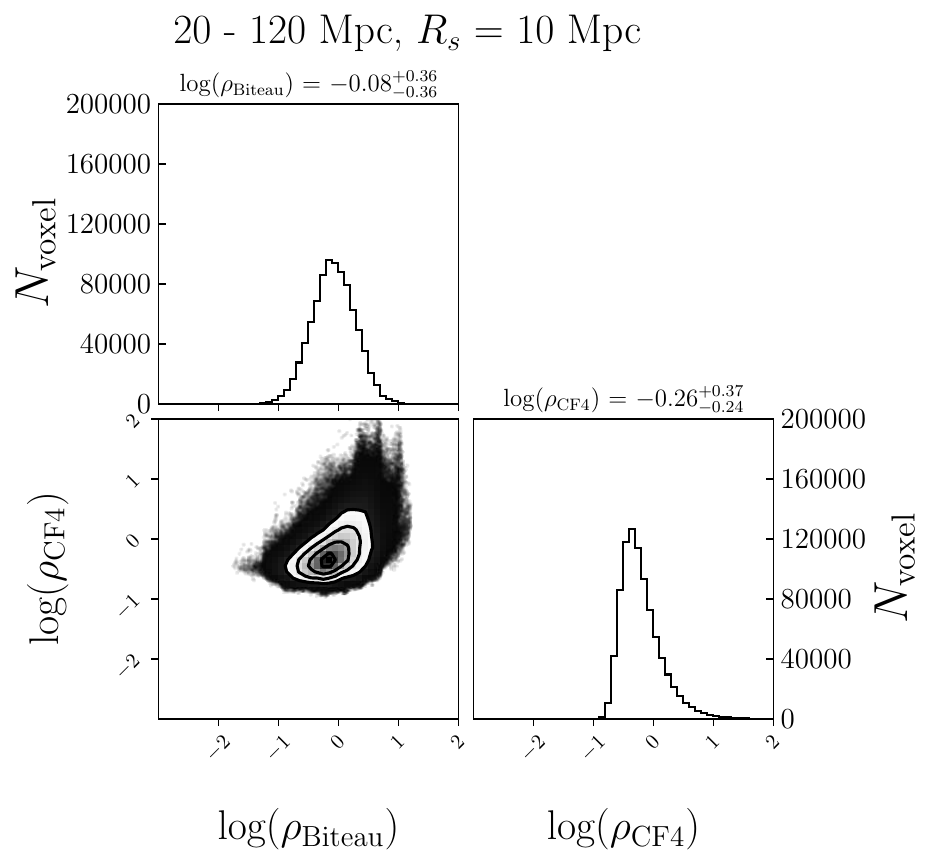}
\includegraphics[width=0.33\textwidth]{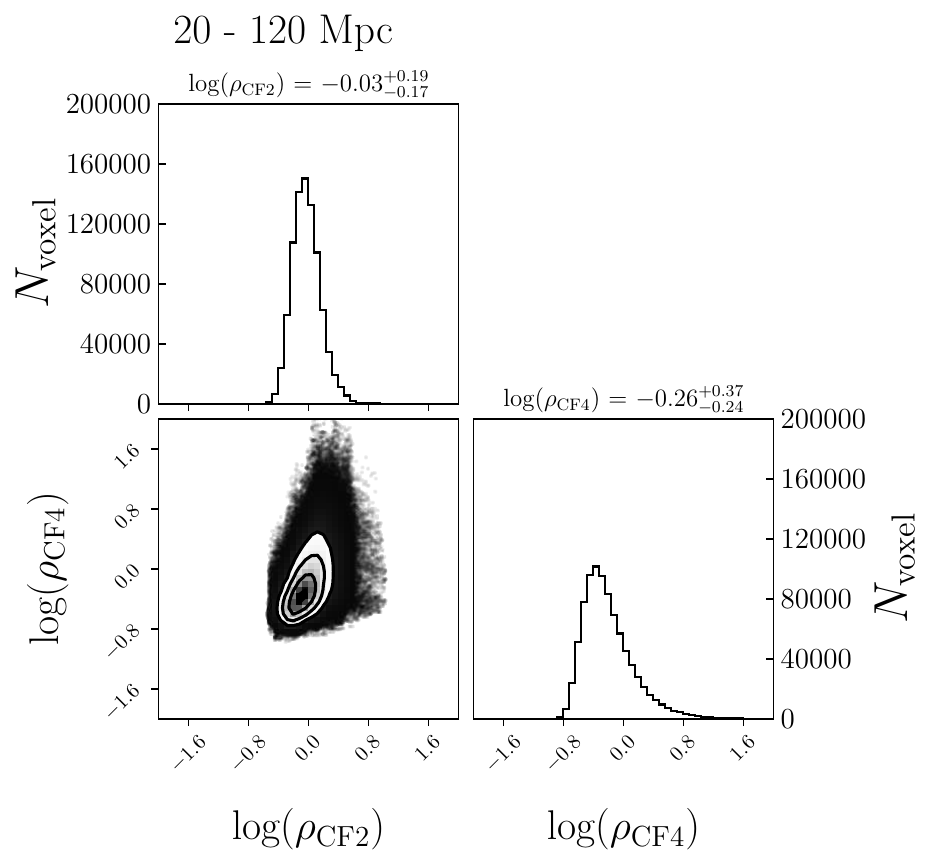}

\includegraphics[width=0.33\textwidth]{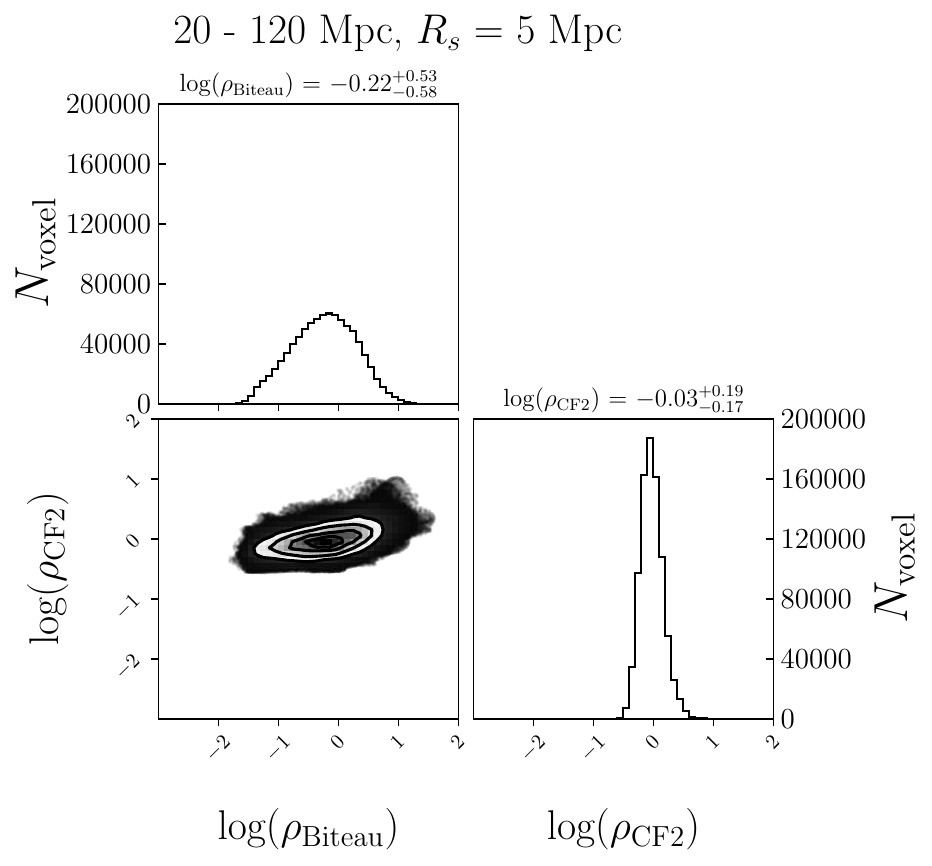}
\includegraphics[width=0.33\textwidth]{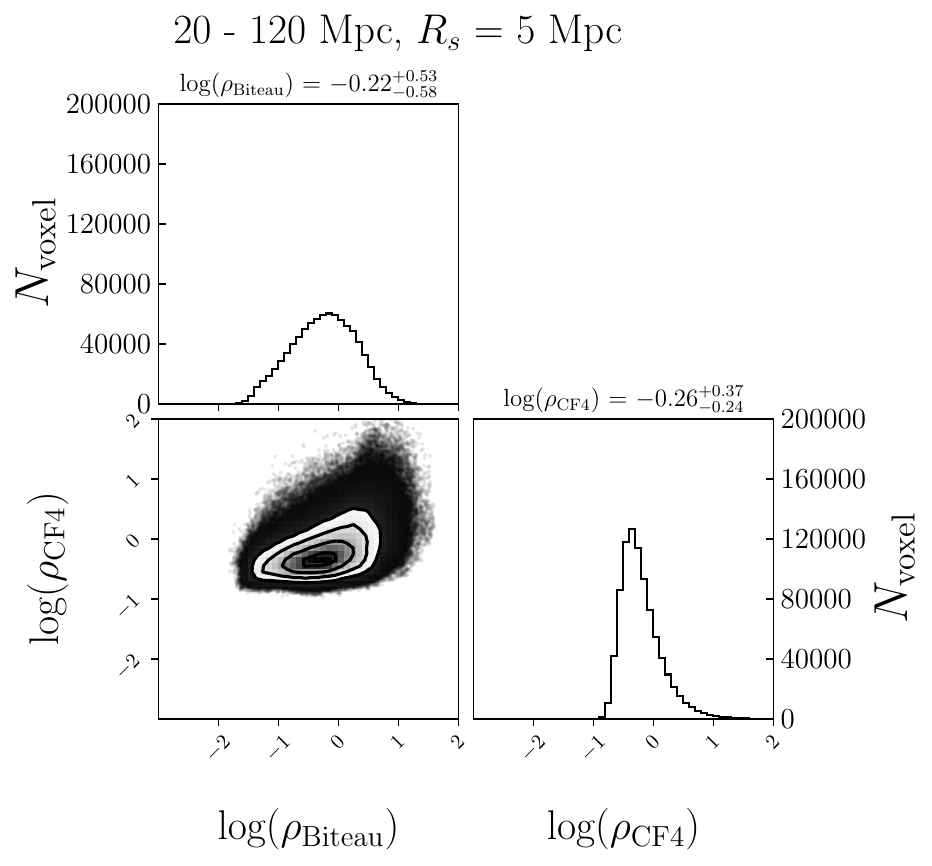}

\caption{Corner plots of the logarithmic mass overdensities in individual voxels. The two axes are log[overdensity] in a given voxel for different combinations of CosmicFlows2 and CosmicFlows4 compared to Biteau's Catalog with smoothing scales of $R_s=5\,\mathrm{Mpc}$ and $R_s=10\,\mathrm{Mpc}$, and CF2 versus CF4 at the far right. Each corner plot includes three subplots: the lower-left shows the 2D correlation between the two fields, while the top-left and bottom-right show the projected distributions of the x-axis or y-axis. The mean values and variances are shown on the top of the distributions. }
\label{corner}
\end{figure}

\begin{figure*}[ht!]
\includegraphics[width=0.245\textwidth]{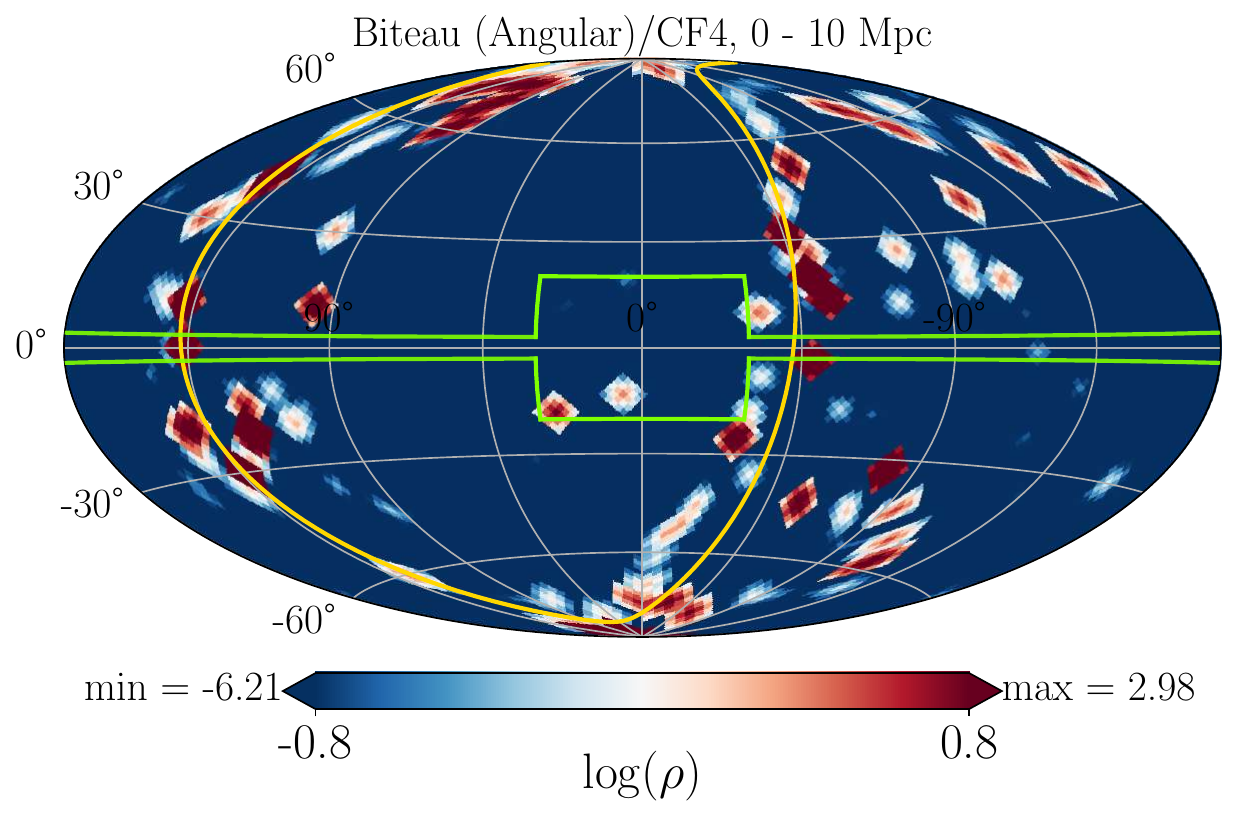}
\includegraphics[width=0.245\textwidth]{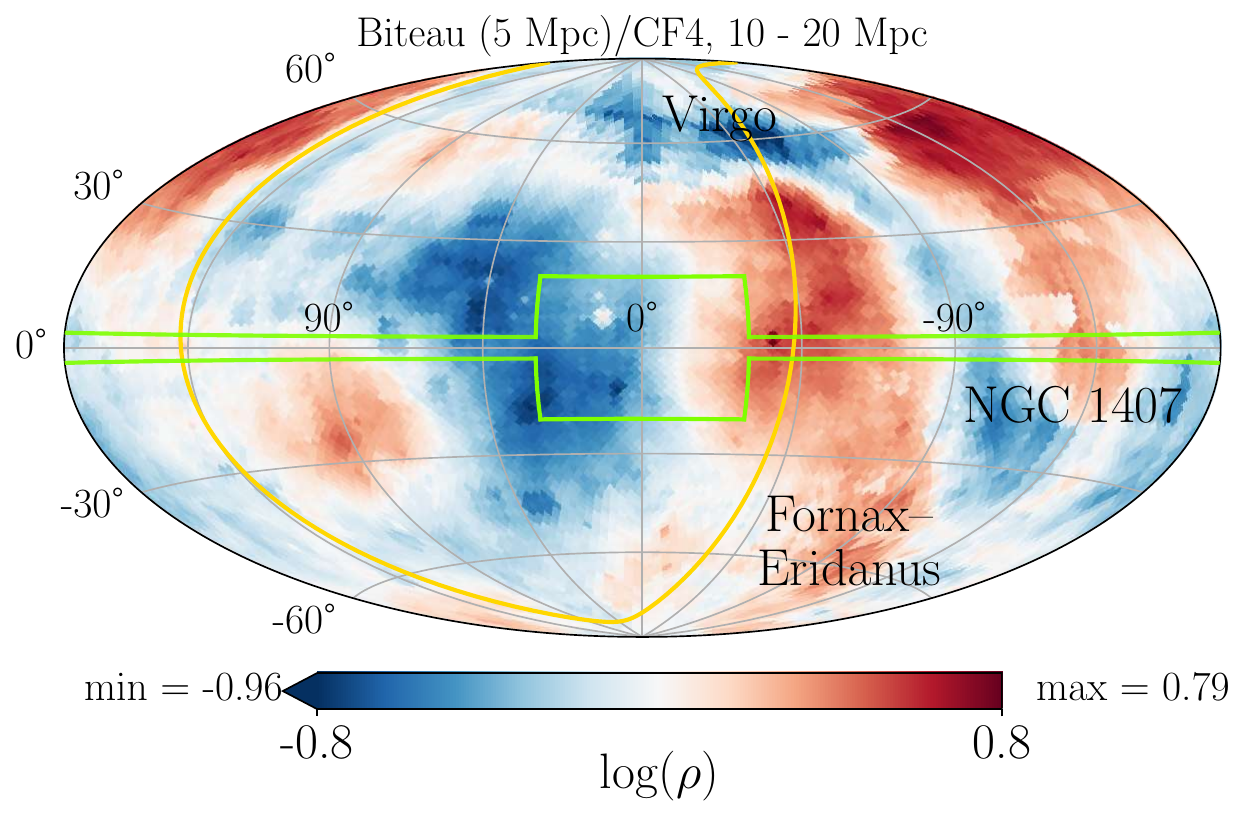}
\includegraphics[width=0.245\textwidth]{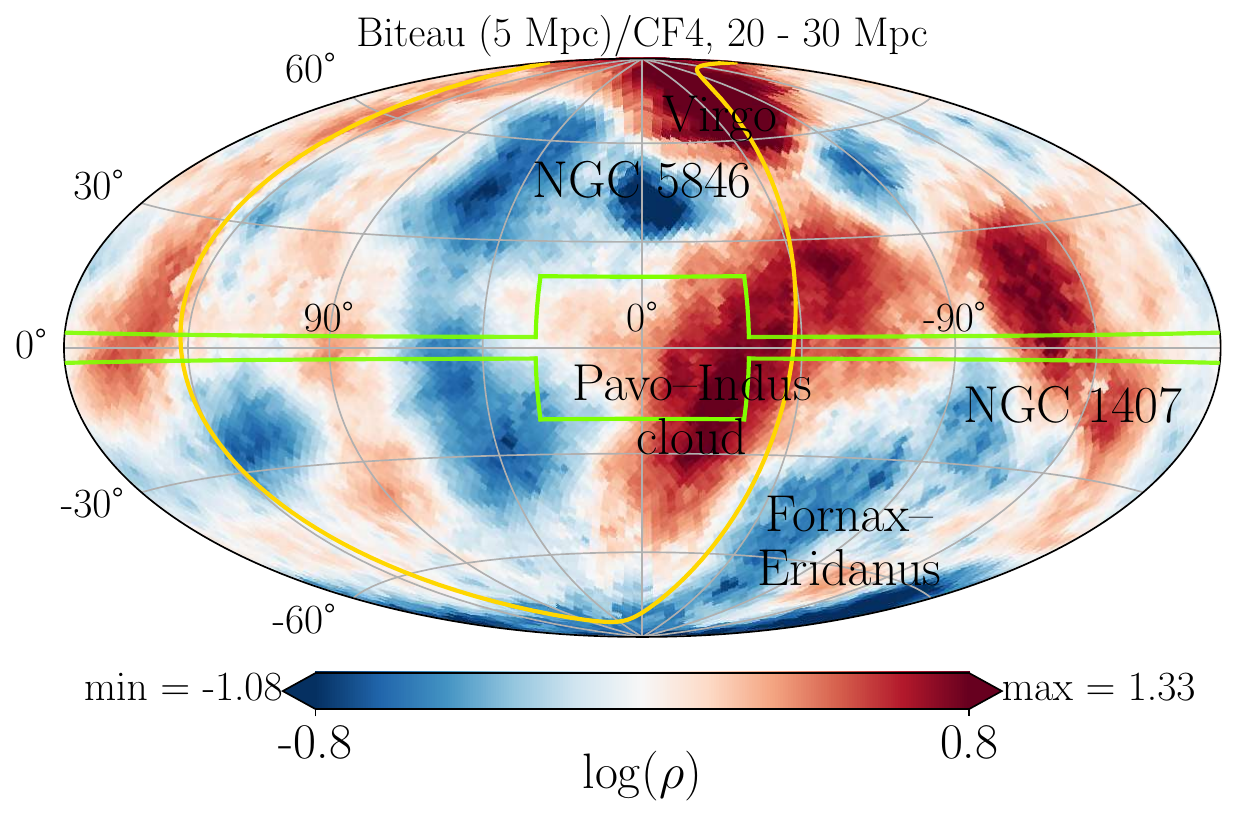}
\includegraphics[width=0.245\textwidth]{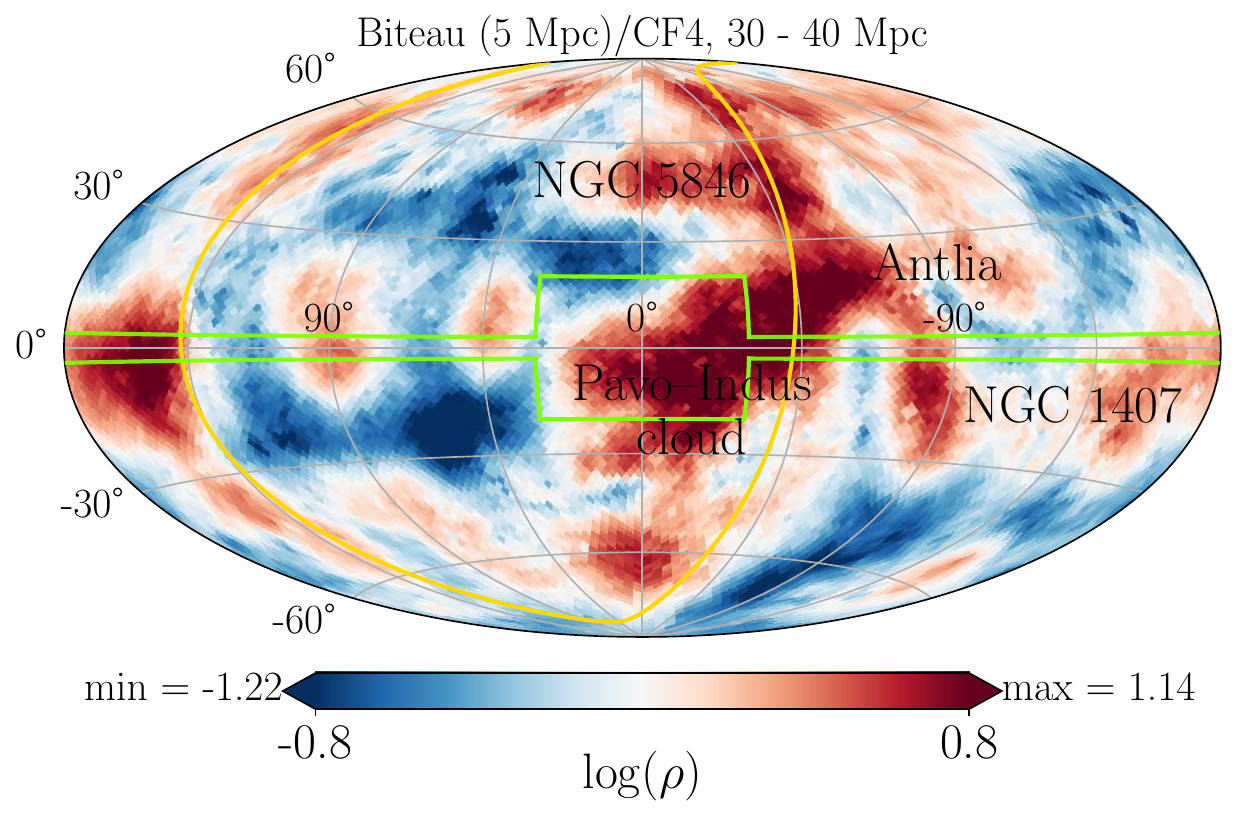}

\includegraphics[width=0.245\textwidth]{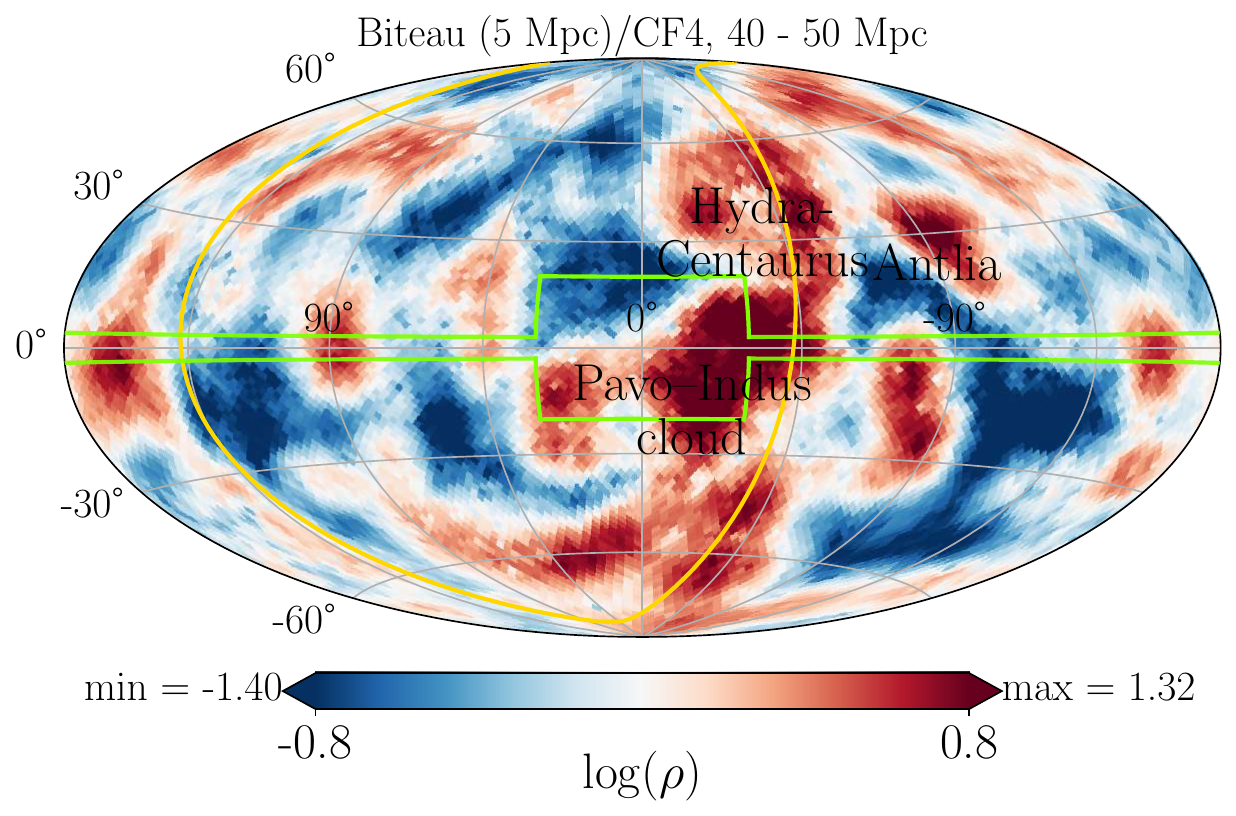}
\includegraphics[width=0.245\textwidth]{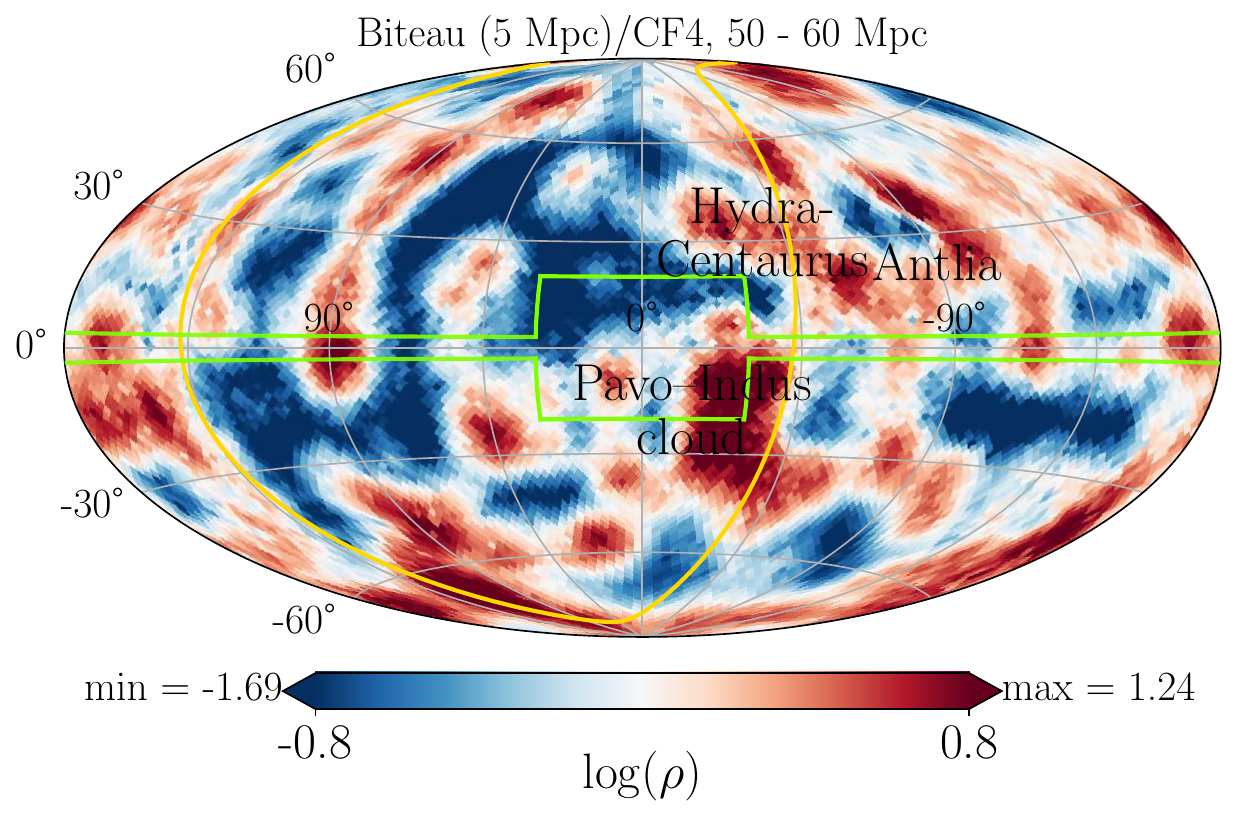}
\includegraphics[width=0.245\textwidth]{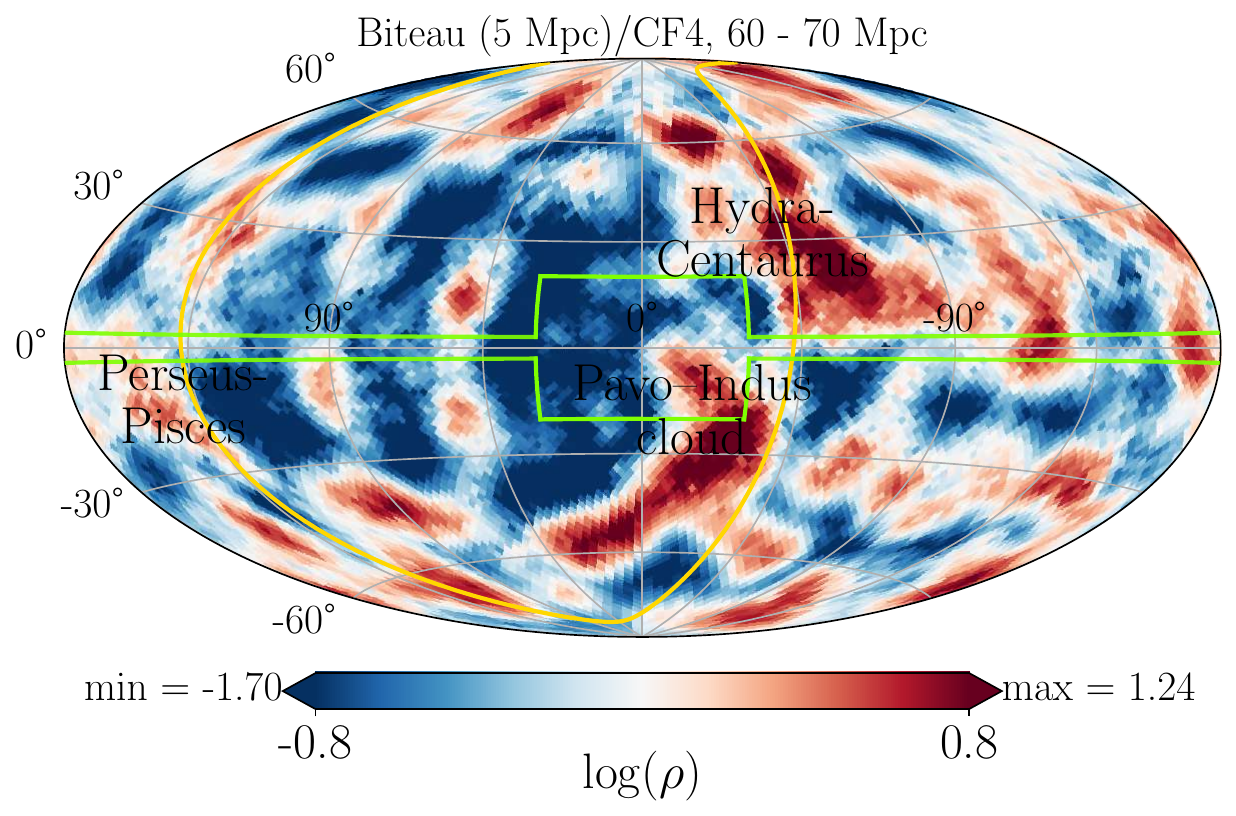}
\includegraphics[width=0.245\textwidth]{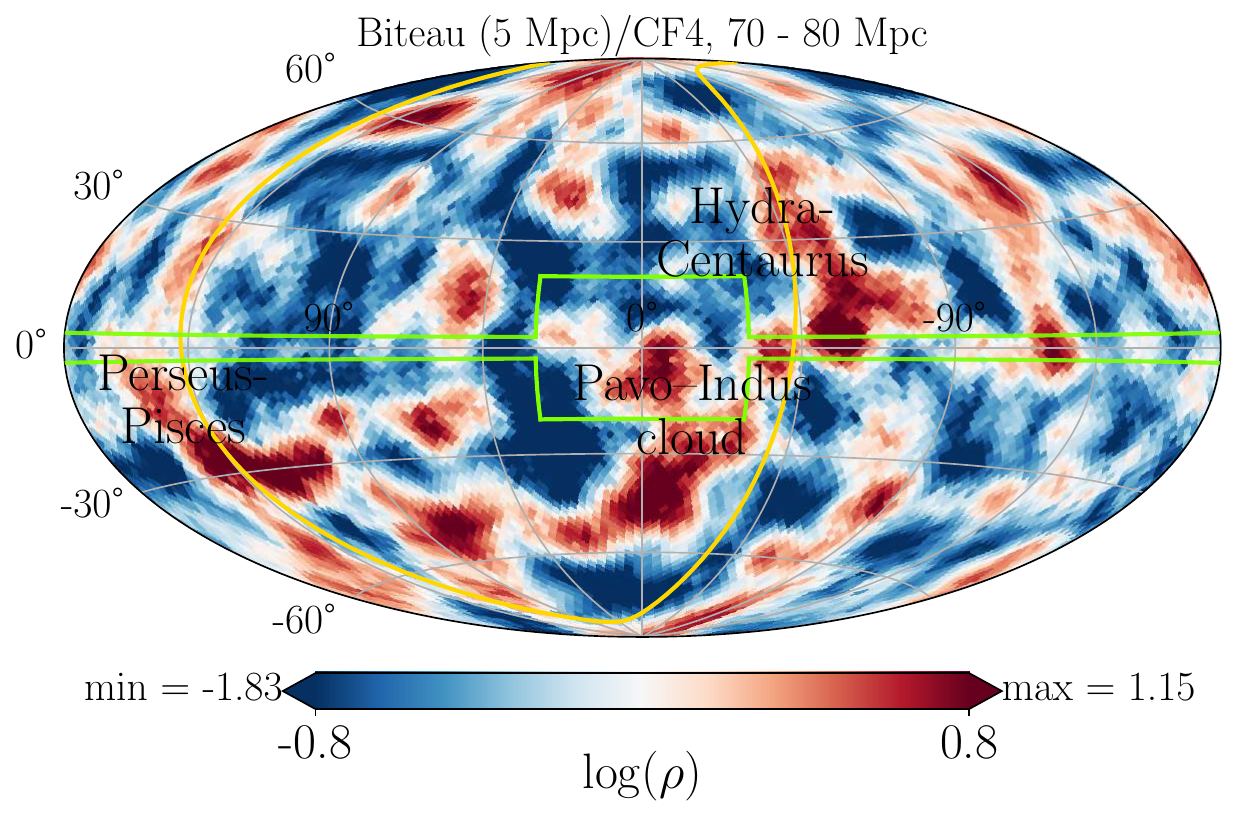}

\includegraphics[width=0.245\textwidth]{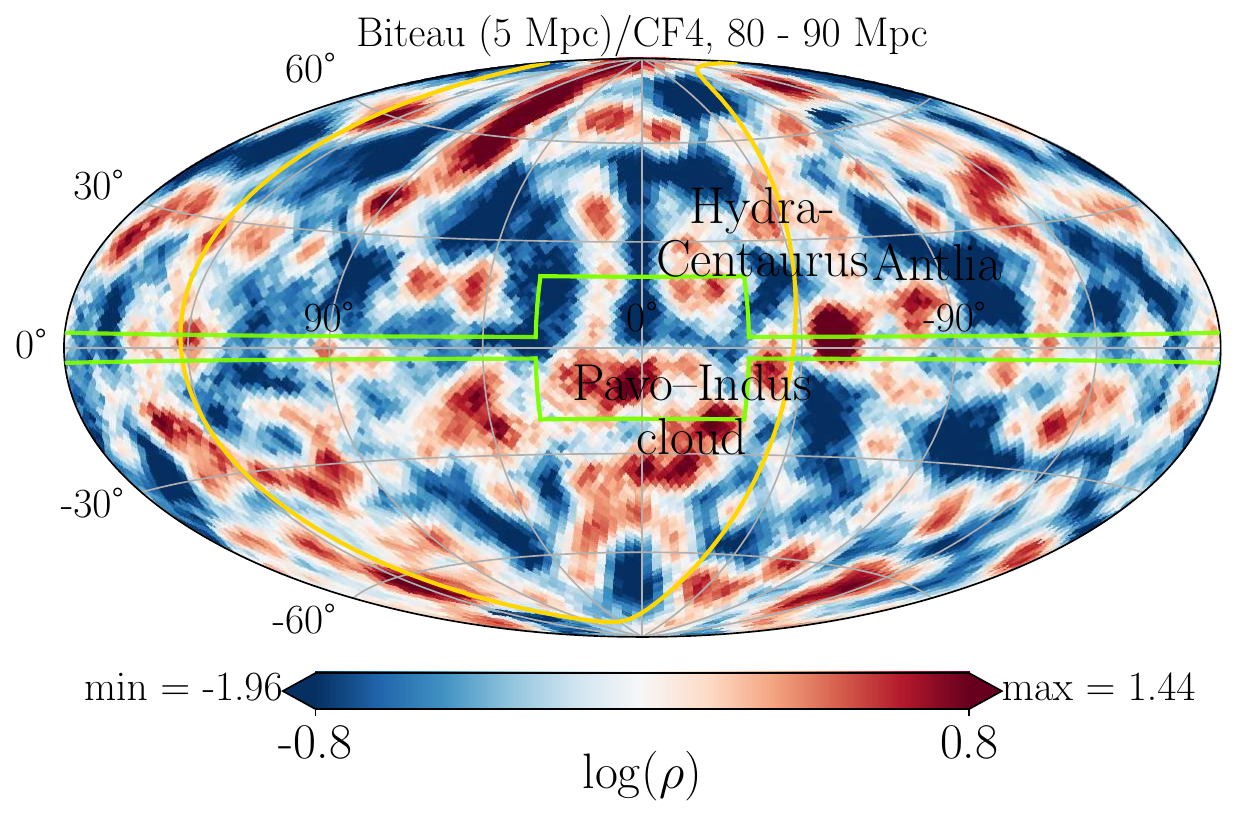}
\includegraphics[width=0.245\textwidth]{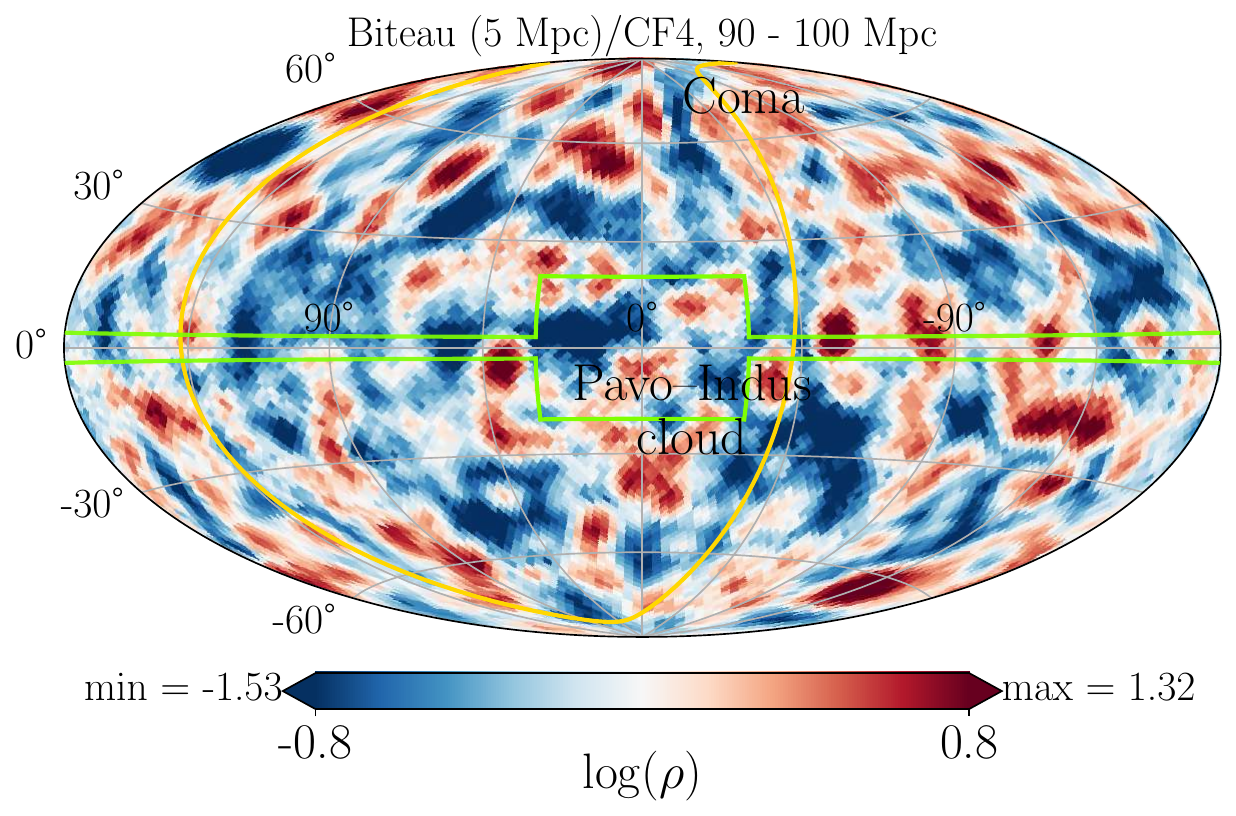}
\includegraphics[width=0.245\textwidth]{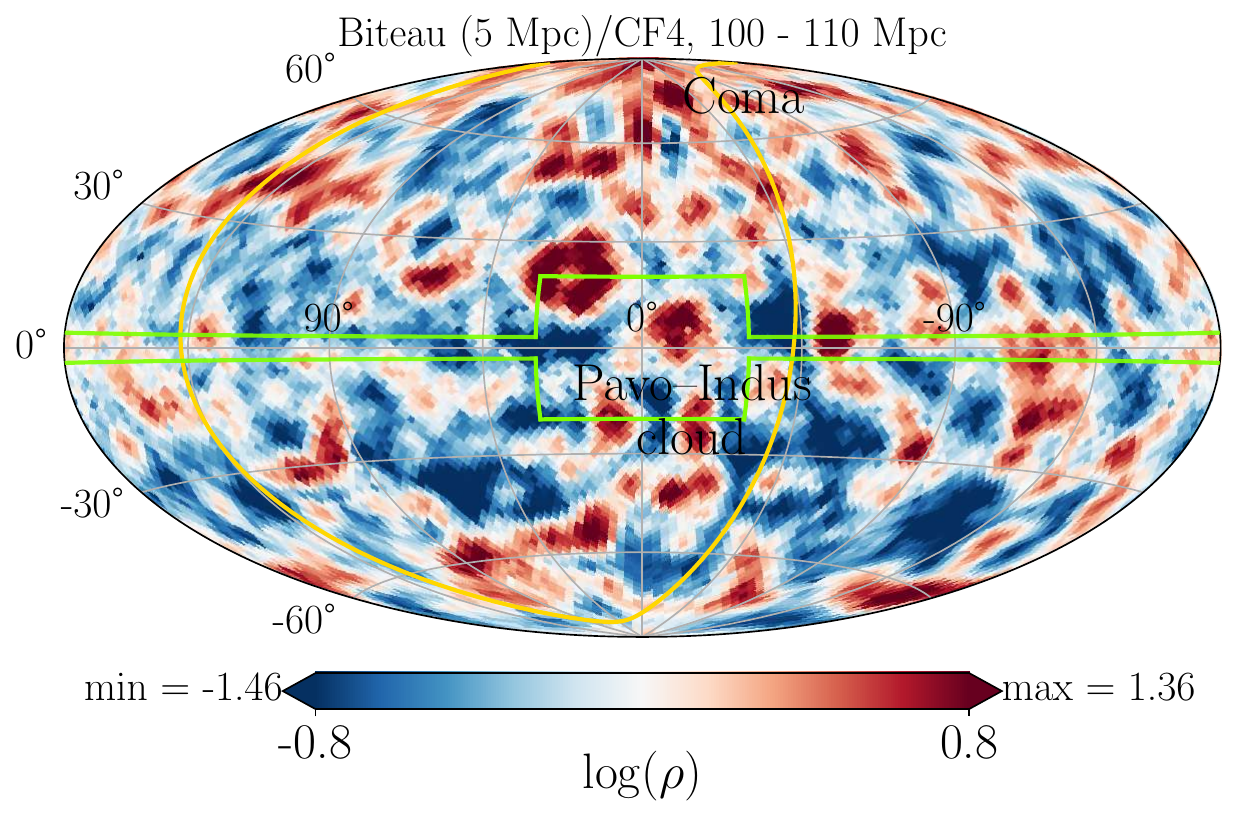}
\includegraphics[width=0.245\textwidth]{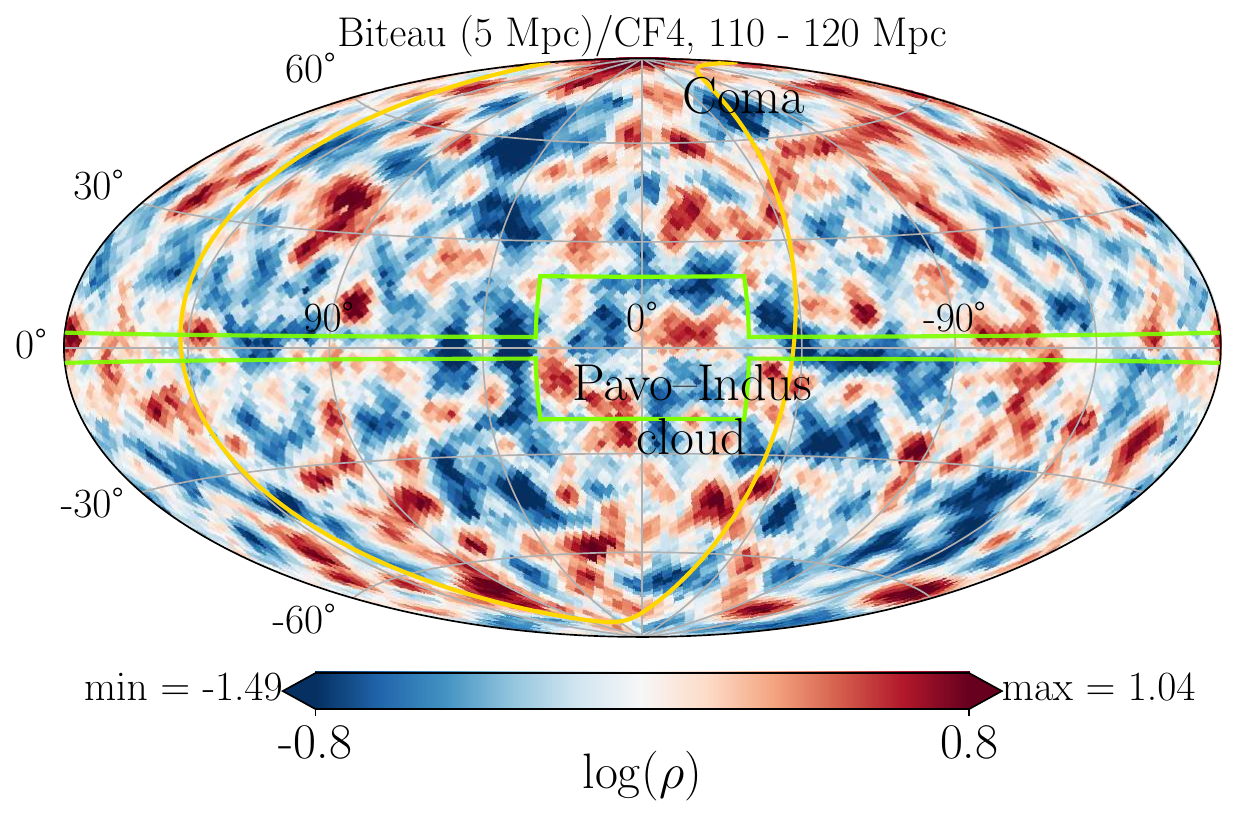}

\includegraphics[width=0.245\textwidth]{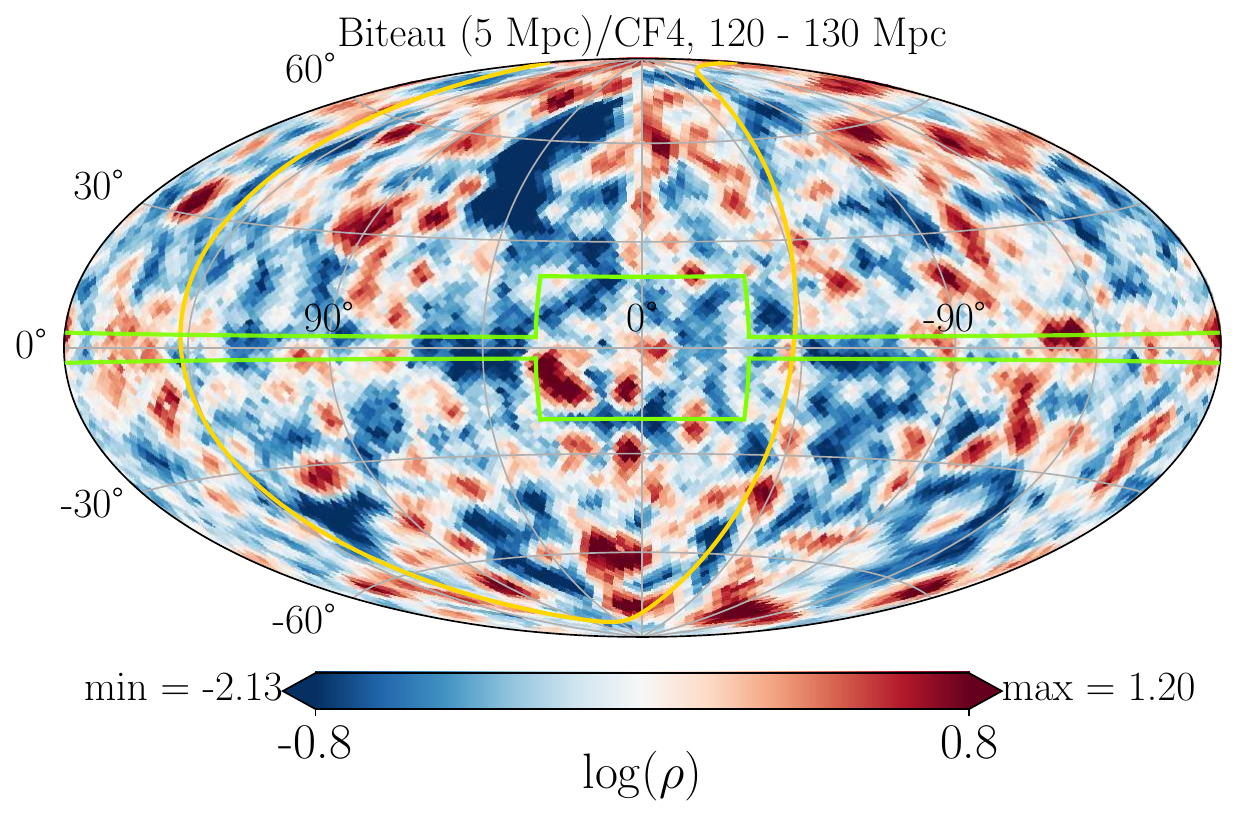}
\includegraphics[width=0.245\textwidth]{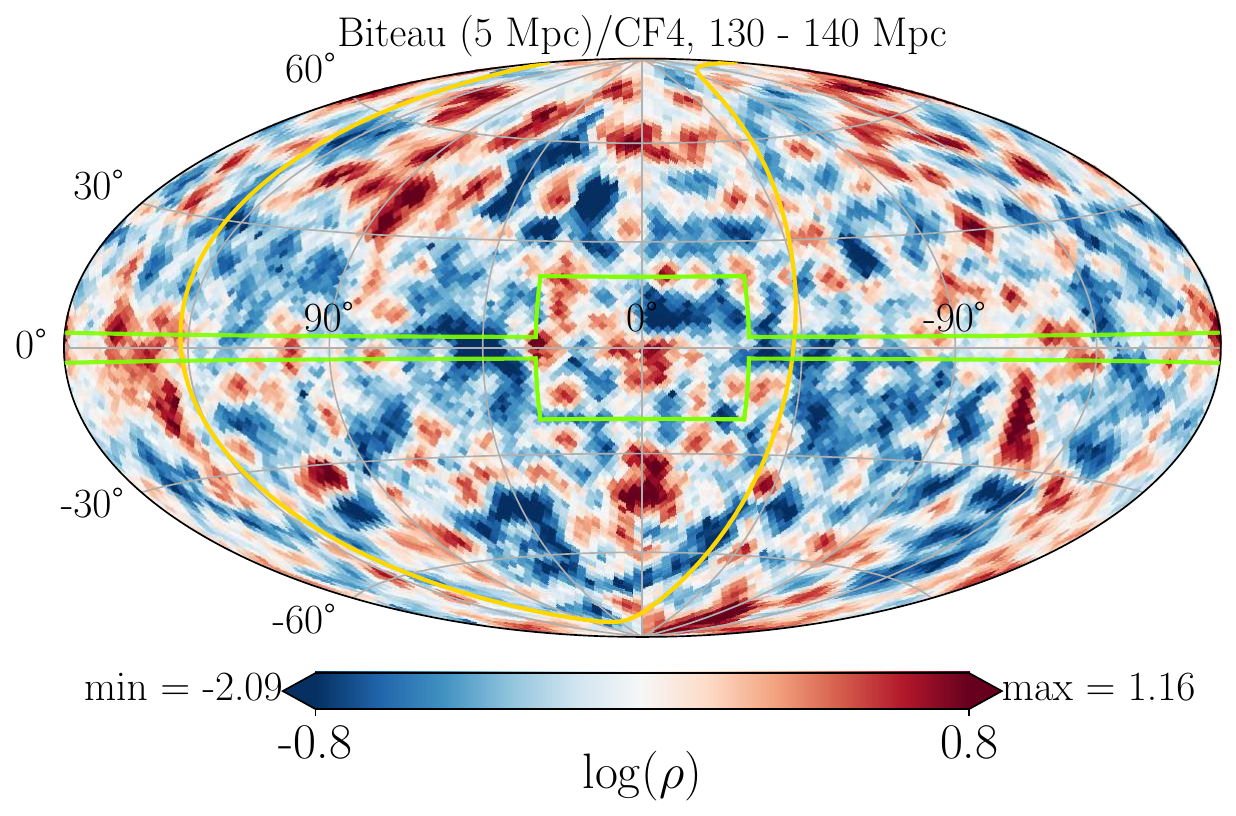}
\includegraphics[width=0.245\textwidth]{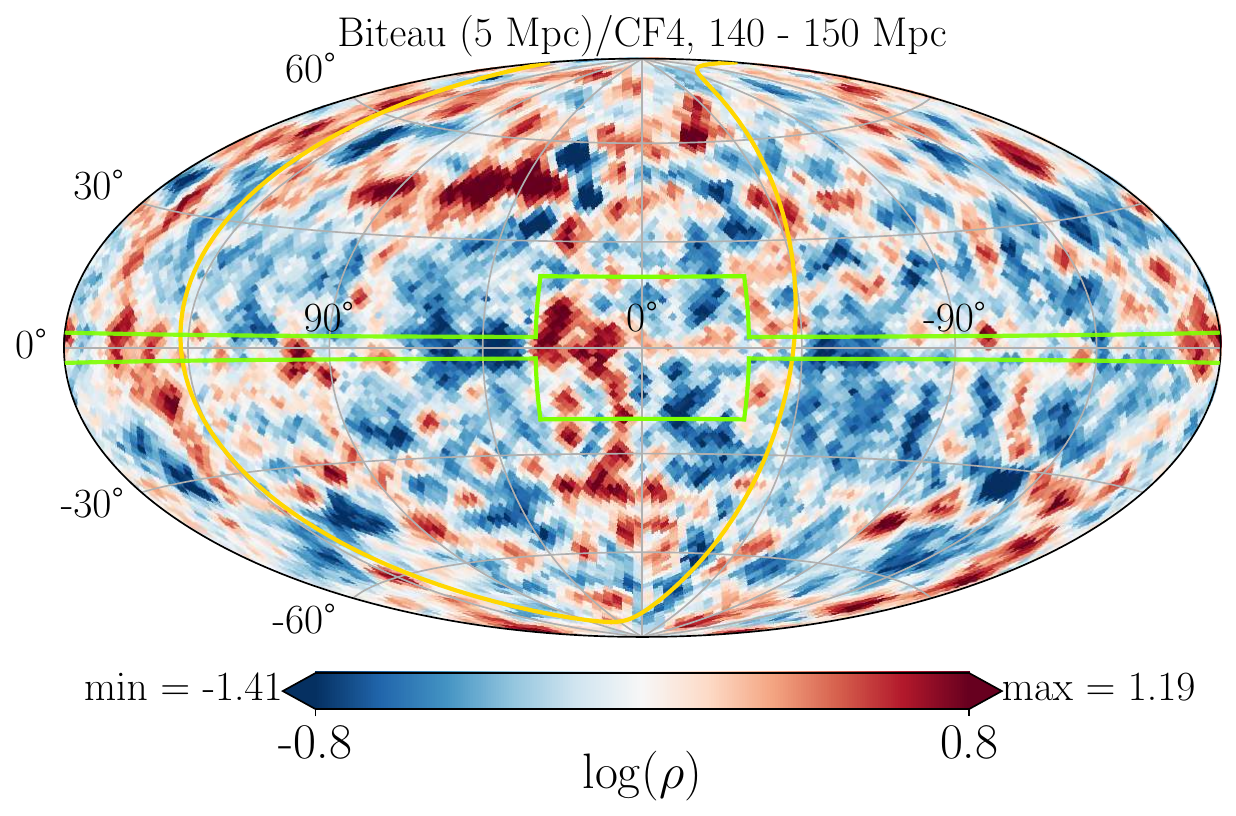}
\includegraphics[width=0.245\textwidth]{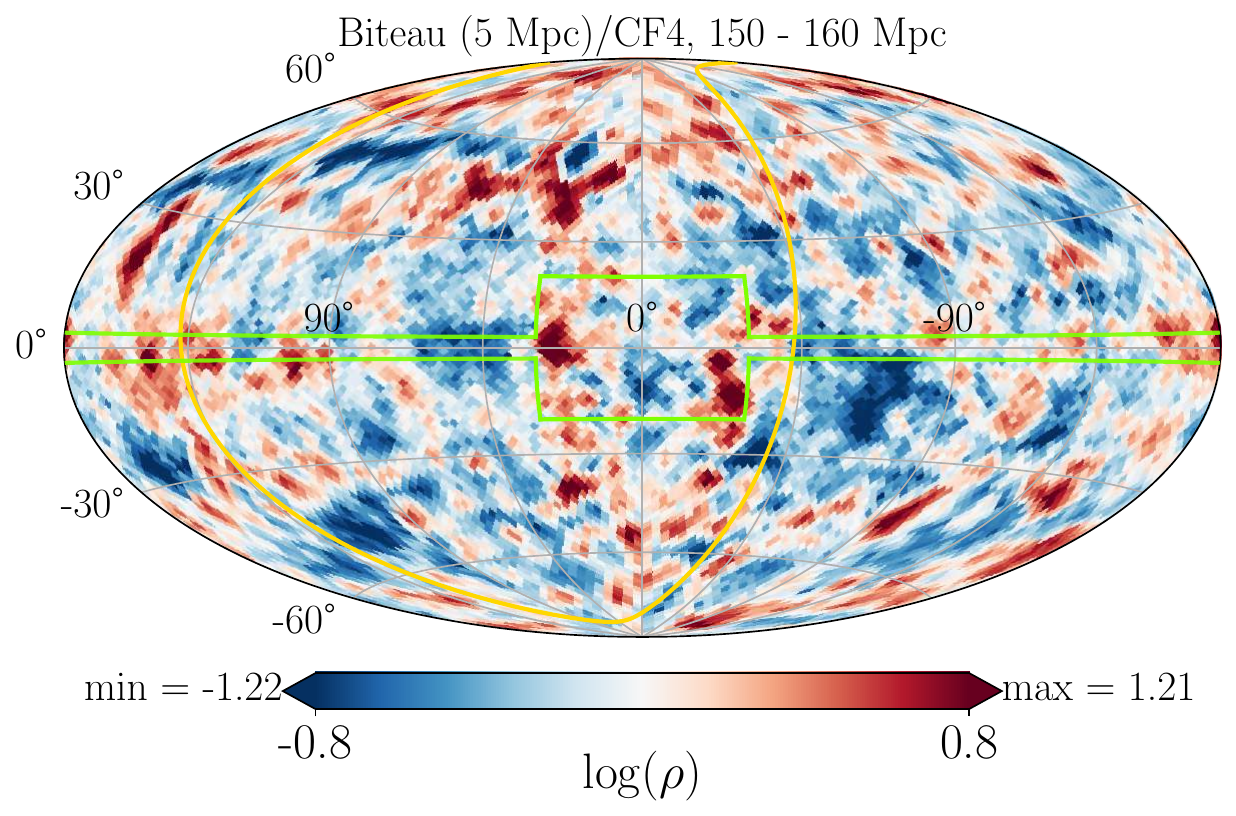}

\includegraphics[width=0.245\textwidth]{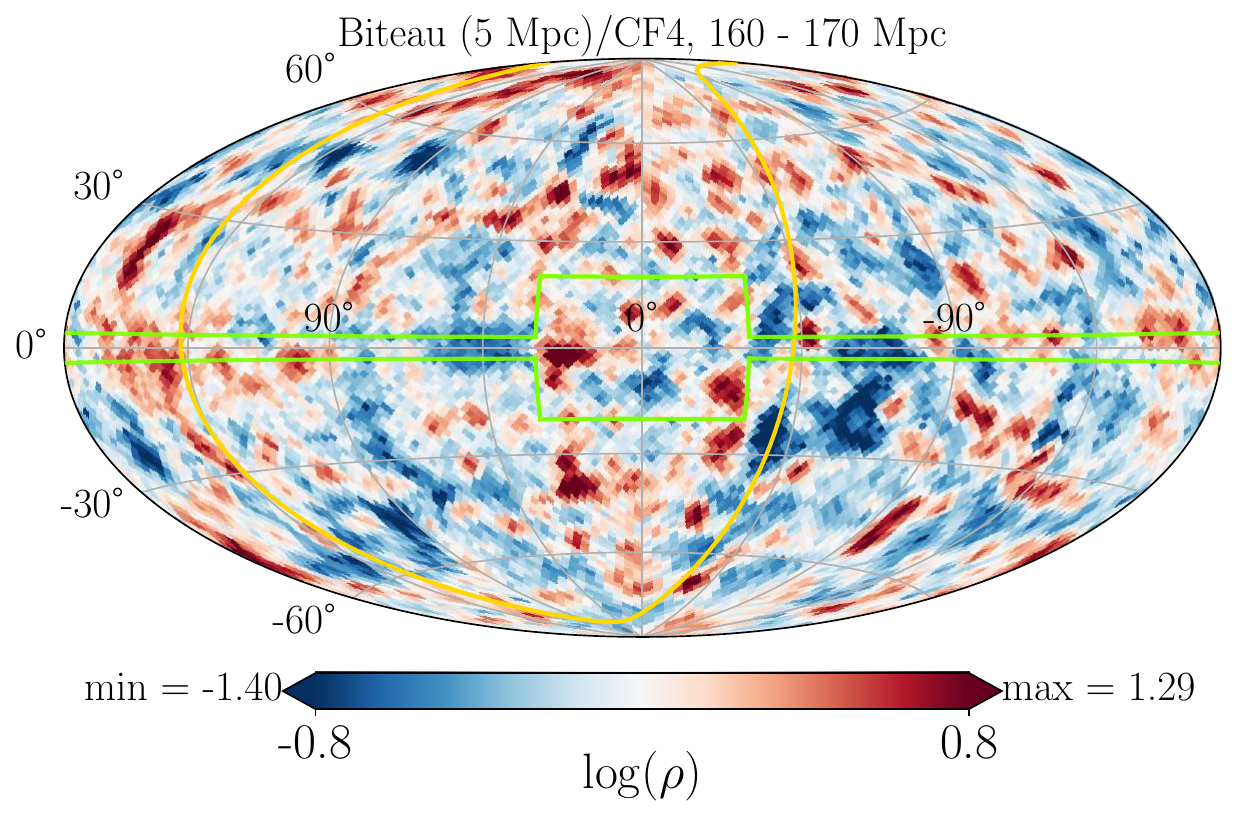}
\includegraphics[width=0.245\textwidth]{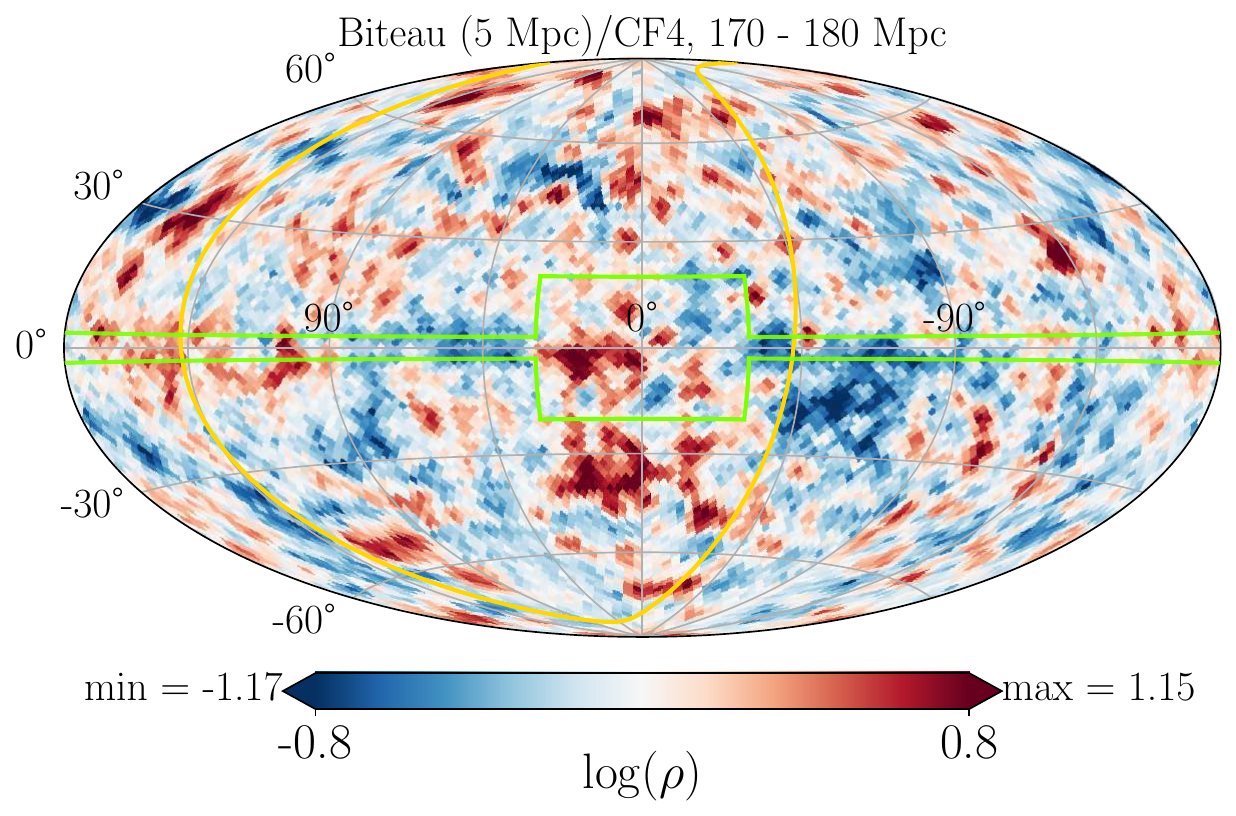}
\includegraphics[width=0.245\textwidth]{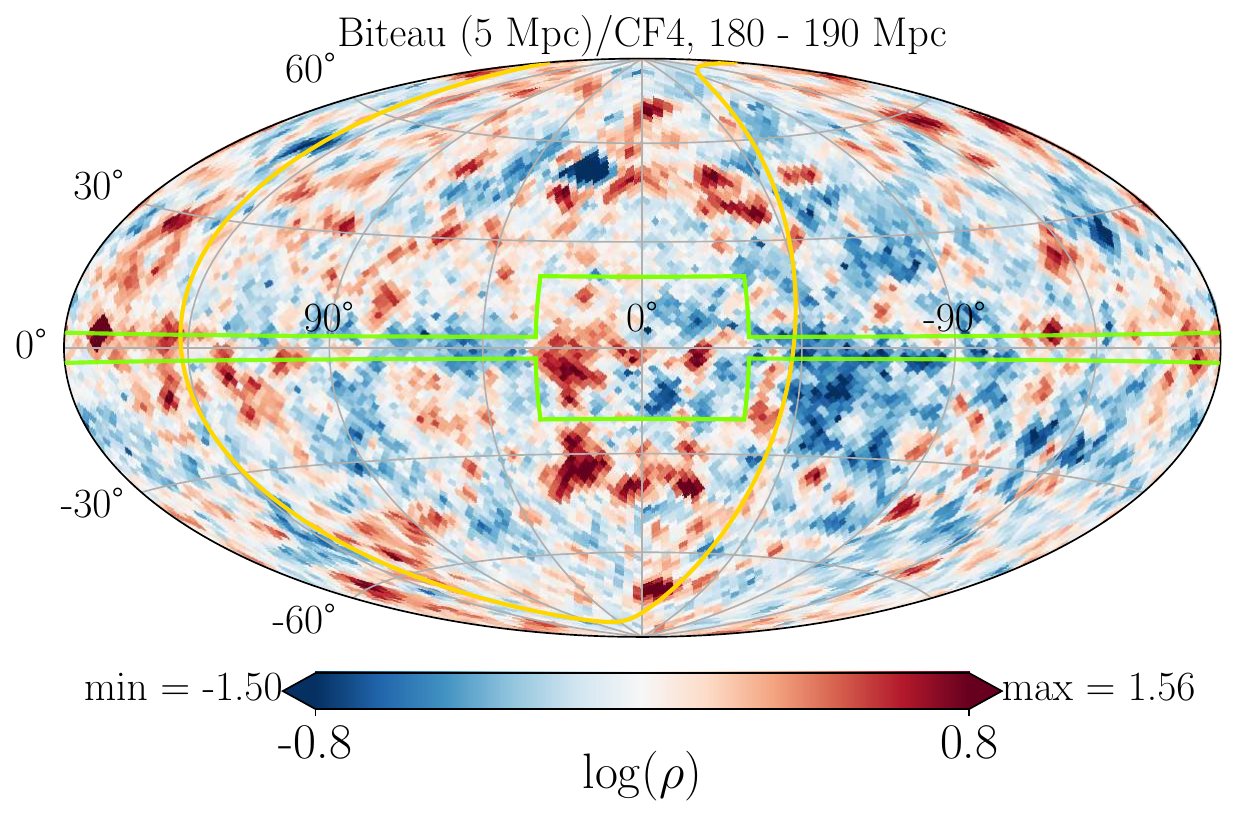}
\includegraphics[width=0.245\textwidth]{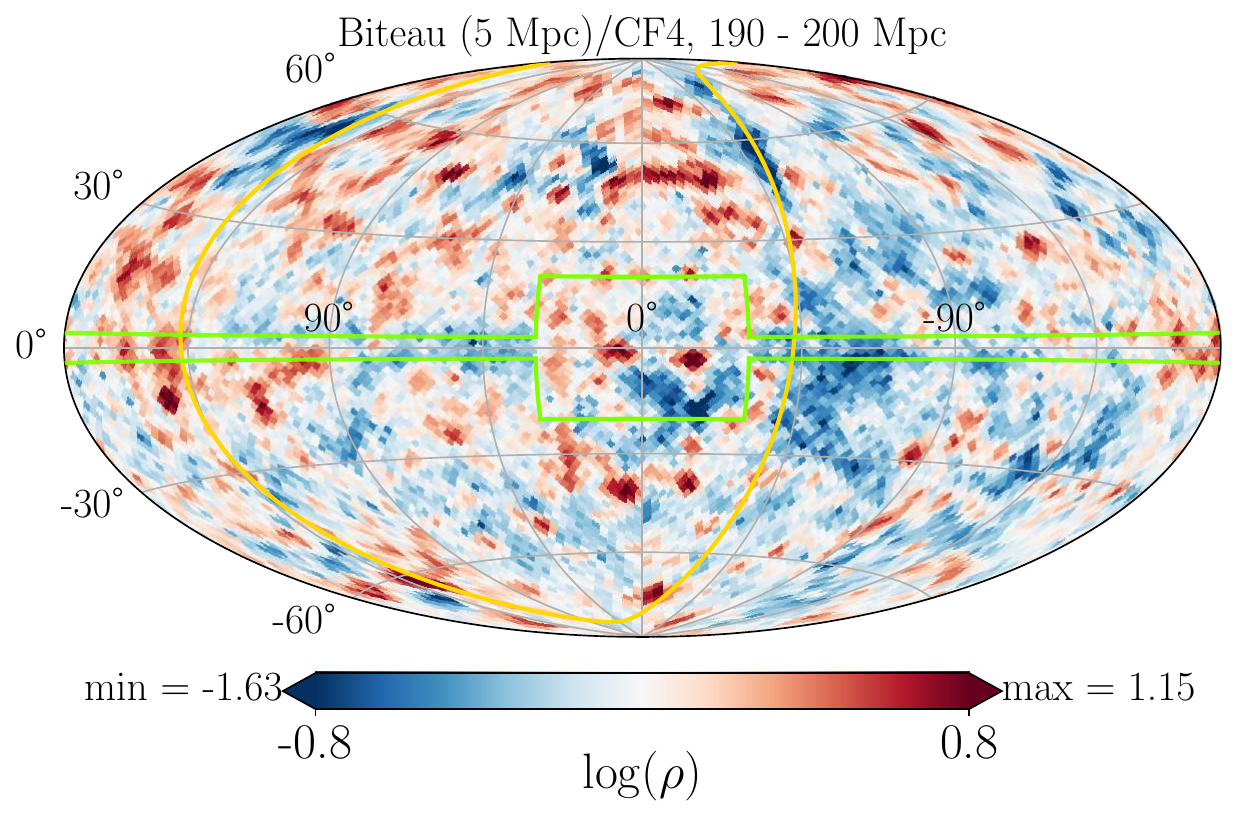}

\includegraphics[width=0.245\textwidth]{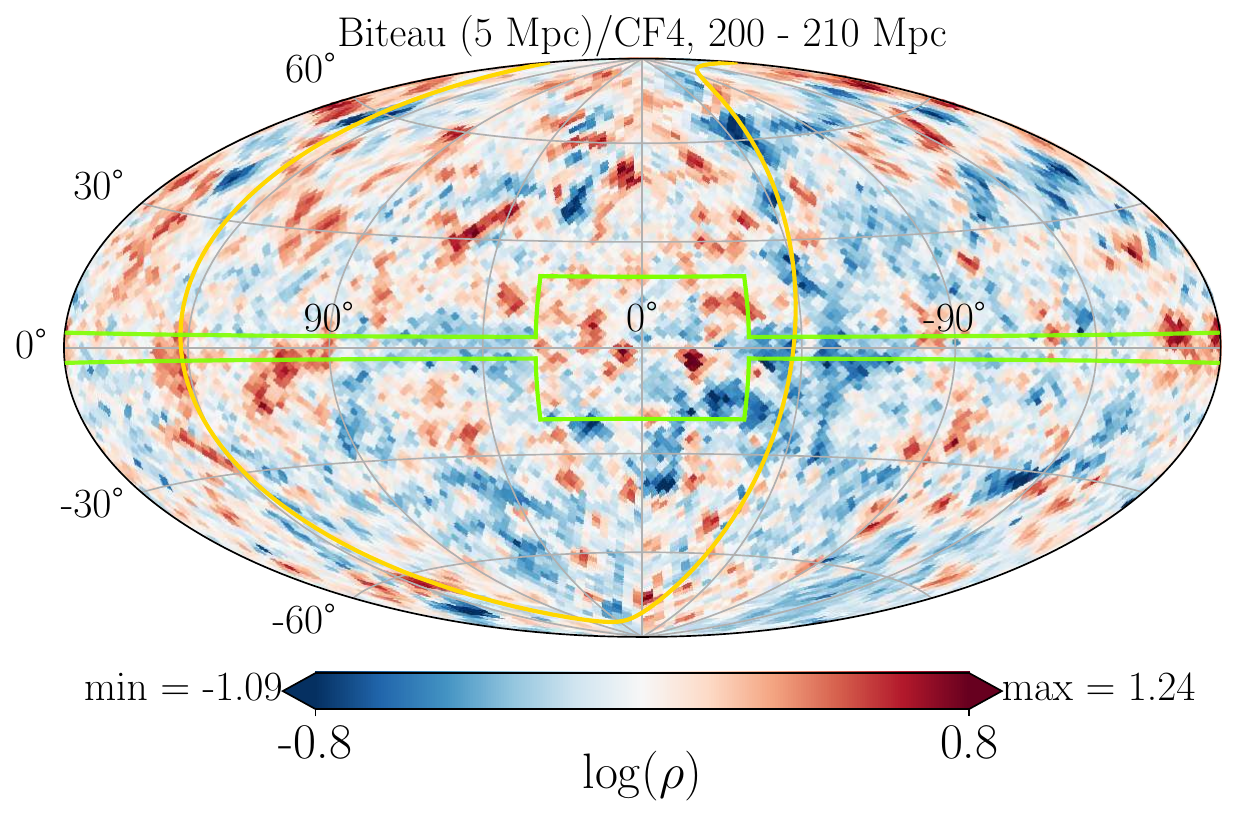}
\includegraphics[width=0.245\textwidth]{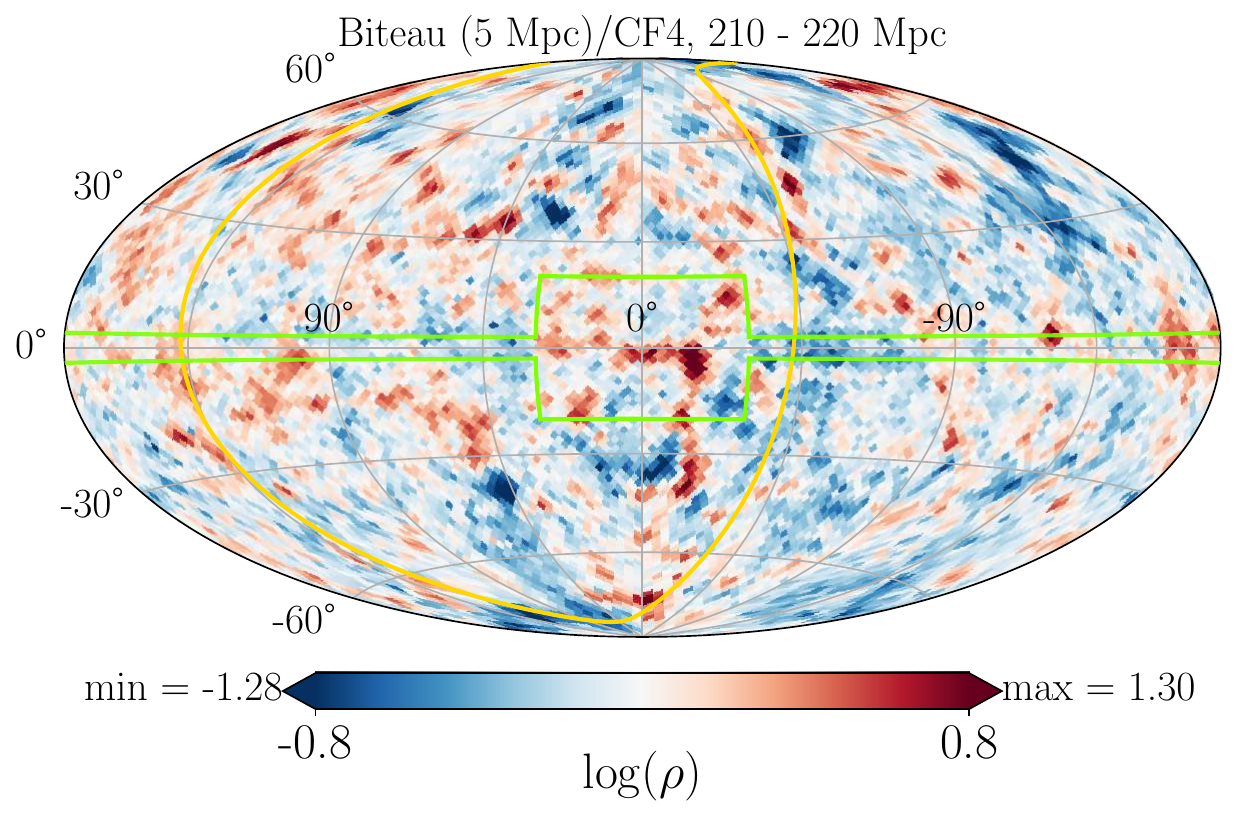}
\includegraphics[width=0.245\textwidth]{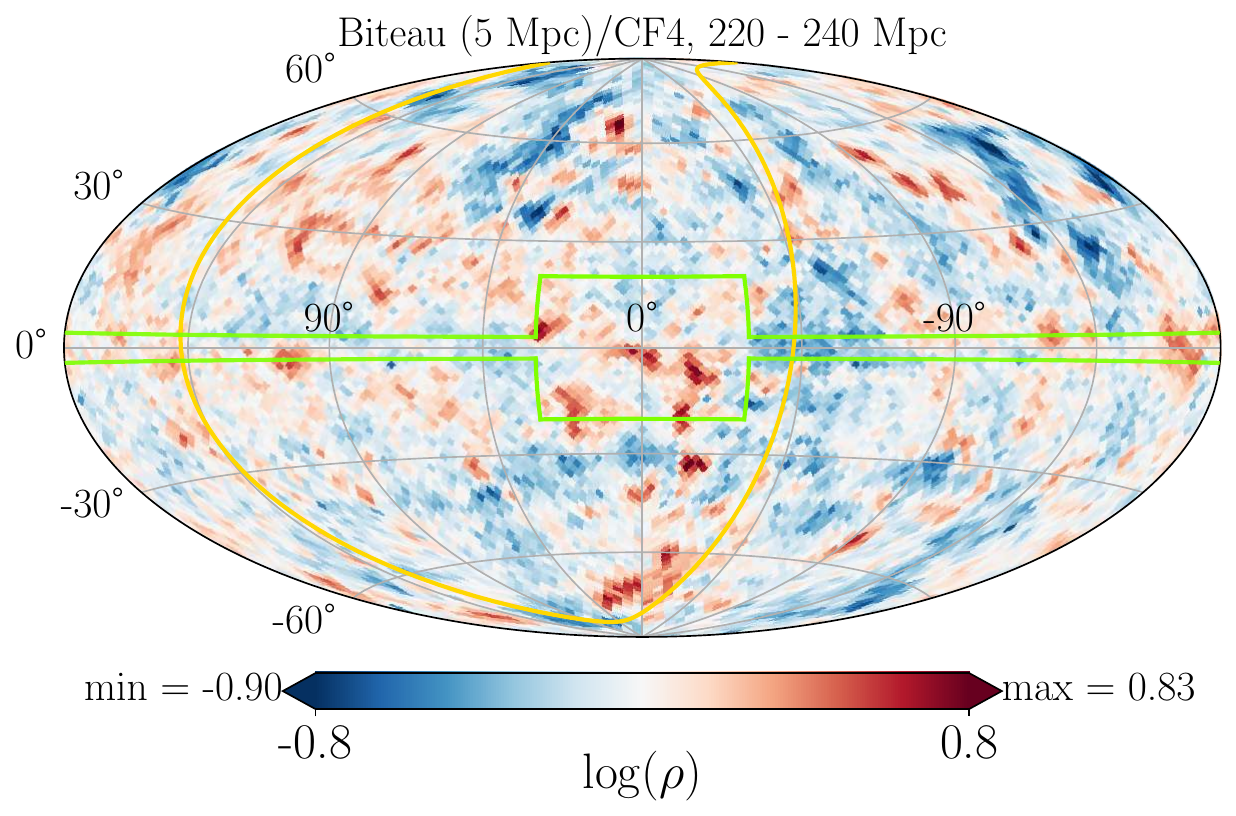}
\includegraphics[width=0.245\textwidth]{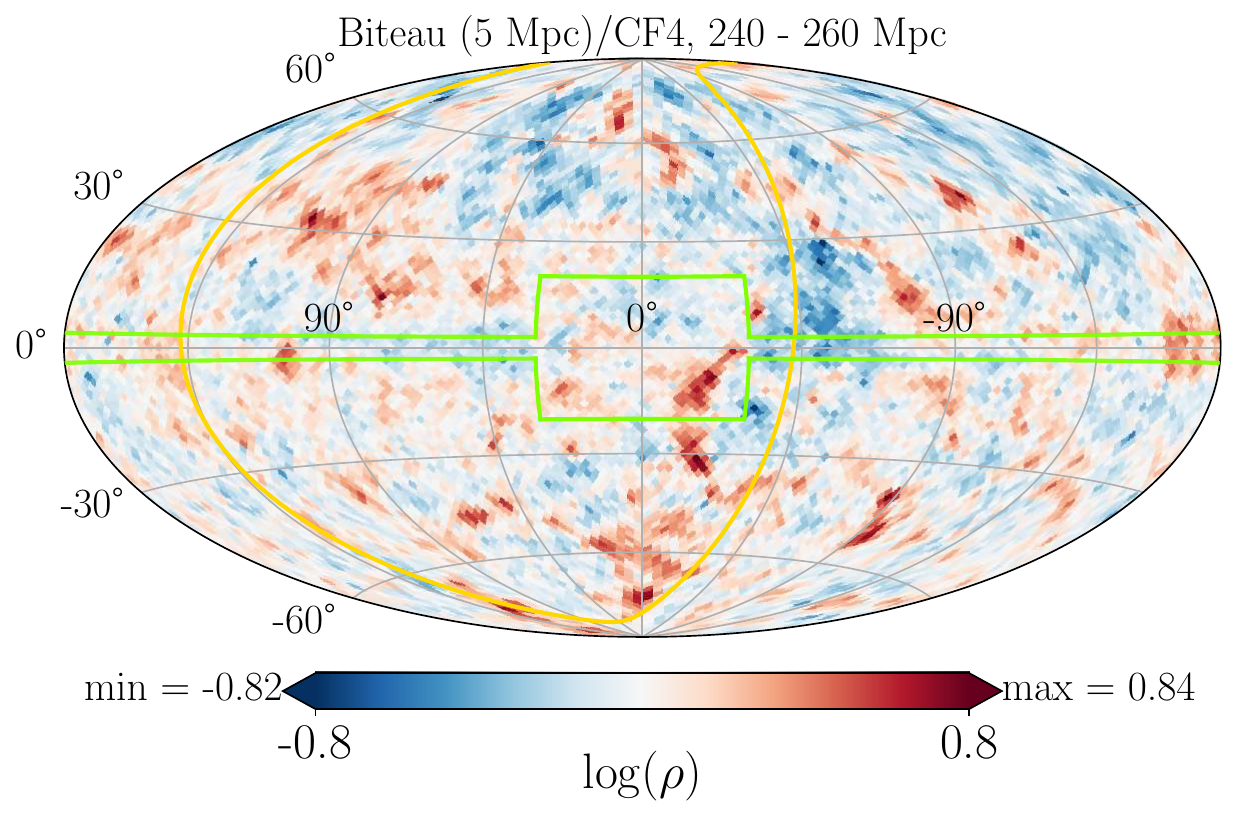}

\includegraphics[width=0.245\textwidth]{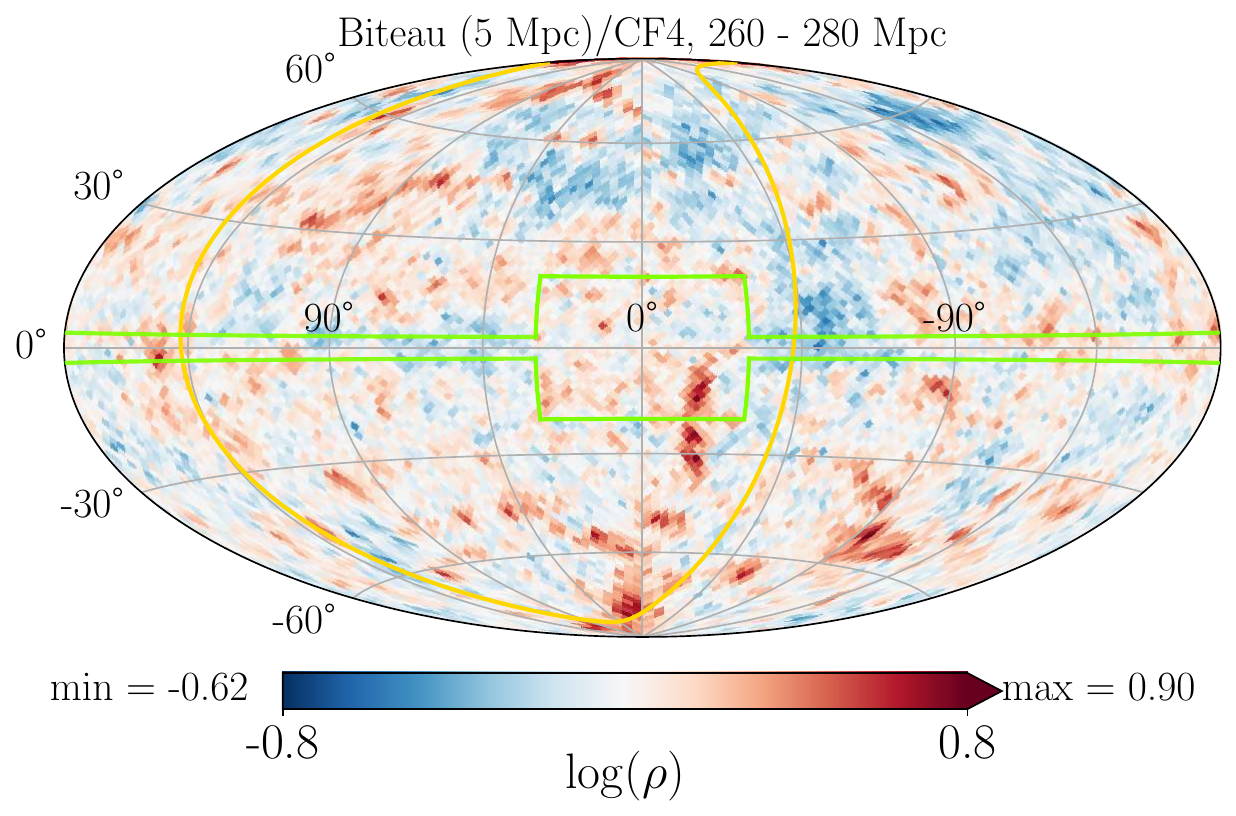}
\includegraphics[width=0.245\textwidth]{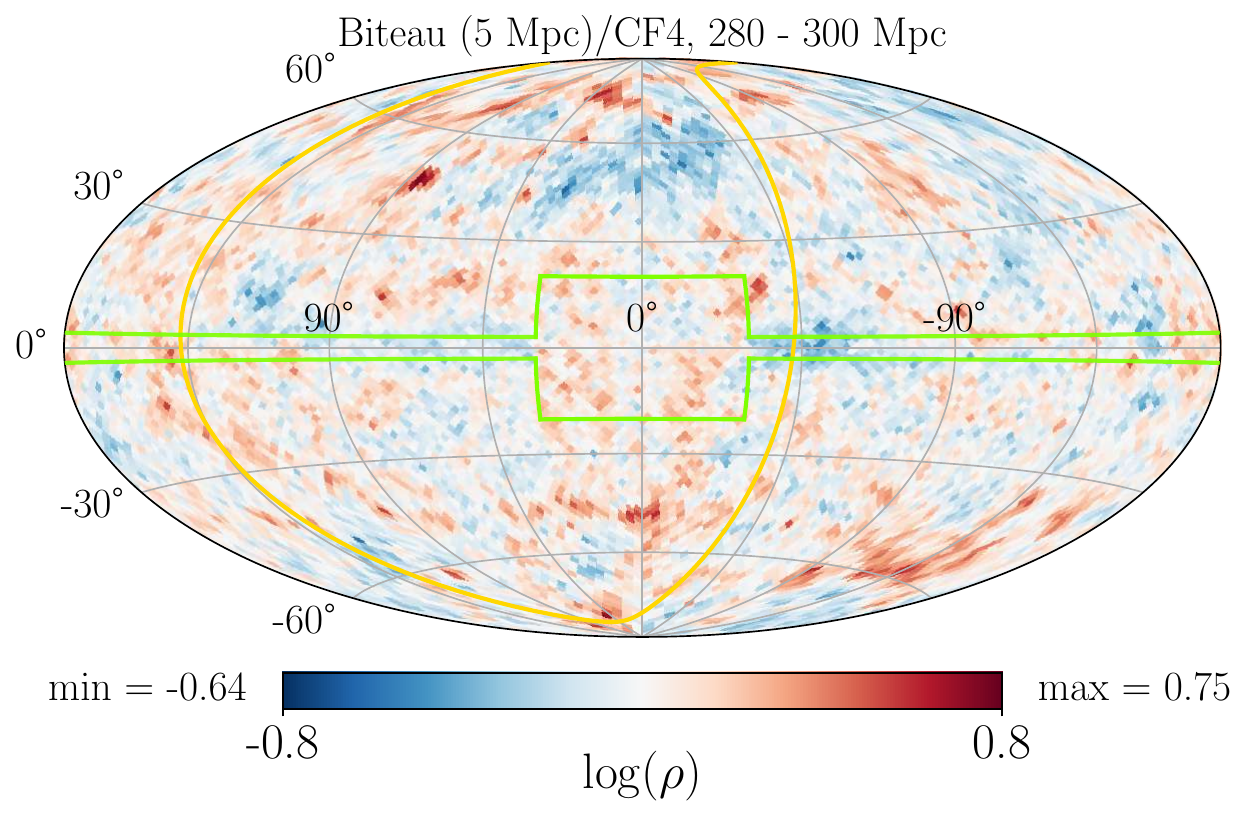}
\includegraphics[width=0.245\textwidth]{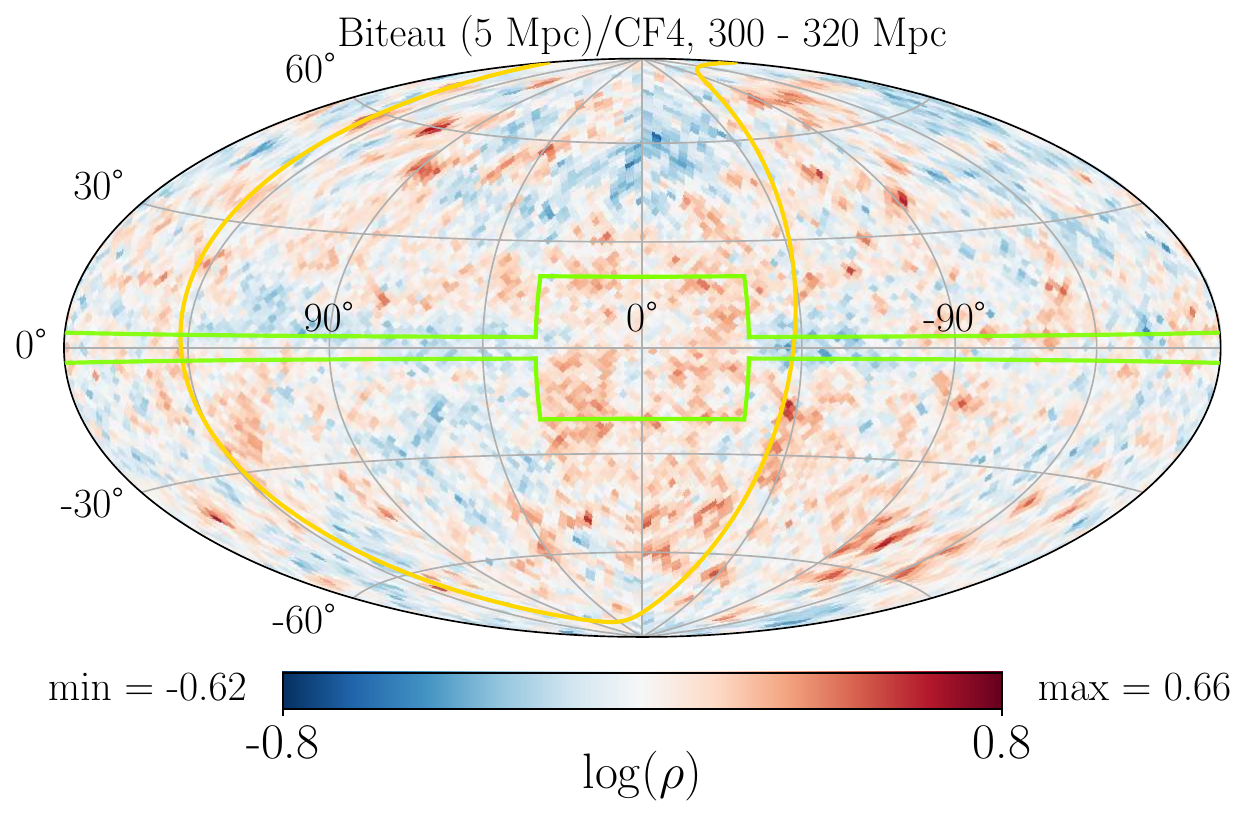}
\includegraphics[width=0.245\textwidth]{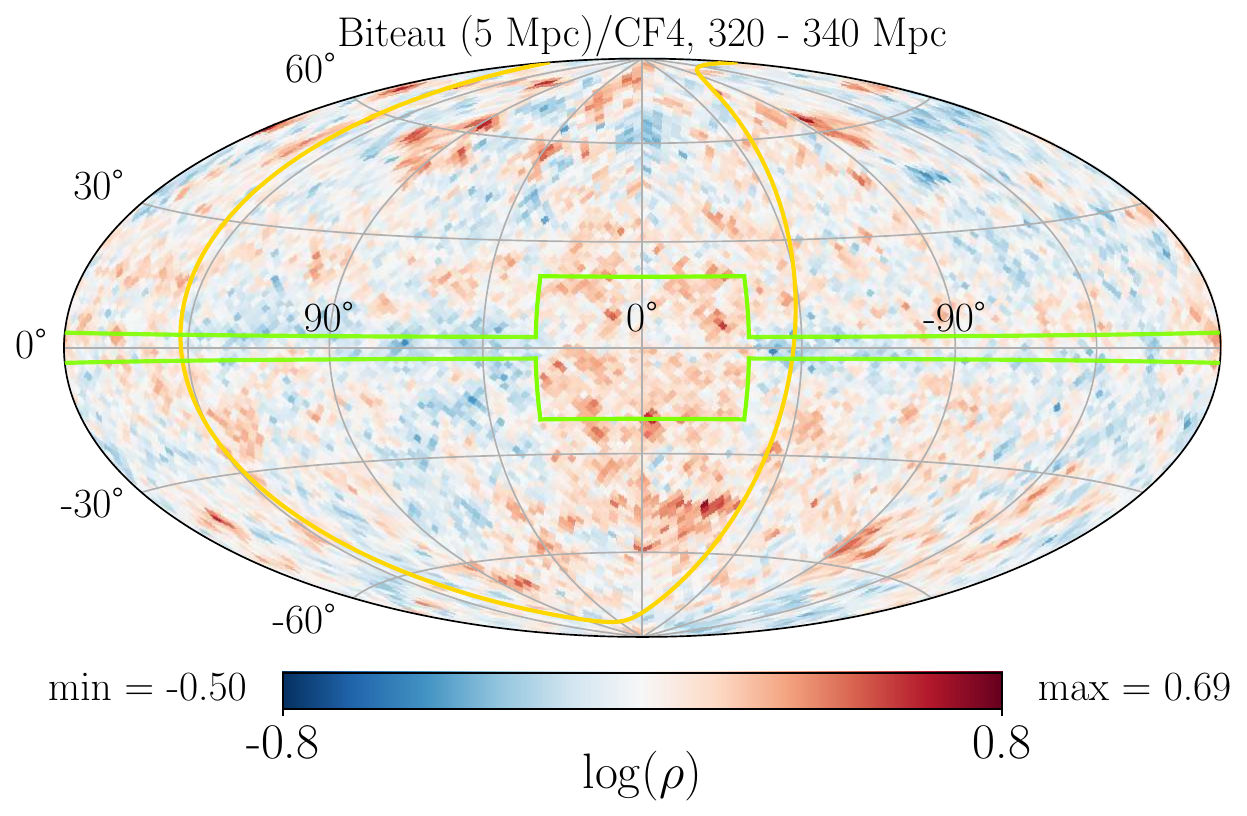}
\caption{Skymaps of the logarithm of the ratio of mass density of Biteau's catalog (5 Mpc) to CosmicFlows4, in 10 Mpc shells out to 220 Mpc and 20 Mpc shells beyond. For the $0-10$ Mpc map, the mass of galaxies in Biteau's catalog is replaced by the angular smoothing ones. Yellow and green lines are as in Fig.~\ref{skymap}.}
\label{skymapratio}
\end{figure*}

\end{CJK*}
\end{document}